\documentclass{aa}  

\usepackage{graphicx}
\usepackage{txfonts}
\usepackage{xcolor}
\usepackage[T1]{fontenc}
\usepackage{ae,aecompl}
\usepackage{bm}		
\usepackage{multirow}
\usepackage{natbib,twoopt}
\bibpunct{(}{)}{;}{a}{}{,}             
\makeatletter
  \newcommandtwoopt{\citeads}[3][][]{\href{http://adsabs.harvard.edu/abs/#3}%
    {\def\hyper@linkstart##1##2{}%
     \let\hyper@linkend\@empty\citealp[#1][#2]{#3}}}
  \newcommandtwoopt{\citepads}[3][][]{\href{http://adsabs.harvard.edu/abs/#3}%
    {\def\hyper@linkstart##1##2{}%
     \let\hyper@linkend\@empty\citep[#1][#2]{#3}}}
  \newcommandtwoopt{\citetads}[3][][]{\href{http://adsabs.harvard.edu/abs/#3}%
    {\def\hyper@linkstart##1##2{}%
     \let\hyper@linkend\@empty\citet[#1][#2]{#3}}}
  \newcommandtwoopt{\citeyearads}[3][][]%
    {\href{http://adsabs.harvard.edu/abs/#3}
    {\def\hyper@linkstart##1##2{}%
     \let\hyper@linkend\@empty\citeyear[#1][#2]{#3}}}
\makeatother

\newcommand{\kms}{\,km\,s$^{-1}$} 
\newcommand{\Msol}{M$_\odot$}
\newcommand{\Rsol}{R$_\odot$}
\newcommand{\Pdot}{$\dot{P}$}

\definecolor{bluegreen}{rgb}{0.0, 0.87, 0.87}
\begin{document}

   \title{On the Effect of Rotation on Populations of Classical Cepheids}

   \subtitle{II. Pulsation Analysis for Metallicities 0.014, 0.006, and 0.002}

   \author{R.I. Anderson\inst{1,2}
   \and H. Saio\inst{3}
   \and S. Ekstr\"om\inst{4}
   \and C. Georgy\inst{4}
   \and G. Meynet\inst{4}
          }

   \institute{Department of Physics \& Astronomy, The Johns Hopkins University, 3400 N Charles St, Baltimore, MD 21218, USA
         \and
             Swiss National Science Foundation Fellow, \email{ria@jhu.edu}
         \and
             Astronomical Institute, Graduate School of Science, Tohoku University, Sendai, Miyagi 980-8578, Japan
         \and
             D\'epartement d'Astronomie, Universit\'e de Gen\`eve, 51 Ch. des Maillettes, 1290 Sauverny, Switzerland
             }

   \date{Received December 22, 2015; Accepted April 18, 2016}

 
\abstract{Classical Cepheid variable stars (from hereon: Cepheids) are high-sensitivity probes of stellar evolution and fundamental tracers of cosmic distances. While rotational mixing significantly affects the evolution of Cepheid progenitors (intermediate-mass stars), the impact of the resulting changes in stellar structure and composition on Cepheids on their pulsational properties is hitherto unknown.

Here we present the first detailed pulsational instability analysis of stellar evolution models that include the effects of rotation, for both fundamental mode and first overtone pulsation.
We employ Geneva evolution models spanning a three-dimensional grid in mass ($1.7 - 15$\,\Msol), metallicity ($Z=0.014$, $0.006$, $0.002$), and rotation (non-rotating, average \& fast rotation). We determine (1) hot and cool instability strip (IS) boundaries taking into account the coupling between convection and pulsation, (2) pulsation periods, and (3) rates of period change. We investigate relations between period and (a) luminosity, (b) age, (c) radius, (d) temperature, (e) rate of period change, (f) mass, (g) the flux-weighted gravity-luminosity relation (FWGLR). We confront all predictions aside from those for age with observations, finding generally excellent agreement.

We tabulate period-luminosity relations (PLRs) for several photometric pass-bands and investigate how the finite IS width, different IS crossings, metallicity, and rotation affect PLRs. We show that a Wesenheit index based on $H$, $V$, and $I$ photometry is expected to have the smallest intrinsic PLR dispersion. We confirm that rotation resolves the Cepheid mass discrepancy. Period-age relations depend significantly on rotation, with rotation leading to older Cepheids, offering a straightforward explanation for evolved stars in binary systems  that cannot be matched by conventional isochrones assuming a single age. We further show that Cepheids obey a tight FWGLR.

Rotation is a fundamental property of stars that has important implications for the study of intermediate-mass stars, intermediate-age clusters, and classical Cepheid variable stars.}

   \keywords{Stars: variables: Cepheids -- supergiants -- Stars: oscillations -- Stars: evolution  -- Stars: rotation -- distance scale}

   \maketitle
%


\section{Introduction}
Classical Cepheid variable stars (from hereon: Cepheids) are crucial objects for stellar astrophysics and the extragalactic distance scale. They are evolved intermediate-mass stars ($M_{\rm{ini}} \sim 3-12$ \Msol) that occupy a well-defined region in the Hertzsprung-Russell diagram (HRD) called the classical instability strip (IS). While their progenitor (B-type) stars exhibit fast Main Sequence (MS) rotation \citep{2010ApJ...722..605H}, Cepheids have very expanded envelopes with radii between approx. $20-200$\,\Rsol , i.e., their surface rotation is low (a typical $v\sin{i}$ is around $10$\kms). 
The observed low surface rotation leads to the common misconception that rotation be relatively unimportant for Cepheids. On the contrary, rotation significantly affects the entire evolutionary path of a star \citep{2000ARA&A..38..143M,2009pfer.book.....M,2012A&A...537A.146E,2013A&A...553A..24G}, with consequences that manifest themselves particularly clearly in the later stages of stellar evolution \citep{2014A&A...564A.100A}. 

Cepheids famously obey a tight statistical relation between their pulsation period and luminosity \citep{1908AnHar..60...87L,1912HarCi.173....1L}, rendering them the most precise standard candles available for calibrating Supernova-based direct measurements of the local Hubble constant \citep[e.g.][]{1998AJ....116.1009R,2001ApJ...553...47F,2011ApJ...730..119R}. This period-luminosity relation (PLR) has been the focus of intense research and it is now understood that moving to near IR wavelengths and/or using Wesenheit relations \citep{1982ApJ...253..575M} has the benefit of reducing observed PLR scatter due to interstellar extinction as well as the intrinsic width of the IS. In \citet{2014A&A...564A.100A}, we argued that rotation may lead to some level of dispersion in the PLR. However, this contribution had thus far not been quantified due to the unavailability of pulsation periods determined based on models with rotation.

Cepheids are excellent targets for testing the effect of rotation on intermediate-mass stars in general. Their location on blue loops in the Hertzsprung-Russell diagram (HRD) avoids the typical confusions of placing red giants on isochrones and means that any change in model input physics will affect predictions more strongly than for other evolutionary phases. In addition, the observable evolution of pulsation periods provides a direct measurement of evolutionary timescales along the blue loop. 
Famously dubbed ``magnifying glasses of stellar evolution'' \citep{1994sse..book.....K}, Cepheids thus provide crucial high-sensitivity tests of stellar models \citep{2015MNRAS.447.2951W}. An important recent discussion has centered around the remaining Cepheid mass discrepancy \citep[e.g.][]{1968QJRAS...9...13C,1969MNRAS.144..461S,1969MNRAS.144..485S,1969MNRAS.144..511S,2006MmSAI..77..207B,2008ApJ...677..483K}, which is defined by systematically overestimated masses inferred from stellar evolution models. Key attempts at resolving this discrepancy included either increasing luminosity at fixed mass by enhancing convective core overshooting 
\citep[e.g.][]{2001ApJ...563..319B,2012ApJ...749..108P} or removing mass at fixed luminosity via enhanced mass loss \citep{2008ApJ...684..569N}. Recently, \citet{2014A&A...564A.100A} showed that a combination of rotational mixing and correct identification of the IS crossing number offers a favorable explanation for this discrepancy. 

Thanks to a period-age relationship, Cepheids are versatile tools for dating star formation events in the Galactic nuclear bulge \citep[e.g.][]{2011Natur.477..188M,2015ApJ...812L..29D}
as well as in other galaxies \citep{2015ApJ...813...31S}, and for tracing Galactic structure 
\citep[e.g.][]{2014Natur.509..342F}. However, dating intermediate-age stellar populations has recently revealed some shortcomings in how ages are inferred from stellar evolution models. For instance, some authors have argued for multiple stellar populations in intermediate-age open clusters in the LMC \citep{2007MNRAS.379..151M,2008ApJ...681L..17M,2009AJ....137.4988G,2011ApJ...737....3G}, in analogy with multiple populations observed in globular clusters. On the other hand, rotation-related effects have been shown to be highly successful in explaining {\it apparent} age spreads in such clusters as a combination of different main sequence lifetimes for stars of different initial rotation rate but identical mass and metallicity \citep{2009MNRAS.398L..11B,2015MNRAS.453.2637D}, and inclination effects such as gravity darkening \citep{2011A&A...533A..43E,2015ApJ...807...25B,2015MNRAS.453.2070N}. As another example, different initial rotation velocities may provide a straightforward explanation for some cases of binary stars that do not both fit on a single isochrone 
\citep[e.g.][]{2015MNRAS.451..651S}. 

Given the diverse effects that rotation can have on individual stars and binaries, as well as entire populations of stars, it is important to test and calibrate stellar models that include treatment of rotational effects. Following our exploratory work on the effect of rotation on stars in the IS \citep{2014A&A...564A.100A}, we here present a detailed pulsational instability analysis of the Geneva stellar evolution models for the mass range $1.7$ to $15$\,\Msol, for three different initial rotation rates (non-rotating, typical, and fast), and for metallicities representative of the Sun, the LMC, and the SMC. 

Following a concise overview of the main model characteristics in Sect.\,\ref{sec:analysis} and the applied method of pulsational instability analysis (Sect.\,\ref{sec:ana:pulsation}), we present our results in Sect.,\,\ref{sec:results}. We present the boundaries of the IS in Sect.\,\ref{sec:res:IS} and from thereon consider all models that lie within these boundaries as Cepheids. We determine PLRs in Sect.\,\ref{sec:res:PLRs} and construct period-age relations in  Sect.\,\ref{sec:res:age}. We further investigate and compare our predictions to empirical results from the literature for a series of relations with period, including period-temperature and period-radius relations (Sect.\,\ref{sec:res:radius}), and rates of period change (Sect.\,\ref{sec:res:pdot}). In Sect.\,\ref{sec:discussion}, we revisit the mass discrepancy (Sect.\,\ref{sec:MassDiscrepancy}), discuss the relative impact of the intrinsic IS width, different crossing numbers, metallicity, and rotation on PLRs (Sect.\,\ref{sec:PLRdependence}) and discuss the applicability of flux-weighted gravity luminosity relations (FWGLR) for Cepheids (Sect.\,\ref{sec:FWGLR}). We summarize our results and conclude in Sect.\,\ref{sec:conclusions}. Additional information is provided in the electronic appendix available online. We note that throughout of this paper, we use 10-base logarithms, i.e., $\log$ means $\log_{10}$.  

\section{Analysis}\label{sec:analysis}
Here we briefly recall the basic properties of the single star (Geneva) models used as basis for the pulsation analysis as well as the basics of the pulsation analysis carried out. We point out relevant references for readers interested in more detail.

\subsection{Geneva Stellar Evolution Models}\label{sec:ana:models}\label{sec:evol}

The models used for the pulsational instability analysis (Sect.\,\ref{sec:ana:pulsation}) are presented in detail by \citet{2013A&A...553A..24G}. The grid employed here spans masses of $1.7$, $2$, $2.5$, $3$, $4$, $5$, $7$, $9$, $12$, and $15$\,\Msol.  We here focus on three of the nine available initial rotation rates, namely : $\omega=\Omega/\Omega_{\rm crit}=0.0$, $0.5$, and $0.9$,  where $\Omega_{\rm crit}$ refers to the first critical angular velocity \citep[we adopt the nomenclature as in][]{2013A&A...553A..24G}. In the remainder of the paper, we refer to these initial rotation rates as slow, average, and fast rotating models. Details about these models and the input physics were presented by \citet{2012A&A...537A.146E} and \citet{2013A&A...553A..24G}. Here, we recall the basics most relevant to the current investigation.

The models are calibrated at Solar metallicity on features that are not directly related to this work \citep[such as initial abundance ratios that reproduce the present-age Sun, as well as mixing length parameters and mass-loss rates on the main sequence, see][]{2012A&A...537A.146E} and no fine-tuning or re-calibration was performed here to improve agreement with Cepheid observations.

The models assume differential rotation based on prescriptions by \citet{1992A&A...265..115Z} for $D_{\rm h}$ and \citet{1998A&A...334.1000M} for $D_{\rm shear}$. The boundaries of the convective core are extended with a mild overshooting of $d_{\rm over}/H_P=0.1$. The initial composition is set as follows: the initial helium abundance is computed assuming a linear variation as a function of metallicity so that $Y(Z) = Y_\text{P} + \frac{\Delta Y}{\Delta Z}Z$. $Y_\text{P}$ is $0.2484$ \citep{2003PhLB..567..227C}, and $\frac{\Delta Y}{\Delta Z} = 1.257$. $Z$ and $Y$ being known, the hydrogen abundance is therefore $X=1-Y-Z$. Among the metals, a solar mixture is assumed \citep[mostly from][see \citet{2012A&A...537A.146E} for more details.]{2005ASPC..336...25A}. Opacities were computed for this particular mixture with the OPAL online tool\footnote{\footnotesize{\url{http://opalopacity.llnl.gov/}}} \citep[see][]{1996ApJ...464..943I} and are complemented at low temperature with opacity computations from \citet{2005ApJ...623..585F}. No adjustments to opacities are made when the mixture departures from the solar mixture. Standard recipes are used to account for mass loss during different evolutionary phases; these are presented in Tab.\,\ref{tab:masslossrates}. Certain models that reach critical rotation speeds shed this excess angular momentum via mechanical mass loss \citep[cf.][Section 2.2]{2013A&A...553A..24G}; this applies to higher-mass models with fast initial rotation ($\omega=0.9$) during their MS evolution.

\begin{table*}
\centering
\begin{tabular}{c|ccc}
phase / mass range & $[1.7-5]\,\text{M}_\odot$ & $]5-7]\,\text{M}_\odot$ & $]7-15]\,\text{M}_\odot$\\\hline
\multirow{2}{*}{MS} & \rule{0mm}{4mm}- & - & \citet{1988AAS...72..259D} \\
 & - & - & $\alpha = 0.5$\\\hline
\rule{0mm}{4mm}Advanced stages & \citet{1975psae.book..229R} & \citet{1975psae.book..229R} & \citet{1988AAS...72..259D}\\
($3.7 \leq \log(T_\text{eff}) < 3.8$) & $\eta = 0.5$, $\alpha = 0.$ & $\eta = 0.6$, $\alpha = 0.$ & $\alpha = 0.5$\\\hline
\rule{0mm}{4mm}Advanced stages & \citet{1975psae.book..229R} & \citet{1975psae.book..229R} & \citet{2001ASSL..264..215C}\\
($\log(T_\text{eff}) < 3.7$) & $\eta = 0.5$, $\alpha = 0.$ & $\eta = 0.6$, $\alpha = 0.$ & $\alpha = 0.$
\end{tabular}
\caption{Mass-loss rate prescriptions used during the different stellar evolution phases. The $\eta$ value for the \citet{1975psae.book..229R} prescription is a multiplying factor applied to the original formulation of the prescription. The $\alpha$ value indicates how the law is scaled with respect to the metallicity: $\dot{M}(Z) = \dot{M}(\text{Z}_\odot)\left(\frac{Z}{\text{Z}_\odot}\right)^\alpha$.}
\label{tab:masslossrates}
\end{table*}

For the purpose of the present instability analysis, we recomputed evolutionary phases when models cross the Hertzsprung gap with a slightly modified version of the code. This was necessary to make available all the quantities required for the pulsational instability analysis throughout the entire star.

\begin{figure*}
\centering
\includegraphics[scale=0.31]{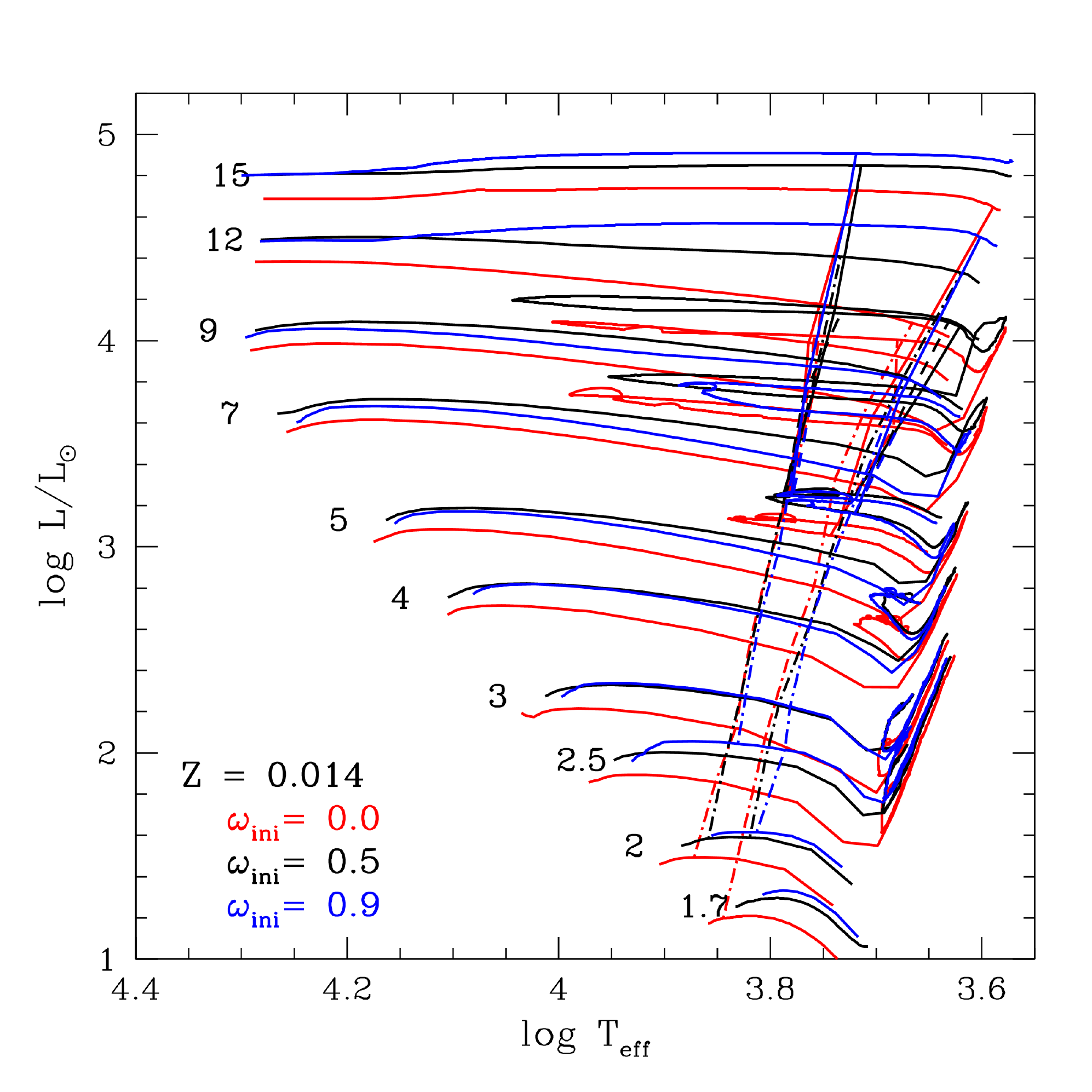}
\hspace{-0.025\textwidth}
\includegraphics[scale=0.31]{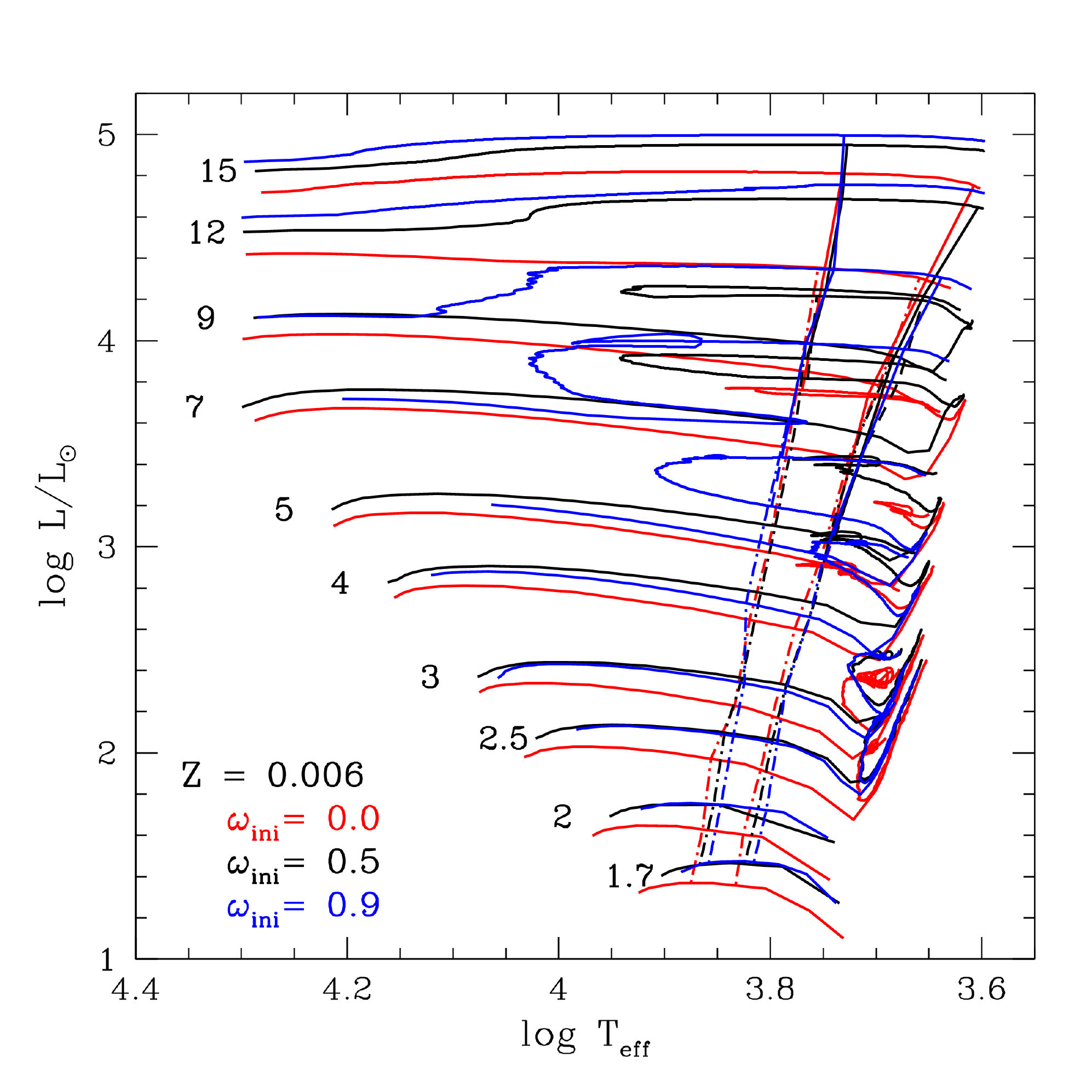}
\hspace{-0.025\textwidth}
\includegraphics[scale=0.31]{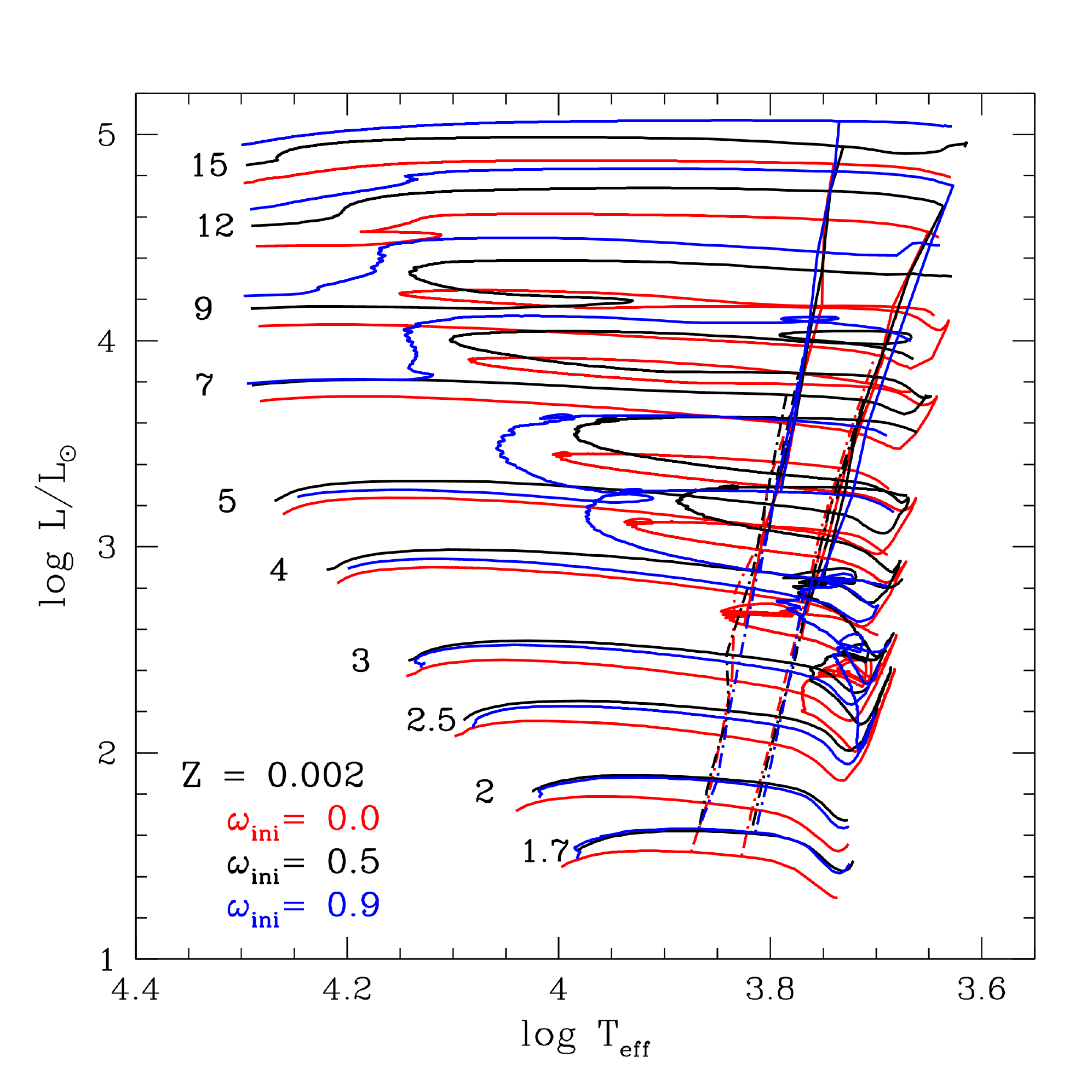}
\caption{Post-MS evolutionary tracks used in this work \citep{2013A&A...553A..24G}, with overplotted IS  boundaries of radial fundamental pulsation determined for the different assumed initial rotation rates. The number at the beginning of each track indicates initial mass in solar units.
Assumed initial rotation rates are color-coded; red lines are for models without rotation ($\omega_{\rm ini}=0$), black lines for models with an initial rotation of 50\% critical rate ($\omega_{\rm ini}=0.5$), and blue lines for an initial rotation of 90\% critical rate.
IS boundaries shown as dash-dotted lines indicate first crossing models, dashed lines the second crossing, and solid lines the third crossing.}
\label{fig:hrd}
\end{figure*}

Figure~\ref{fig:hrd} shows the post MS parts of evolutionary tracks up to the end of core-helium burning, as well as theoretical blue (hot) and red (cool) boundaries of the Cepheid IS. 
As discussed in \citet{2014A&A...564A.100A}, intermediate-mass stars ($10 M_\odot \gtrsim M \gtrsim 3 M_\odot$, depending on metallicity and initial rotation rate) cross the IS rapidly for the first time during core contraction phase after turning off from the MS and before core helium ignition; this is called the first crossing. Second and third crossings occur during the core-helium burning phase during the so-called blue loop evolution. Since the evolutionary timescale of the first crossing is much faster than that of subsequent crossings, \emph{nearly all observed Cepheids are generally considered to be evolving along blue loops} and undergoing core He burning. 

We note that some models exhibit multiple loops near the tip of the blue loop evolution. Such ``He-spikes'' lead to formal 4th and 5th crossings of the IS that are unlikely to be real. These He-spikes are linked with the uncertainty regarding the location of the convective core's boundary, and the mixing occurring through this interface. With our current implementation of convection \citep[instantaneous penetrative overshoot and Schwarzschild criterion, see e.g.][]{2016arXiv160101572C}, a small displacement of the boundary can bring a significant amount of fresh helium into the core, particularly when the central helium mass fraction becomes low. This provides a strong boost to the core's energy generation, changing luminosity and the radiative gradient rapidly and pushing outward the convective core's boundary \citep[see also][]{1985ApJ...296..204C}. It is so far not clear whether this feature is real or not \citep[see also the discussion in][]{2015MNRAS.452..123C}. The main relevance of these ``He-spikes'' is that they can significantly alter the lifetime of 2nd or 3rd crossings, implying systematic uncertainties for population synthesis and arguments based on the frequency of Cepheids observed with negative or positive rates of period change. 

Figure\,\ref{fig:hrd} illustrates the different evolutionary paths of Solar, LMC, and SMC metallicity models, illustrating the how the extent of blue loops changes with mass and metallicity. In general, the extension of the blue loop to higher temperatures increases with mass until it eventually disappears. Towards lower masses, blue loops extend less and less blueward. At a certain point,  blue loops enter the IS near the red edge and turn back toward the red giant branch before passing the region where the blue IS boundary is to be expected based on extrapolation from higher-mass models or the analysis first IS crossings at similar luminosity. Consequently, we cannot formally determine a blue boundary from such models, although the red boundary remains accessible. At even lower mass, blue loops become even shorter and do not enter the IS at all, although predictions for first IS crossings can still be made.  Blue loops are very sensitive to input physics, such as convective core overshooting, which tends to reduce the extent of blue loops and thereby increases the minimal mass of models entering the IS \citep[e.g.][Fig. 1]{2014A&A...564A.100A}. Metallicity is another defining factor for the extension of blue loops: the lower the metal content (within the range of models considered here), the farther blue loops extend for low-mass models. The consequences of these effects on the predicted period range are discussed in Sec.\,\ref{sec:periodrange}.

Contrary to the lower mass-stars discussed above, stars with $M_{\rm{ini}} \gtrsim 10$M$_\odot$ (this mass limit becomes smaller for lower metallicity and higher rotation rates) cross the instability strip only once, and {\it after} igniting the core-helium burning. Although this is technically a first IS crossing, we here refer to this evolutionary phase as third crossings, since the evolutionary timescales are those of Cepheids in the process of core He burning.
As a result, no clear-cut maximum luminosity or maximum period of Cepheids is expected. However, in practice such stars will be rarely observed due to their high mass and short lifetime as Cepheids (the timescale of core He-burning accelerates with mass).
The onset of core He burning in such massive stars leads to luminosity increases around $\log T_{\rm eff} \sim 4.1$, cf. Fig.\,\ref{fig:hrd}.

\subsection{Pulsation Analysis}\label{sec:ana:pulsation}
In this paper, we perform linear non-adiabatic radial pulsation analysis for evolutionary models to determine the stability of pulsation modes and their periods. The method described in \citet{1983ApJ...265..982S} was extended to include the effect of pulsation-convection coupling, which is required for predicting the red edge of the Cepheid instability strip (IS).

Our linear non-adiabatic analysis solves linearized energy conservation and energy flux equations simultaneously with linearized momentum and mass conservation equations. The latter equations, common to adiabatic analysis, determine pulsation period, while the former equations determine the stability (i.e., excitation or damping) of pulsation. These thermal equations take into account perturbation (pulsational variation) of local luminosity, $\delta L_r$, which consists of radiative and convective luminosity,\,i.e., $\delta L_r = \delta L_{\rm rad}+ \delta L_{\rm conv}$. The equation for $\delta L_{\rm rad}$ is obtained by linearizing the diffusion equation of radiation. Obtaining $\delta L_{\rm conv}$, however, requires a time-dependent convection (TDC) theory. In this paper, we treat the pulsation-convection coupling by deriving $\delta L_{\rm conv}$ from TDC theory as described by \citet{1967PASJ...19..140U} and extended by \citet{2005A&A...434.1055G}. This allows us to predict the red edge of the instability strip. For comparison, we also determine the blue IS edge by neglecting $\delta L_{\rm conv}$ and refer to this as the frozen convection (FC) assumption. Under this assumption, the red IS edge cannot be predicted. Throughout this paper, we use IS boundaries predicted by taking into account pulsation-convection coupling. 

We note that turbulent pressure is not included in the evolutionary models nor in our pulsation analysis. Convection velocities from MLT indicate that the ratio of turbulent to thermal pressure in the second He ionization zone ($\log T \sim 4.6$) is always very small ($\lesssim 0.01$) in the Cepheid models. Although the ratio can be larger than $\sim 0.1$ in the H and first He ionization zone ($3.9 \lesssim \log T \lesssim 4.2$), such layers are confined into a thin zone of less than $\sim 1$\,\% of radius just below the photosphere. This strongly suggests that turbulent pressure hardly affects the properties of Cepheids.

It is known that TDC calculations based on mixing-length theory yield rapid spatial oscillations of the flux perturbation near the bottom of the convection zone \citep[e.g.][]{1979ApJ...234..232B,1980A&A....84..304G,2005A&A...434.1055G} if the convection turn-over-time associated with the mixing length is longer than the pulsation period. These spatial oscillations are suppressed based on the argument about the possible effect of small-scale eddies \citep{1980ApJ...240..685S}, which is technically identical to setting the parameter $\beta$ of \citet{2005A&A...434.1055G} to unity. 

We thus determine the stability of all radial pulsation modes and calculate their corresponding periods. The most relevant model parameters at the instability boundaries are listed in Appendix\,\ref{app:tables}, cf. Tabs.\,\ref{tab:models_z14}, \ref{tab:models_z06}, and \ref{tab:models_z02} for fundamental mode pulsation, and in Tabs.\,\ref{tab:models_z14_FO}, \ref{tab:models_z06_FO}, and \ref{tab:models_z02_FO} for first overtone pulsation.

Although the models that form the basis of the pulsational instability include the evolutionary effects of rotation, we do not account for mechanical effects of rotation on the radial pulsation itself. This is justified, since radial pulsation is not affected by centrifugal deformation \citep[e.g.][]{1981ApJ...244..299S}, and since the leading effects of rotation on our predictions are already accounted for by the evolutionary models (change in luminosity and radius). Any remaining effects would arise from second-order Coriolis force terms, which we estimate to be approximately $(P/P_{\rm rot})^2$ with pulsation period $P$ and rotation period $P_{\rm rot}$, leading to an impact of at most $\sim 1$\% for some extreme cases and much smaller for most other cases.

\section{Results}\label{sec:results}
This section presents the results from our pulsational instability analysis and compares these to empirical data. In the following, we usually treat separately Cepheid models of different $\omega$ (initial rotation), $Z$, crossing number, and those on the blue and red edge of the IS. In certain cases, it is useful to fit together models with different rotation or on different IS crossings, and we refer to the resulting relations as rotation or crossing \emph{averages}. We focus primarily on 2nd and 3rd crossings, since first crossings are rarely observed. Unless otherwise stated, all physical parameters of Cepheids considered are pulsation-averaged values, i.e., not pulsation phase-dependent.

To facilitate comparisons with observations, we translate predicted luminosities and temperatures to magnitudes in photometric passbands using the calibrations given by \citet{2011ApJS..193....1W} and $M_{\rm{bol,\odot}} = 4.74$\,mag. To this end, we employ photometric filters $B$, $V$, $I$, and $H$, where $B$, $V$, and $I$ are in the Johnson-Cousins system, and $H$ is close to the near-IR system of \citet{1988PASP..100.1134B}, which is very close to the F160W filter on the {\it Hubble Space Telescope} WFC3 \citep{2001AJ....121.2851C,2011wfc..rept...15R}. 

We compare predictions from our Solar metallicity models ($Z=0.014$) to Galactic Cepheids, and those with $Z=0.006$ and $Z=0.002$ to LMC and SMC Cepheids, respectively. For Galactic Cepheids, we rely on stars with directly measured absolute magnitudes \citep[e.g.][]{2007AJ....133.1810B,2007MNRAS.379..723V,2011A&A...534A..94S}. For Magellanic Cloud Cepheids, we adopt distance moduli determined from eclipsing binaries, namely $18.494$ \citep{2013Natur.495...76P} for the LMC and $18.965$ \citep{2014ApJ...780...59G,2016ApJ...816...49S} as an average distance for the SMC. 

To enable precise comparisons with observations, we employ Wesenheit relations \citep{1982ApJ...253..575M} from the literature due to their insensitivity to extinction. Specifically, we compute  
\begin{equation}
W_{\rm{VI}} = I - 1.55\cdot(V-I)
\label{eq:W_VI}
\end{equation}
and
\begin{equation}
W_{\rm{H,VI}} = H - 0.41\cdot(V-I)
\label{eq:W_HVI}
\end{equation}
following \citet{2008AcA....58..163S} and \citet{2011ApJ...730..119R}, thus assuming a \citet{1989ApJ...345..245C} reddening law with slope $R_V = 3.1$. $W_{H,VI}$ is of particular interest for the extragalactic distance scale and in this context is usually based on the {\it Hubble Space Telescope} WFC3 filters F555W, F814W (both UVIS\footnote{\url{http://www.stsci.edu/hst/wfc3/documents/handbooks/currentIHB/c06\_uvis06.html}}), and F160W (IR channel\footnote{\url{http://www.stsci.edu/hst/wfc3/documents/handbooks/currentIHB/c07\_ir06.html}}).

\subsection{Blue and Red Edges of Classical Instability Strip}\label{sec:res:IS}
\begin{figure*}
\centering
\includegraphics[scale=0.42]{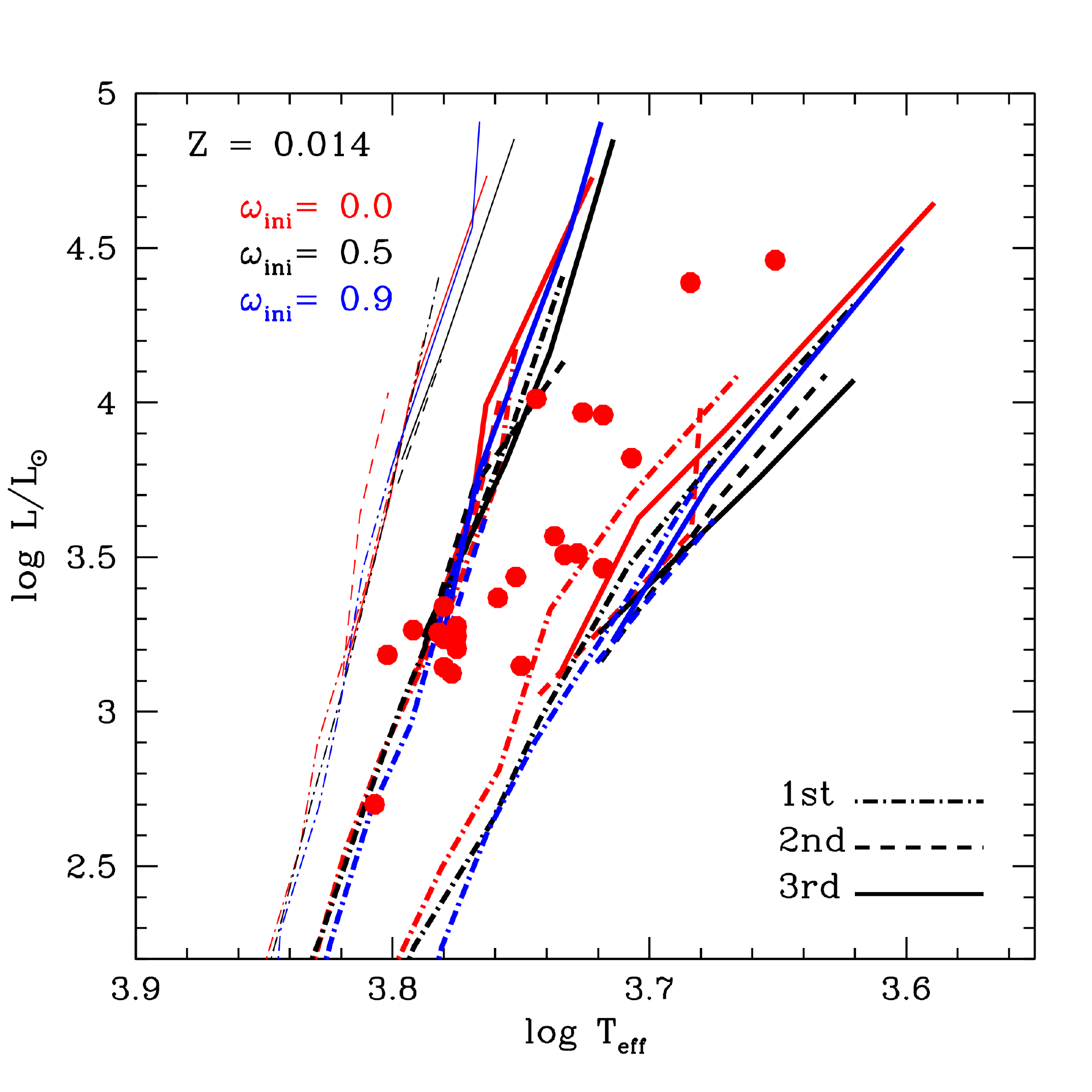}
\includegraphics[scale=0.42]{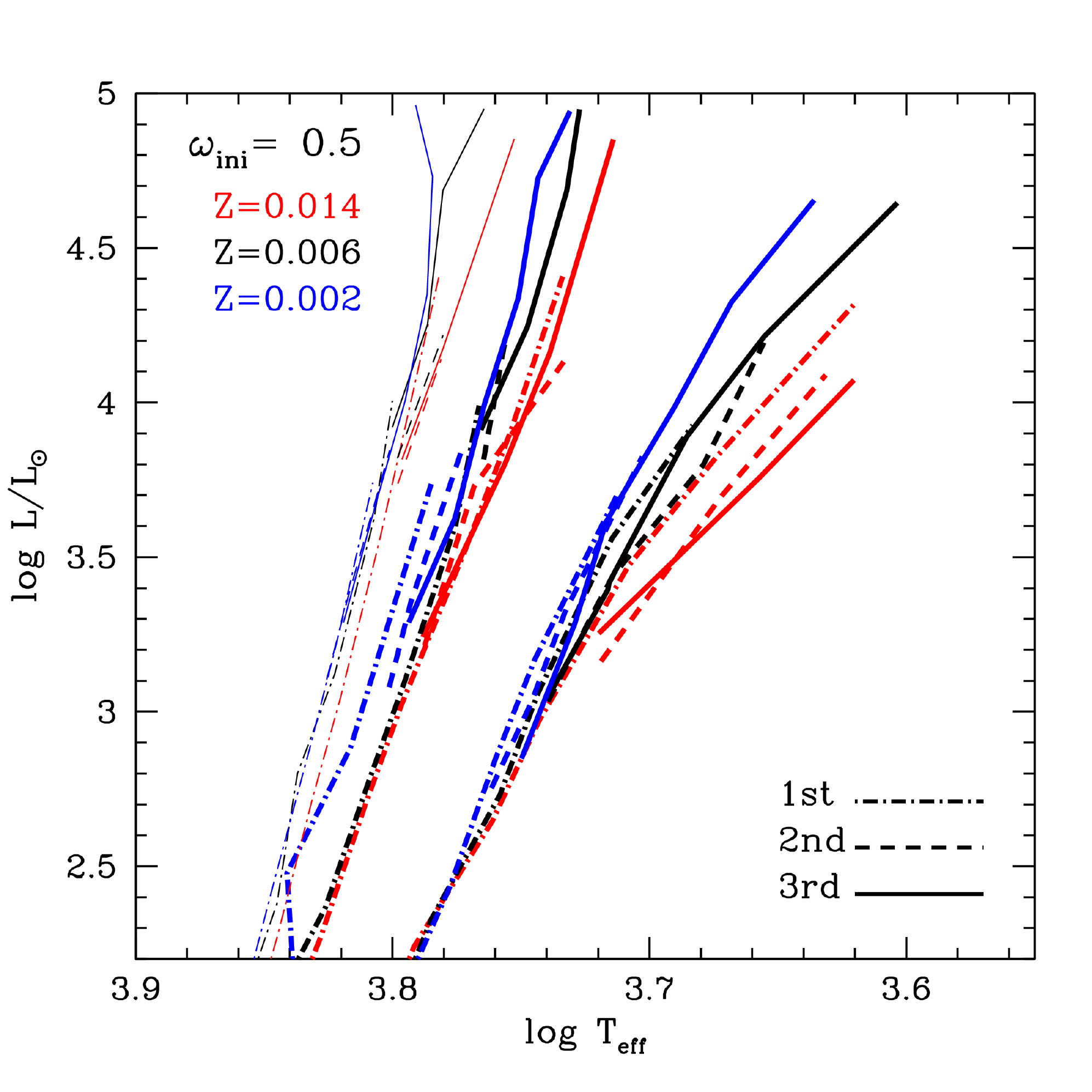}
\caption{Locations of the blue and red edges for fundamental mode pulsation. The left panel is for models of the solar metallicity with various initial rotation  rates, while the right panel is for models of an initial rotation of $\omega_{\rm ini}=0.5$ with various metal abundances. The thinner lines show the locations of  blue edges obtained with the frozen convection assumption, while the thicker lines show blue and red edges obtained by taking into account the perturbation of convection flux.
The left panel also shows the positions of some Galactic Cepheids whose parameters are taken from \citet{2002AJ....124.2931T}, \citet{2012AJ....144..187T,2013ApJ...772L..10T}, and \citet{2013OAP....26..115T}.}
\label{fig:res:FBE}
\end{figure*}

We show IS boundaries for fundamental mode pulsation predicted for different metallicities and initial rotation rates in Fig.\,\ref{fig:res:FBE}. 
We indicate locations of blue and red IS boundaries obtained by taking into account pulsation-convection interaction with thicker lines. Blue edges obtained using the `frozen convection' (FC) assumption (cf. Sec.\,\ref{sec:ana:pulsation} are shown as thinner lines. No red-edge is obtained in the analysis with the FC assumption, since the red boundary occurs due to pulsation-convection coupling.

\begin{figure*}
\centering
\includegraphics[scale=0.29]{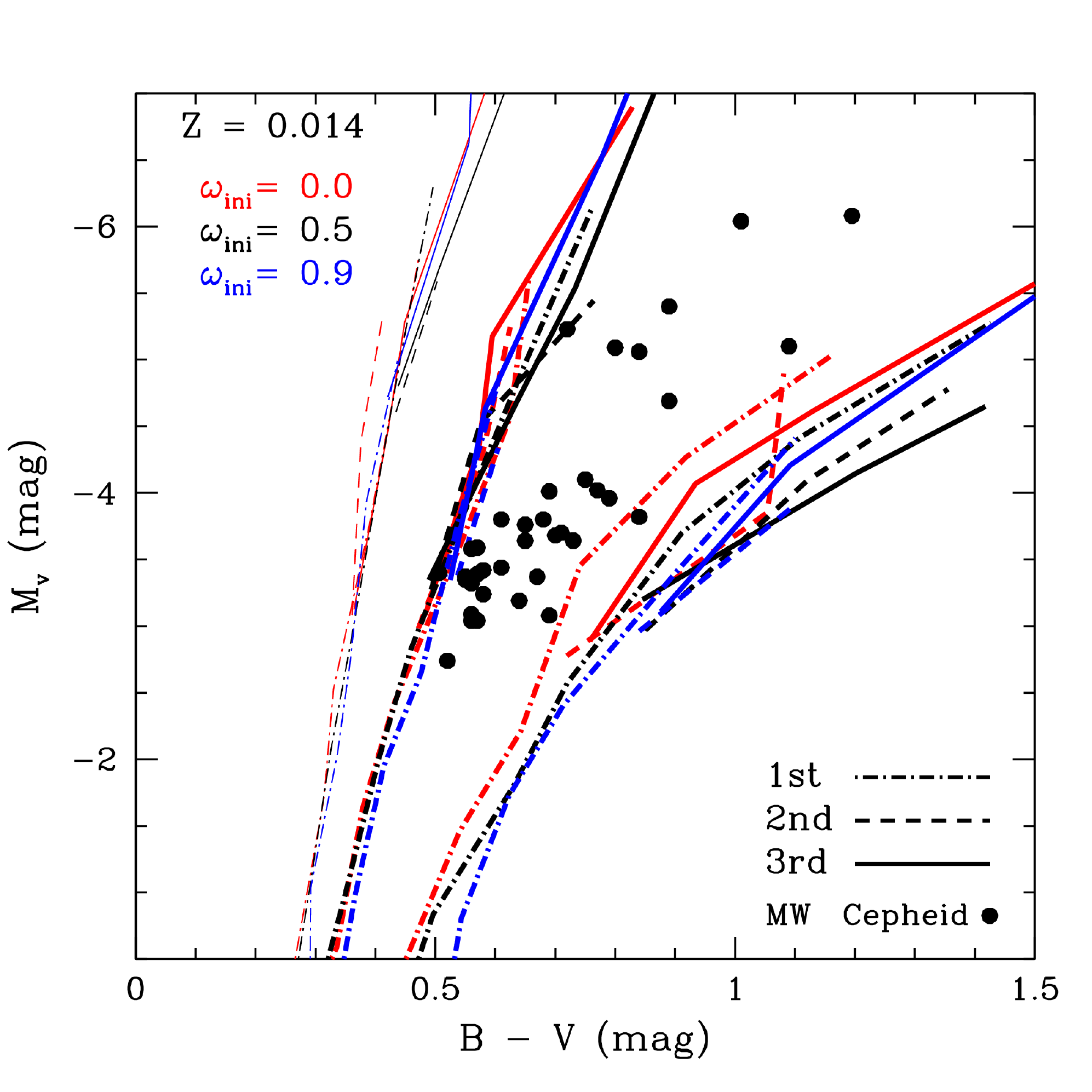}
\includegraphics[scale=0.29]{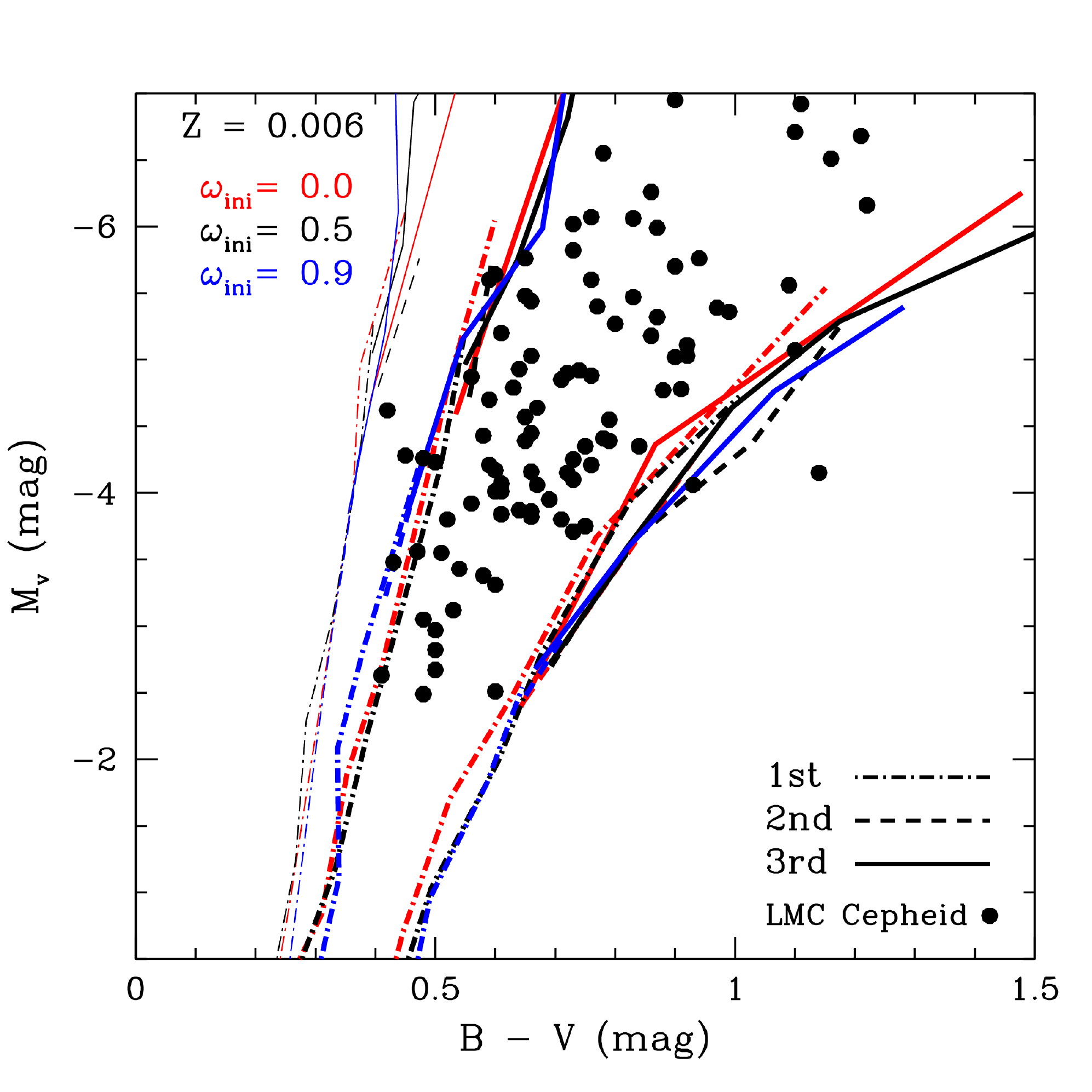}
\includegraphics[scale=0.29]{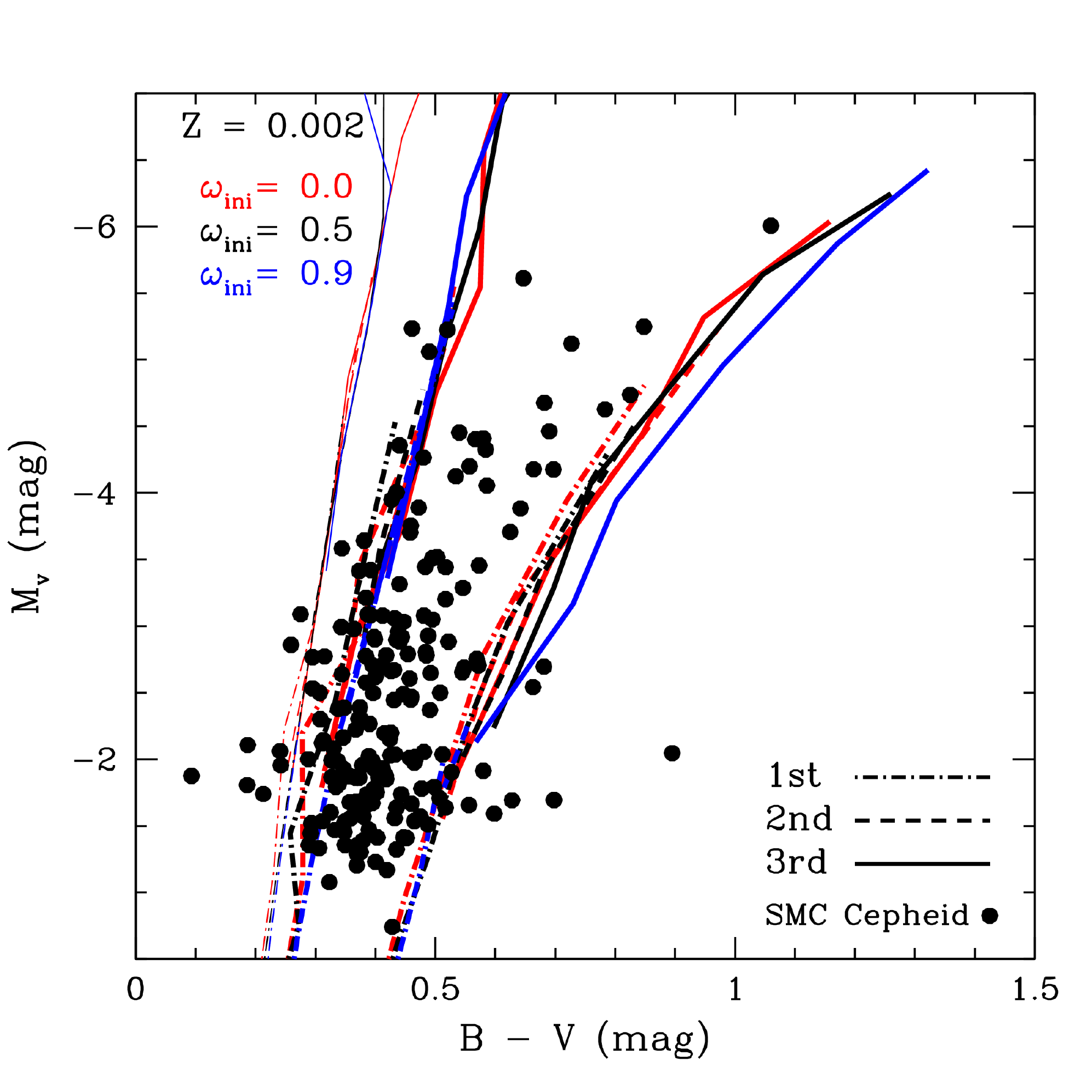}
\caption{Locations of the blue and red edges of the fundamental pulsation in the color-magnitude diagram for  models of various metallicities and initial rotation rates. As in Fig.\,\ref{fig:res:FBE}, thinner lines show locations of  blue edges obtained with the frozen convection assumption, while thicker lines show blue and red edges obtained by taking into account the perturbation of convection flux.
For Milky-Way Cepheids (left-most panel), data obtained by \citet{2013AJ....146...93R} are included in addition to the data shown in Fig.\,\ref{fig:res:FBE}.
Data for LMC Cepheids (middle panel) are taken from \citet{2004A&A...424...43S} adopting a distance modulus of 18.494\,mag \citep{2013Natur.495...76P}. Data for SMC Cepheids in the right-most panel are taken from \citet{1999AcA....49..437U}, in which only fundamental pulsators with reddening of $E(B-V) < 0.08$\,mag are chosen using distance modulus $18.965$ for the SMC \citep{2014ApJ...780...59G,2016ApJ...816...49S}.}
\label{fig:res:FBE_CM}
\end{figure*}

It is often assumed that the FC assumption can be used to accurately predict the location of blue edge. Interestingly, Fig.\,\ref{fig:res:FBE} shows that pulsation-convection coupling significantly affects the predicted blue boundary. This effect had previously been described by \citet{1980A&A....84..304G}. Moreover, Fig.\,\ref{fig:res:FBE_CM} shows that the IS boundaries including pulsation-convection coupling are much more consistent with the observed distribution of the Galactic Cepheids than if this coupling is neglected. 
The figure also shows that our predicted red edges are also consistent with the observed distributions in the HR diagram. This excellent consistency with observations is crucial for having confidence in the properties of Cepheid variable stars.

\begin{figure*}
\centering
\includegraphics{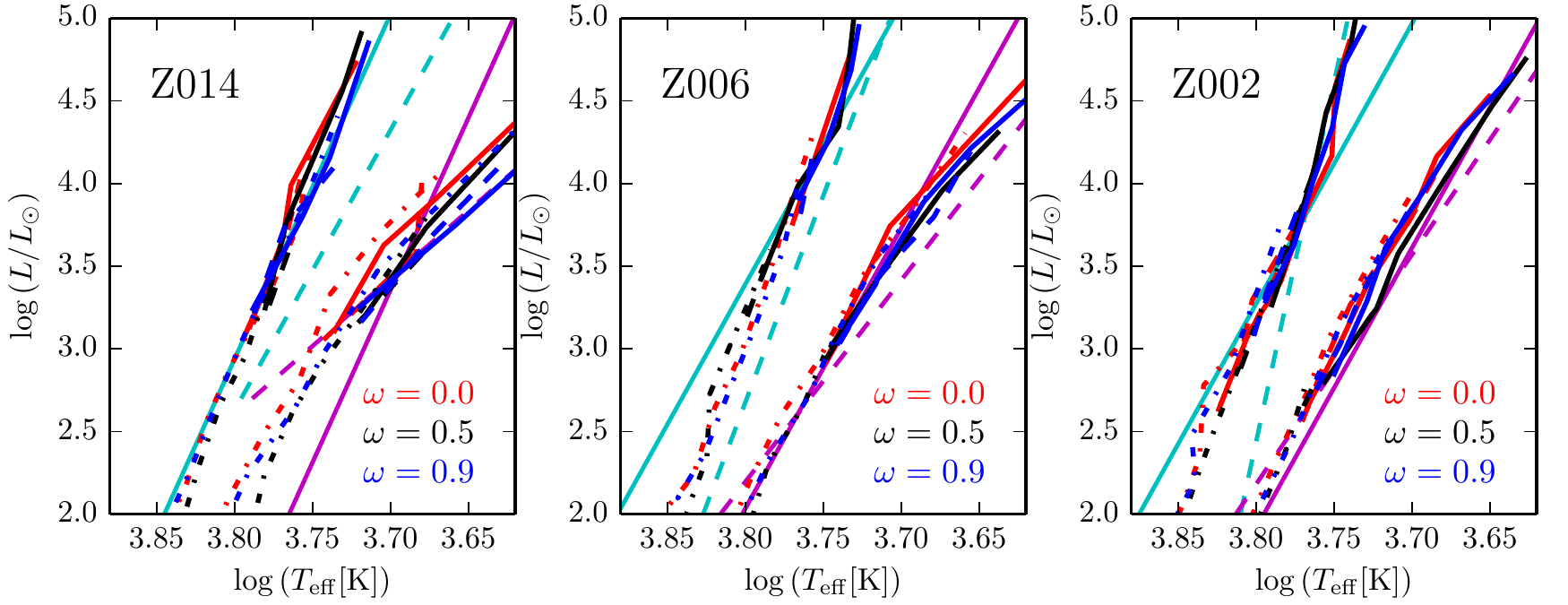}
\includegraphics{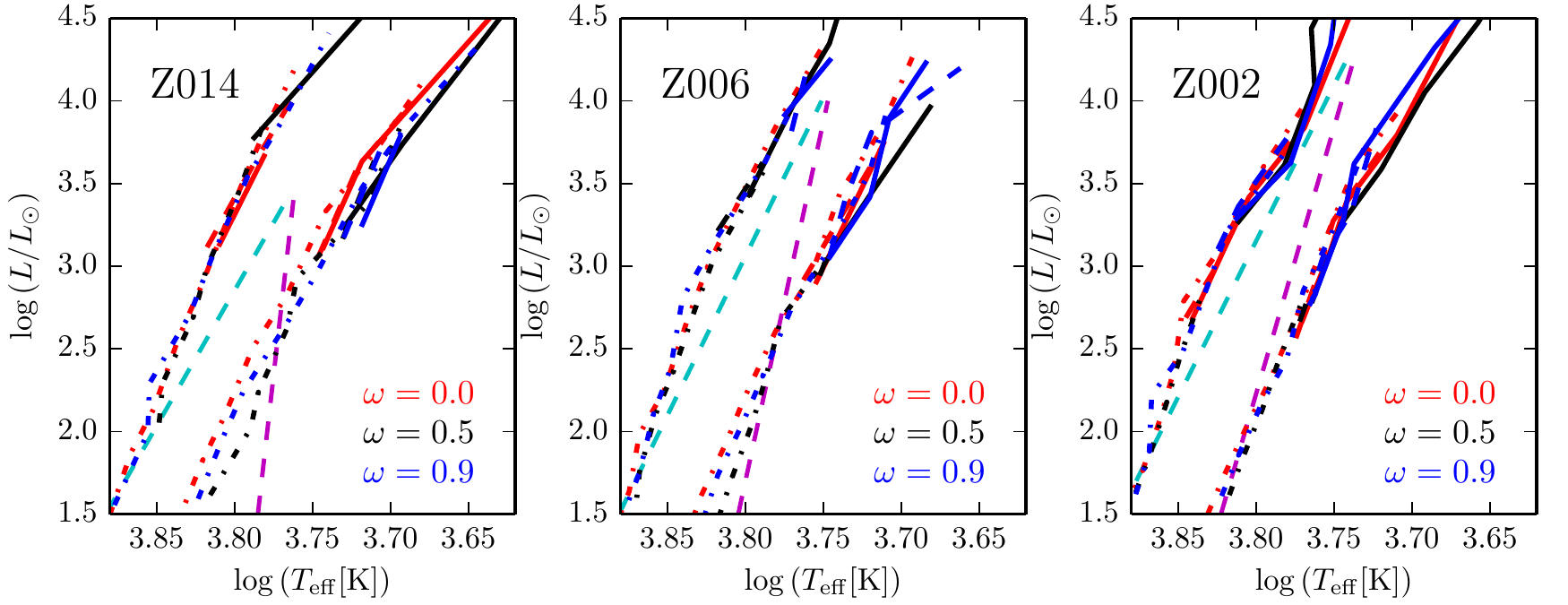}
\caption{Comparison of our IS boundaries with the literature for both fundamental mode Cepheids (upper panels) and first overtone Cepheids (lower panels). The blue edge shown is computed without the frozen convection assumption. Metallicity decreases over the panels from left to right, as indicated. Our IS boundaries are drawn in red, black, and blue for $\omega=0.0$, $0.5$, and $0.9$, respectively. First crossings are drawn as dash-dotted, second crossings as dashed, and third crossings as solid lines.  Theoretically computed IS boundaries by \citet[][{\it noncanonical} models computed for Z=0.02, 0.008, 0.004]{2005ApJ...621..966B} are shown as cyan and magenta dashed lines for the hot and cool edge, and empirical boundaries by \citet[fundamental modes only]{2003A&A...404..423T} are shown as solid cyan and magenta lines for hot and cool edges.}
\label{fig:res:ISboundaries}
\end{figure*}

Figure\,\ref{fig:res:FBE} illustrates the effect of metallicity, crossing number, and rotation on IS boundaries. We find a weak dependence of the blue edge position on metallicity, with lower metallicity models having slightly higher temperature. Qualitatively, the same effect was shown by \citet{2000ApJ...529..293B} and \citet{2002ApJ...576..402F}, albeit with a stronger dependence of IS boundaries on metallicity (see Fig.\,\ref{fig:res:ISboundaries}). The difference between our results and these literature relations can be explained by the different initial helium abundances adopted \citep[cf. Sec.\,\ref{sec:ana:models} and][Fig.1]{2005ApJ...632..590M} as well as the overall lower metal content for solar metallicity (we use $Z=0.014$ instead of $Z=0.02$). 
The different IS boundaries have important implications for the dependence of period on metallicity, see Sec.\,\ref{sec:periodrange} below. Figure\,\ref{fig:res:ISboundaries} also provides a comparison with empirical IS boundaries determined by \citet{2003A&A...404..423T}. For the Galactic metallicity, the \citet{2003A&A...404..423T} hot edges are in near perfect agreement, and so are the cool edges for the lower metallicities.

The blue edge does not significantly depend on crossing number or initial rotation rate. This behavior shows the dependence of the pulsational instability on luminosity and temperature: metallicity changes temperature, e.g. via line blanketing, while rotation and crossing numbers mainly shift Cepheids in luminosity without affecting temperature, cf. Fig.\,\ref{fig:hrd}. On the other hand, the cool (red) IS edge is significantly affected by metallicity, crossing number, and rotation. The main difference is due to the dependence of stellar structure on rotation and crossing number, both of which affects the size and average density of the star as well as the pressure scale height. This changes the way convection operates in the star, leading to a noticeable effect on the cool IS boundaries. The change in the red IS boundary location between first and later crossings is further significantly affected by the higher helium abundance in the outer envelope following the dredge-up that occurs after the first crossing.

Figure\,\ref{fig:res:FBE_CM} shows a more extensive comparison with observations for low-extinction ($E(B-V) < 0.08$\,mag) Galactic, LMC, and SMC Cepheids in color-magnitude diagrams. The excellent agreement  indicates that our implementation of the pulsation-convection interaction is on the right track. In particular, the predicted widening of the IS at high luminosities is a good match to observations, although the scarcity of long-period (high-L) Cepheids results in a less densely populated region. 

We here do not provide analytic relations representing the IS boundaries due to the wedge-shaped nature of the IS. IS boundaries can be easily reproduced from Tabs.\,\ref{tab:models_z14} through \ref{tab:models_z02} for fundamental modes and Tabs.\,\ref{tab:models_z14_FO} through \ref{tab:models_z02_FO} for first overtones given in Appendix\,\ref{app:tables}.

\subsection{Range of predicted periods}\label{sec:periodrange}
\begin{figure*}
\includegraphics[]{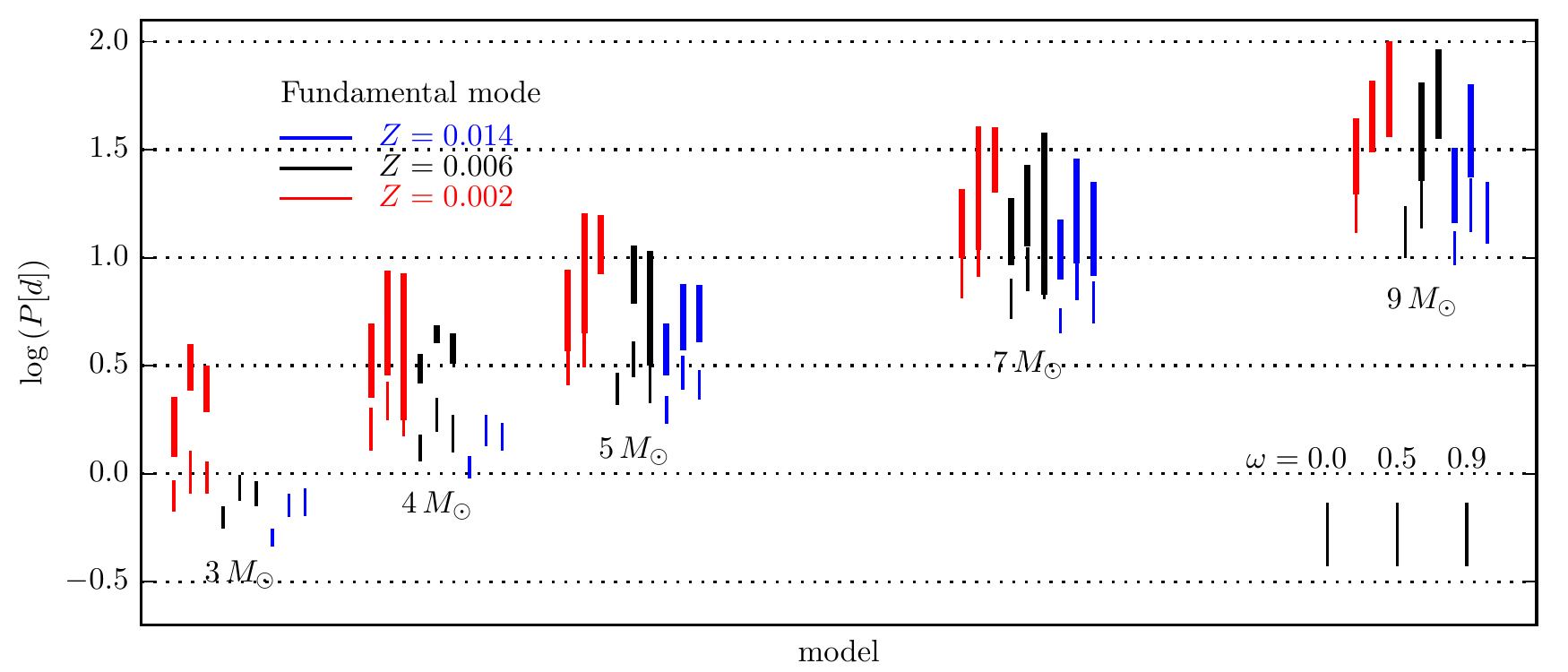}\\
\includegraphics[]{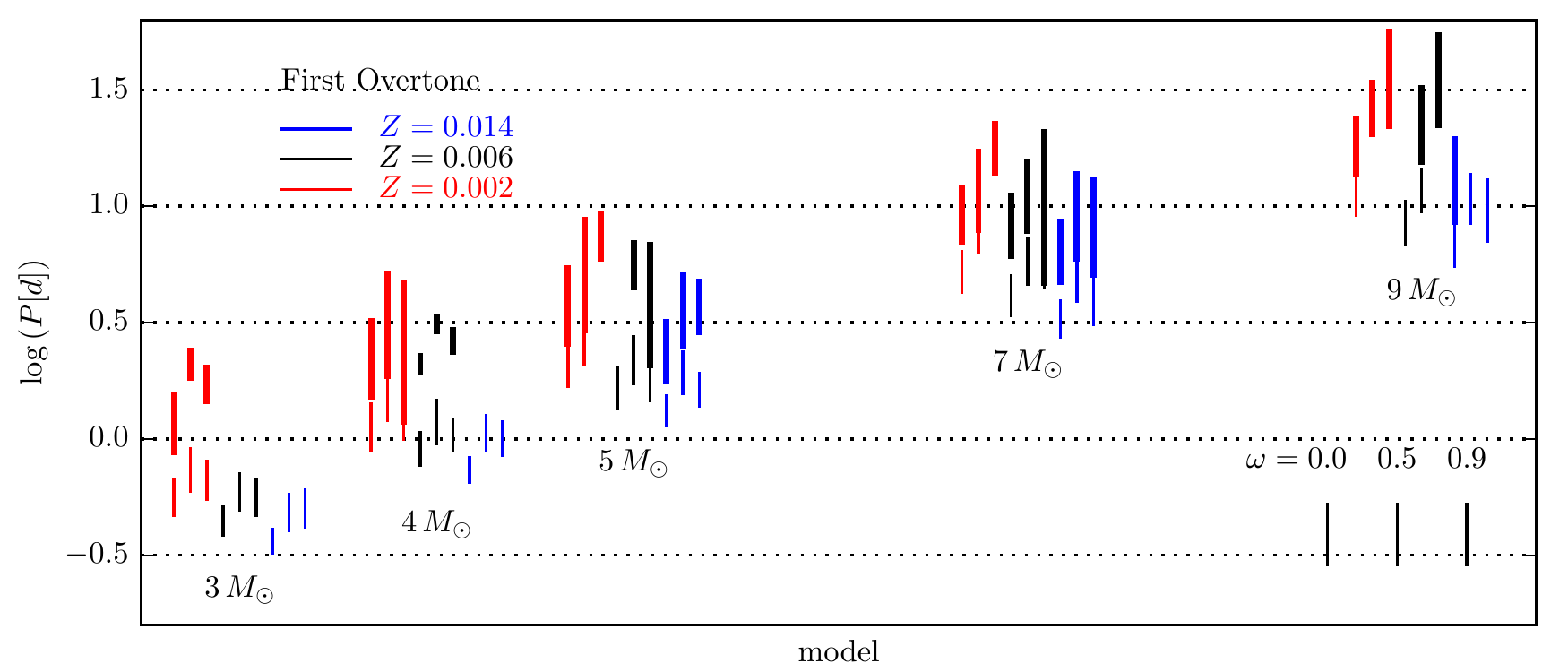}
\caption{Predicted period ranges for fundamental mode Cepheids (top) and first overtone Cepheids (bottom) as function of mass, metallicity, and initial rotation rate. Period ranges on the first crossing are shown with thinner lines than period ranges on 2nd and 3rd crossings. For each mass, we plot results for three different rotation rates (increase left to right) and three metallicities (increase left to right, see colors). For a given model mass, lower metallicity tends to increase pulsation period. Lower metallicity models, however, predict shorter minimum periods for Cepheids on 2nd and 3rd crossings due to longer blue loops.}
\label{fig:periodrange}
\end{figure*}
Here we present the range of pulsation periods predicted by our models. The finite resolution of our model grid (especially in mass) limits the precision with which longest and shortest periods can be predicted. Particularly the lower limit on periods is not fully sampled, since this depends sensitively on the extension of blue loops to hotter temperatures (for Cepheids during core He burning, i.e., on 2nd or later crossings). Yet, inspecting the range of periods predicted sheds light on some important tendencies with metallicity and initial rotation rate.

Figure\,\ref{fig:periodrange} shows the range of periods predicted for Cepheids pulsating in fundamental mode and the first overtone as a function of initial rotation rate, mass, and metallicity. We do not distinguish between second and third crossings in this figure, plotting simply the full range of periods predicted. First crossings are shown as thinner lines, since this evolutionary phase is rarely observed, cf. Sec.\,\ref{sec:res:pdot} and references therein.

\begin{figure}
\centering
\includegraphics[]{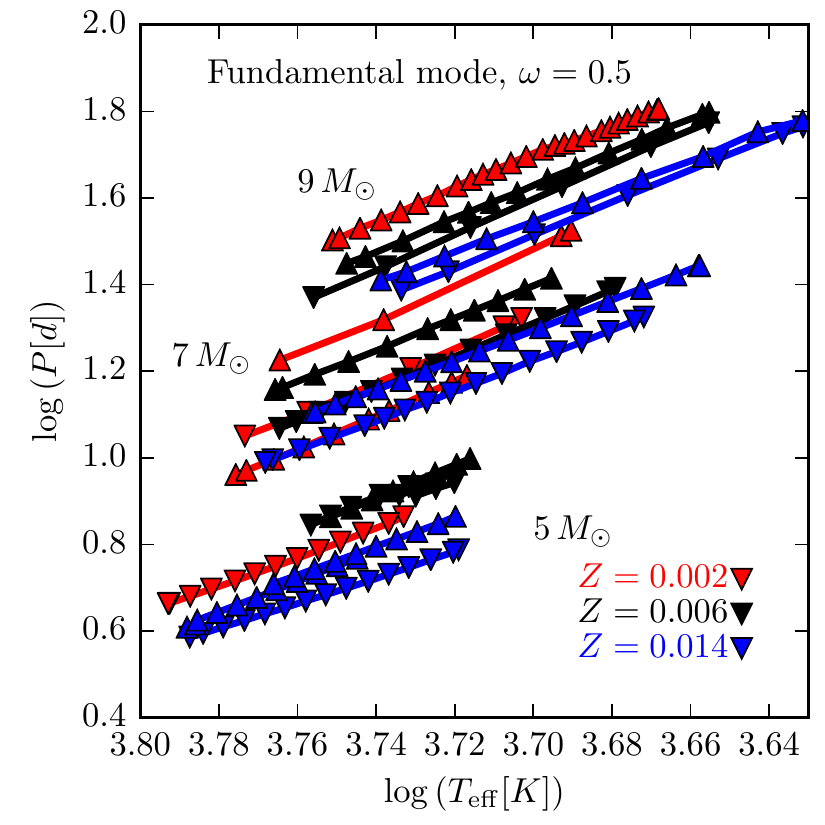}
\caption{Period against effective temperature for $5$, $7$, and $9$\,\Msol\ fundamental mode models with average rotation and different metallicities, as indicated. Downward triangles represent models on the 2nd crossing, upward triangles those on the 3rd crossing.}
\label{fig:PeriodTeffMetallicity}
\end{figure}

For a given model mass our models predict pulsation period to decrease with increasing metallicity and vice versa, cf. Fig.\,\ref{fig:PeriodTeffMetallicity}. This dependence is driven by the higher luminosity of lower metallicity models. For instance, our $7$\,\Msol, $\omega=0.5$ models on the 3rd crossing at the blue edge boundary predict $P=12.7$\,d, $14.4$\,d, and $16.8$\,d for $Z = 0.014$, $0.006$, and $0.002$, respectively. We note that our result contradicts the result by \citet{2000ApJ...529..293B} who concluded that periods should increase with increasing metallicity since their predicted IS boundaries depended sensitively to metallicity, being shifted to significantly lower temperatures for higher metallicity models. Our IS boundaries, however, exhibit a much weaker dependence on metallicity due to a lower helium abundance, cf. Sec.\,\ref{sec:res:IS} and Fig.\,\ref{fig:res:ISboundaries}. As a result, the primary effect of metallicity on Cepheids is to vary luminosity at nearly fixed temperature, with lower metallicity models having higher luminosity and thus, longer period. The excellent agreement of our predicted IS boundaries with empirical data (Sec.\,\ref{sec:res:IS}) corroborates this conclusion.

Rotation modifies the Cepheid mass-luminosity relation \citep{2014A&A...564A.100A}, resulting in different radii for a model of a given mass. Since the dependence of luminosity on rotation is not monotonous, the same is true for pulsation periods. However, the range of periods predicted is very sensitive to the exact shape of the blue loops, cf. Fig.\,\ref{sec:ana:models}). Assuming that periods follows luminosity, the expected result would be in analogy to the dependence of the mass-luminosity relation with rotation: non-rotating models would have the shortest periods, whereas models with average rotation would have the highest luminosity and hence the longest periods. Indeed, the $7$\,\Msol\ models mentioned above with $\omega=0.0$ (non-rotating) have predicted $P=8.7$\,d, $10.3$\,d, and $11.8$\,d (compare to above models with $\omega=0.5$). Fast rotating models are less luminous than those with average rotation due to  competing hydrostatic and mixing effects \citep{2014A&A...564A.100A} and hence exhibit intermediate periods.

Figure\,\ref{fig:periodrange} furthermore clearly shows that the long-period end of the period distribution increases with decreasing metallicity for both fundamental mode and first overtone Cepheids. A similar result was previously obtained for overtone Cepheids by \citet{2002ApJ...574L..33B}.

Figure\,\ref{fig:periodrange} reveals another important effect of metallicity on populations of Cepheids: low-metallicity Cepheid populations are predicted to contain shorter-period Cepheids than higher-metallicity populations\footnote{Neglecting Cepheids on the first IS crossing, since these are not frequent enough to be relevant for this discussion}, despite the tendency of increasing period with decreasing metallicity discussed above. This is because low-metallicity blue loops extend to hotter temperatures for a given mass, leading to more compact Cepheids with shorter periods. As a side effect, the minimum mass of a model that enters the IS also decreases with decreasing metallicity, which also decreases the expected minimal period. An interesting consequence of the blue loop extension within the IS is that low-mass short-period Cepheids are expected to cluster near the red edge of the IS. 

In summary, our results predict that period distributions of lower-metallicity populations extend to both shorter and longer periods than those of higher-metallicity populations. For a Cepheid of a given mass, rotation rate, effective temperature, and crossing, however, a decrease in metallicity results in an increase in period due to the associated change in luminosity.

The OGLE-III sample of Magellanic Cepheids \citep{2008AcA....58..163S,2010AcA....60...17S} clearly shows that these predictions are confirmed by the difference in period distribution among Cepheids in the LMC and SMC (shorter minimal periods in the SMC). This important empirical fact corroborates the predicted dependence of the blue loop extension to metallicity (see also Sec.\,\ref{sec:res:PLRs}) and suggests that the blue loop effect on the short-period is indeed the dominating factor to the short-period end of the distribution. 

Our ability to investigate the minimal period predicted by our models are currently limited by the finite resolution of the model grid. A more detailed investigation regarding the shortest predicted periods would require a significant additional computational effort (a fine grid of different masses, rotation rates, and metallicities), which is considered out of scope for this paper.

\subsection{Period-Luminosity Relations}\label{sec:res:PLRs}

\begin{figure*}
\centering
\includegraphics[scale=0.4]{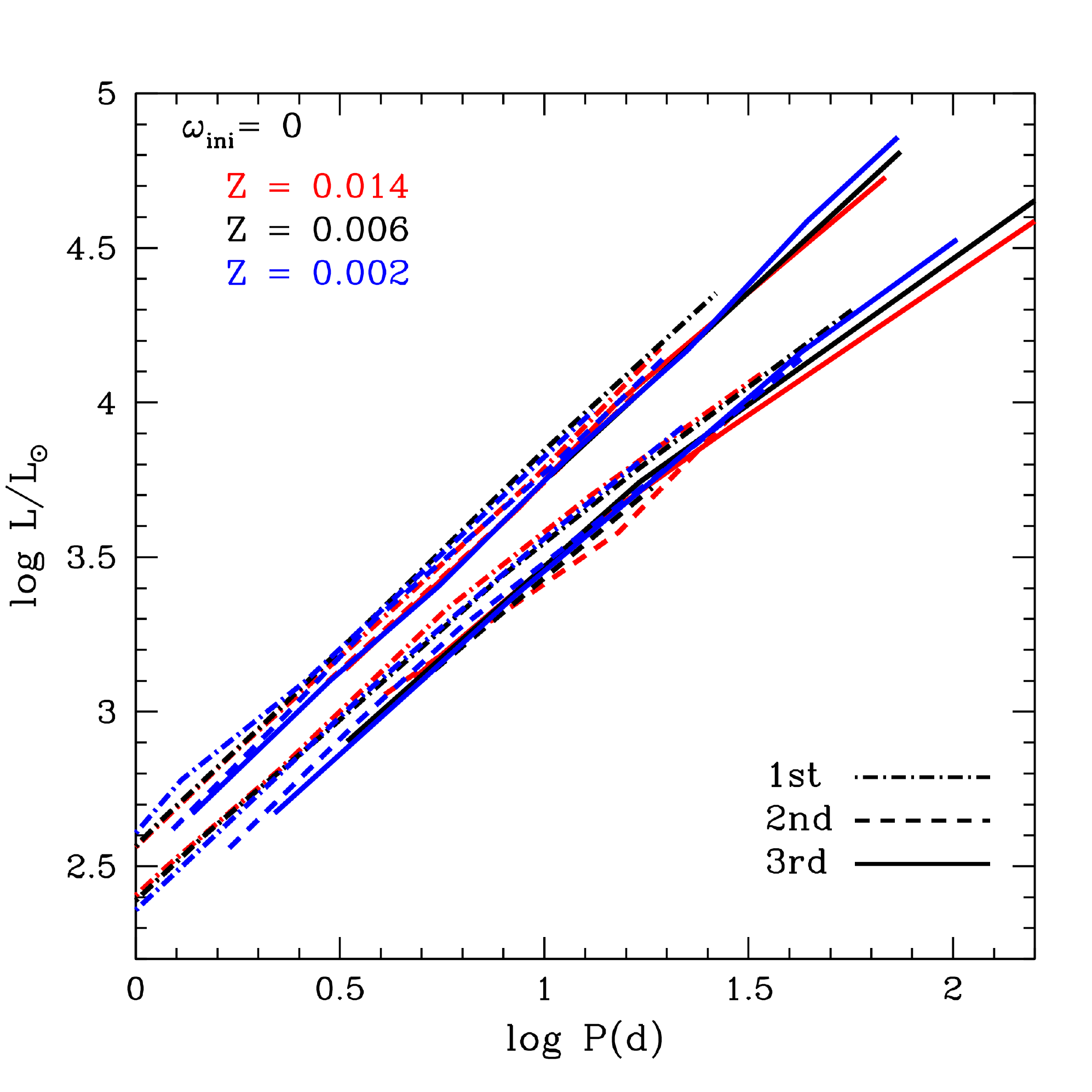}
\includegraphics[scale=0.4]{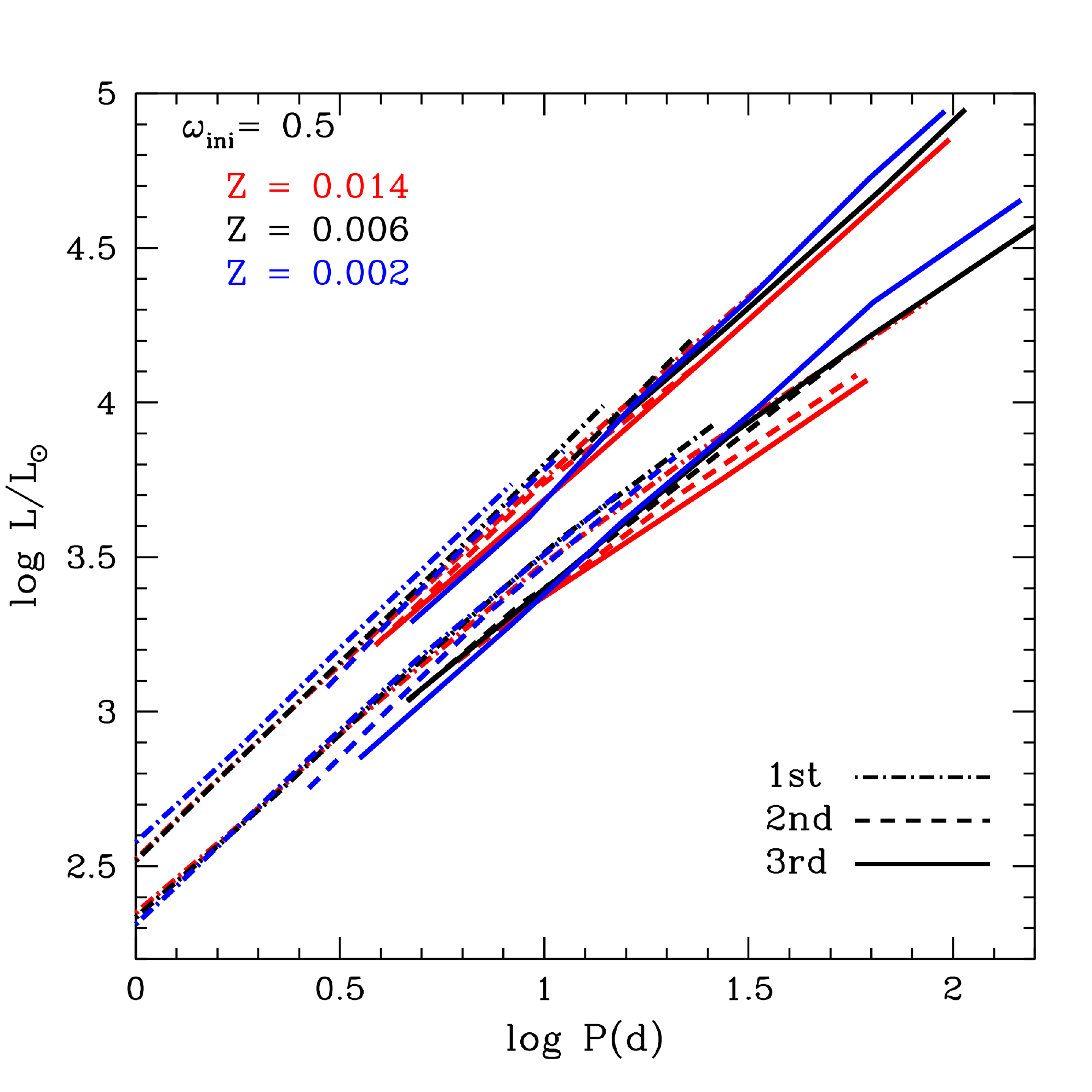}
\caption{PLRs for fundamental mode pulsation at blue and red IS boundaries for models without rotation (left panel) and models with $\omega_{\rm ini} = 0.5$ (right panel). Metallicity is color coded, and different line types correspond to different crossings, see legend.}
\label{fig:res:PLR}
\end{figure*}

We show PLRs predicted separately for the different crossings and each blue and red edge in Fig.\,\ref{fig:res:PLR}. The left panel compares the effect of metallicity for non-rotating models, and the right panel shows the same for models with average rotation ($\omega=0.5$). From the figure it is clear that crossing numbers and rotation influence the position of each PLR, and that rotation tends to slightly broaden the width of the IS, particularly at long periods. The relative effects of the intrinsic width of the IS, crossing numbers, metallicity, and rotation are discussed in more detail in Sect.\,\ref{sec:PLRdependence}.

\begin{figure*}[!htp]
\centering
\includegraphics[scale=0.78]{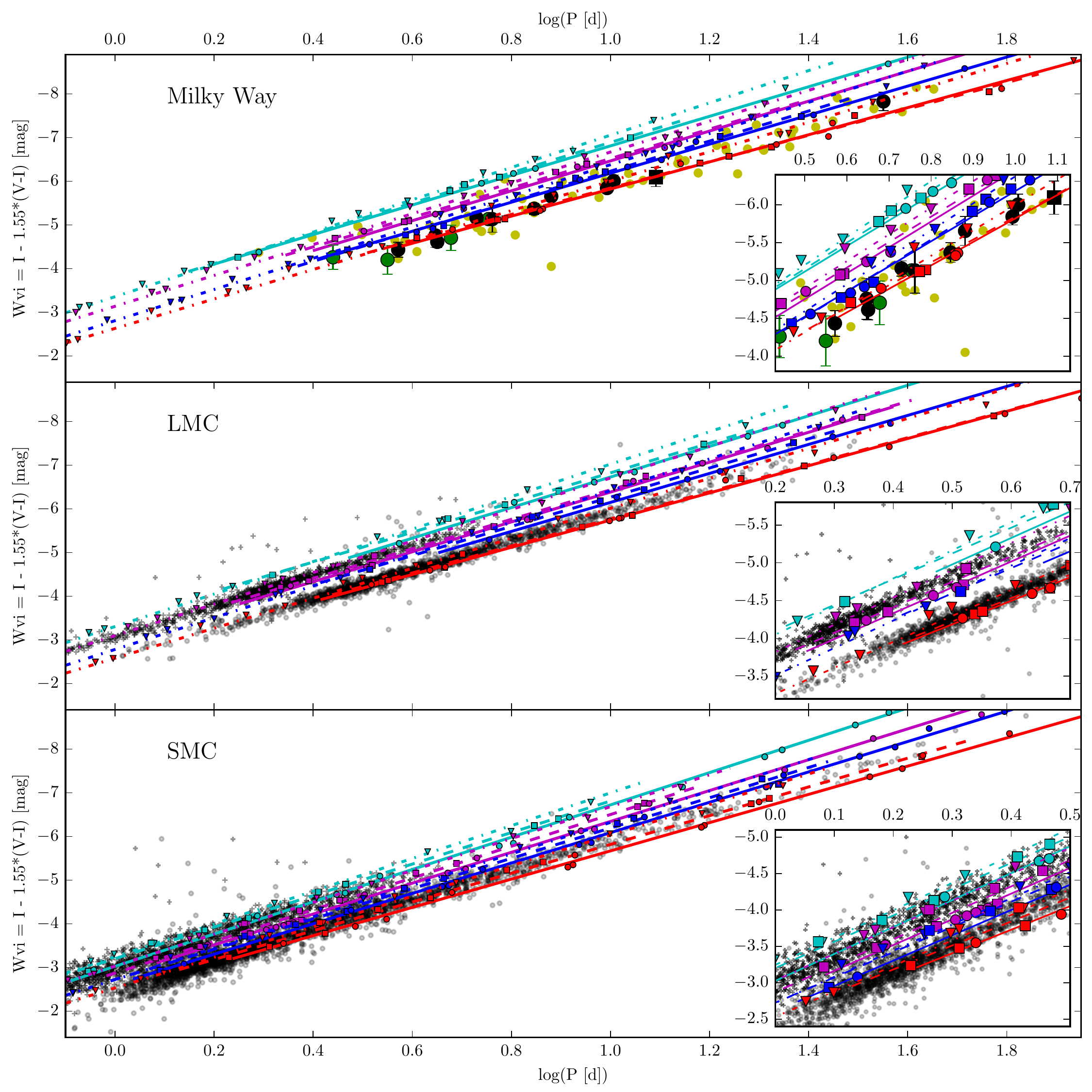}
\caption{Comparison between predicted PLRs and observed PL distributions. We show separate theoretical PLRs near blue and red edges, and for the three crossings. Each PLR is derived by fitting models of all rotation velocities together, cf. Tab.\,\ref{tab:PLR_phot}. We use the ``reddening-free'' Wesenheit index $W_{\rm{VI}}$ to compare the theoretical relations to Milky Way, LMC, and SMC Cepheids (top to bottom panels as labeled). First, second, and third crossing predictions are shown as triangles, squares, and circles and fitted relations are drawn as dash-dotted, dashed, and solid lines over a slightly longer period range ($\pm 0.1$ in $\log{P}$) for clarity. Hot and cool edges are drawn in blue and red for fundamental mode models, and in cyan and magenta for first overtone models. {\it Top panel:} Empirical data shown are from \citet[black circles with errors]{2007AJ....133.1810B}, \citet[green circles with errors]{2007MNRAS.379..723V}, \citet[yellow scatter points]{2011A&A...534A..94S}, and \citet[black square with errors]{2015arXiv151209371C}. {\it Center and bottom panels:} OGLE-3 LMC and SMC Cepheids \citep{2008AcA....58..163S,2010AcA....60...17S}, for which absolute magnitudes are calculated using distance moduli of 18.494 and 18.965 based on eclipsing binary distances \citep{2013Natur.495...76P,2014ApJ...780...59G}. In the LMC and SMC panels, fundamental mode Cepheids are shown as filled circles, and first overtone Cepheids as filled pluses. 
Predictions for short-period fundamental mode Cepheids are limited by the neglected curvature of the red IS edge (cf. Sec.\,\ref{sec:res:IS}) and the finite mass resolution of the model grid (cf. Sec.\,\ref{sec:periodrange}). The increased scatter among SMC Cepheids due to depth effects is clearly evident.}
\label{fig:res:MWvi-logP}
\end{figure*}

We determine theoretical PLRs on the red and blue IS edges (subscripts $r$ and $b$) by fitting absolute magnitudes computed in $V$, $H$, $W_{VI}$, and $W_{H,VI}$ filters linearly:
\begin{equation}
M_{r,b} = \alpha_{r,b} \cdot \log{(P\ [\rm{d}])} + \beta_{r,b} \, .
\label{eq:PLRred}
\end{equation}

The curvature of the IS's red edge (cf. Fig.\,\ref{fig:res:ISboundaries}) could be taken into account by adopting a quadratic term. However, this curvature depends on both IS edge (red shows more curvature) and photometric band pass, becoming less noticeable at wavelengths longer than $V$, as well as in $W_{VI}$. For consistency and in keeping with the literature, we therefore use linear PLRs for both IS edges. We tabulate rotation-averaged PLRs in Table\,\ref{tab:PLR_phot}. Rotation-averaged here indicates that predictions for $\omega=0.0, 0.5$, and $0.9$ were all fit together for each filter, crossing number, IS edge, and metallicity.

Figure\,\ref{fig:res:MWvi-logP} compares predicted rotation-averaged
PLRs for Cepheids in the Milky Way, LMC, and SMC to observed PL distributions using the ``reddening-free'' $W_{VI}$ magnitudes (Eq.\,\ref{eq:W_VI}). The sample of Galactic Cepheids consists of data from \citet{2007AJ....133.1810B}, \citet{2007MNRAS.379..723V}, \citet{2011A&A...534A..94S}, and \citet{2015arXiv151209371C}. We recomputed the absolute $W_{VI}$ magnitudes for these stars using the published parallaxes, apparent magnitudes, and, where available, the stated Lutz-Kelker corrections \citep{1973PASP...85..573L} to ensure that the same definition for $W_{VI}$ was employed for all stars as well as the models. The LMC and SMC samples comprise all fundamental mode and first overtone Cepheids listed in the OGLE-3 catalog of variable stars \citep{2008AcA....58..163S,2010AcA....60...17S} with no cuts applied. 

The out-of-the box agreement between predictions and observations is remarkable. Most Galactic Cepheids with parallax measurements fall squarely within the IS boundaries. Although they are all consistent to within the stated uncertainties, the most deviant from the predicted relations are FF\,Aql,  BG\,Cru, and DT\,Cyg. Galactic Cepheids with Baade-Wesselink distances exhibit larger scatter, but also reproduce the predicted relations very well. For the LMC, the distribution of first overtone Cepheids is particularly well confined within the predicted boundaries. Short-period Cepheids in the LMC tend to cluster near the predicted red edge PLR, which could be an indication of the shorter blue loop extent predicted for lower-mass models and additionally affected by the neglected curvature of the red IS edge. The observed scatter among longer period fundamental mode Cepheids is consistent with the predicted PL width. The finite mass resolution of our models sets the short-period limit of the predicted PLRs plotted here. The true short-period is, in practice, defined by a model mass that falls within our grid's resolution (cf. Sec.\,\ref{sec:periodrange}). For instance, the expected minimal mass for $Z=0.014$ is approx. $4.5$\,\Msol\ \citep{2014A&A...564A.100A}. For the SMC, the depth effect \citep[e.g.][]{2016ApJ...816...49S} is very noticeable.

\begin{table*}
\begin{tabular}{@{}ll|rrrr|rrrr|rrrr@{}}
\hline
& & \multicolumn{4}{c}{{\bf Z=0.014}} & \multicolumn{4}{c}{{\bf Z=0.006}} & \multicolumn{4}{c}{{\bf Z=0.002}} \\
Band & Xing & \multicolumn{2}{c}{Blue Edge} & \multicolumn{2}{c}{Red Edge} & \multicolumn{2}{c}{Blue Edge} & \multicolumn{2}{c}{Red Edge} & \multicolumn{2}{c}{Blue Edge} & \multicolumn{2}{c}{Red Edge} \\
& & $\alpha_b$ & $\beta_b$ & $\alpha_r$ & $\beta_r$ & $\alpha_b$ & $\beta_b$ & $\alpha_r$ & $\beta_r$ & $\alpha_b$ & $\beta_b$ & $\alpha_r$ & $\beta_r$ \\
\hline
V & 1st &-3.040&-1.600&-2.602&-1.037&-3.162&-1.617&-2.791&-1.003&-3.107&-1.609&-2.943& -0.984\\
V & 2nd &-2.878&-1.619&-2.008&-1.487&-2.836&-1.793&-2.281&-1.265&-2.978&-1.607&-2.642& -1.043\\
V & 3rd &-2.816&-1.664&-1.761&-1.766&-2.859&-1.655&-2.077&-1.465&-3.046&-1.435&-2.461& -1.063\\
\hline
H & 1st &-3.483&-2.570&-3.316&-2.353&-3.543&-2.549&-3.389&-2.300&-3.503&-2.499&-3.452& -2.248\\
H & 2nd &-3.389&-2.519&-2.981&-2.540&-3.278&-2.670&-3.086&-2.362&-3.387&-2.490&-3.276& -2.218\\
H & 3rd &-3.286&-2.579&-2.927&-2.580&-3.279&-2.549&-3.026&-2.391&-3.421&-2.352&-3.212& -2.133\\
\hline
W$_{\rm{VI}}$ & 1st &-3.570&-2.809&-3.371&-2.632&-3.635&-2.782&-3.431&-2.581&-3.599&-2.722&-3.499& -2.533\\
W$_{\rm{VI}}$ & 2nd &-3.442&-2.789&-3.020&-2.825&-3.344&-2.924&-3.096&-2.675&-3.466&-2.727&-3.292& -2.525\\
W$_{\rm{VI}}$ & 3rd &-3.324&-2.861&-3.106&-2.716&-3.320&-2.828&-3.114&-2.629&-3.478&-2.609&-3.234& -2.435\\
\hline
W$_{\rm{H,VI}}$ & 1st &-3.569&-2.765&-3.440&-2.609&-3.619&-2.737&-3.492&-2.554&-3.582&-2.678&-3.541& -2.497\\
W$_{\rm{H,VI}}$ & 2nd &-3.480&-2.707&-3.143&-2.755&-3.360&-2.852&-3.217&-2.588&-3.466&-2.670&-3.381& -2.456\\
W$_{\rm{H,VI}}$ & 3rd &-3.367&-2.772&-3.143&-2.733&-3.353&-2.738&-3.193&-2.578&-3.491&-2.541&-3.337& -2.354\\
\hline
\end{tabular}
\caption{PLRs ($M = \alpha\log{P} + \beta$) of fundamental mode Cepheids for $Z=0.014$, $0.006$, and $0.002$, determined separately on blue and red IS edges as well as for each of the three IS crossings in four photometric bands and Wesenheit indices (cf. Eqs.\,\ref{eq:W_VI} and \ref{eq:W_HVI}). Each PLR is obtained as an average over the three initial rotation speeds. All magnitudes are determined using the interpolation program and data by \citet{2011ApJS..193....1W}. The $W_{VI}$ relations are shown in Fig.\,\ref{fig:res:MWvi-logP}
.}
\label{tab:PLR_phot}
\end{table*}

\begin{table*}
\centering
\begin{tabular}{@{}ll|rrrr|rrrr|rrrr@{}}
\hline
& & \multicolumn{4}{c}{{\bf Z=0.014}} & \multicolumn{4}{c}{{\bf Z=0.006}} & \multicolumn{4}{c}{{\bf Z=0.002}} \\
Band & Xing & \multicolumn{2}{c}{Blue Edge} & \multicolumn{2}{c}{Red Edge} & \multicolumn{2}{c}{Blue Edge} & \multicolumn{2}{c}{Red Edge} & \multicolumn{2}{c}{Blue Edge} & \multicolumn{2}{c}{Red Edge} \\
& & $\alpha_b$ & $\beta_b$ & $\alpha_r$ & $\beta_r$ & $\alpha_b$ & $\beta_b$ & $\alpha_r$ & $\beta_r$ & $\alpha_b$ & $\beta_b$ & $\alpha_r$ & $\beta_r$ \\
\hline
V & 1st &-3.145&-2.230&-2.806&-1.489&-3.195&-2.201&-3.030&-1.495&-3.216&-2.178&-3.118& -1.457\\
V & 2nd &-2.764&-2.423&-2.677&-1.546&-2.716&-2.434&-2.584&-1.599&-2.892&-2.239&-3.072& -1.314\\
V & 3rd &-2.596&-2.487&-2.305&-1.757&-2.849&-2.225&-2.734&-1.409&-3.111&-1.997&-2.703& -1.416\\
\hline
H & 1st &-3.593&-3.133&-3.479&-2.856&-3.614&-3.086&-3.558&-2.812&-3.640&-3.033&-3.596& -2.744\\
H & 2nd &-3.351&-3.203&-3.361&-2.834&-3.290&-3.206&-3.364&-2.747&-3.473&-3.003&-3.603& -2.567\\
H & 3rd &-3.320&-3.174&-3.337&-2.786&-3.437&-2.971&-3.408&-2.651&-3.610&-2.801&-3.506& -2.509\\
\hline
W$_{\rm{VI}}$ & 1st &-3.697&-3.358&-3.531&-3.138&-3.717&-3.306&-3.609&-3.098&-3.751&-3.246&-3.650& -3.032\\
W$_{\rm{VI}}$ & 2nd &-3.467&-3.422&-3.371&-3.149&-3.401&-3.423&-3.374&-3.062&-3.609&-3.203&-3.629& -2.872\\
W$_{\rm{VI}}$ & 3rd &-3.384&-3.426&-3.421&-3.044&-3.502&-3.227&-3.420&-2.965&-3.696&-3.027&-3.532& -2.812\\
\hline
W$_{\rm{H,VI}}$ & 1st &-3.682&-3.315&-3.596&-3.121&-3.698&-3.264&-3.652&-3.070&-3.726&-3.204&-3.682& -2.997\\
W$_{\rm{H,VI}}$ & 2nd &-3.464&-3.364&-3.472&-3.092&-3.400&-3.365&-3.491&-2.982&-3.588&-3.158&-3.692& -2.817\\
W$_{\rm{H,VI}}$ & 3rd &-3.447&-3.325&-3.517&-2.993&-3.542&-3.132&-3.518&-2.901&-3.704&-2.967&-3.639& -2.733\\
\hline
\end{tabular}
\caption{Same as Tab.\,\ref{tab:PLR_phot} for first overtone Cepheids.}
\label{tab:PLR_FO_phot}
\end{table*}

Adopting different combinations of models, our analysis allows to test for the impact of the IS width, crossing number, metallicity, and rotation. We discuss these various effects and their implications for distance measurements in  Sect.\,\ref{sec:PLRdependence} below.

\subsection{Period-Age Relations from Rotating Models}\label{sec:res:age}
Longer-period Cepheids tend to have higher initial masses, and hence tend to be younger than Cepheids of shorter periods. The concept of Cepheid period-age relations \linebreak 
\citep{2000ApJ...529..293B}  has thus sparked interest in terms of dating star-forming regions, e.g. in the nuclear bulge of the Galaxy \citep[e.g.][]{2015ApJ...812L..29D} or the Magellanic system \citep{2016arXiv160209141J}. Of course, pulsation period also increases towards the red edge of the IS for a star of a given mass, leading to period-age-color relations, which are often approximated by period-age relations \citep{2005ApJ...621..966B}. Crossing numbers have similar effects on period-age relations. Taken together, a fairly large scatter in age at a fixed period can thus be expected even for models that do not include the rejuvenating effects of rotation.

Rotation affects period-age and period-age-color relations via mixing processes that supply the core with fresh material; this is particularly important during hydrogen burning on the MS. Contrary to the competing hydrostatic and mixing impacts on luminosity, rotation affects MS lifetimes monotonously: the faster the initial rotation of a model with fixed mass, the longer its MS lifetime. At the time when a star of a given mass finally crosses the IS, its age as a Cepheid therefore depends on its MS lifetime, which is related to its rotational history (expressed as $\omega$ in our models). However, period is proportional to luminosity, and luminosity does not increase monotonously with rotation. Therefore, we expect a fairly complex dependence of the period-age relation on rotation. Differently stated, a given luminosity can be reached by lower-mass Cepheid models with rotation as well as higher-mass Cepheids with very slow rotation (within limits). Since these lower-mass (rotating) stars have longer MS lifetimes, one can expect the rotating models to yield a higher age than the non-rotating models for a given luminosity and thus, period.
In the following, we determine period-age relations (linear in $\log{P}$) from our results separately depending on initial rotation rate, IS crossing, IS position, and metallicity. Table\,\ref{tab:res:PeriodAge} tabulates the different scenarios for $\omega=0.5$ (average rotation).

\begin{figure*}
\centering
\includegraphics[]{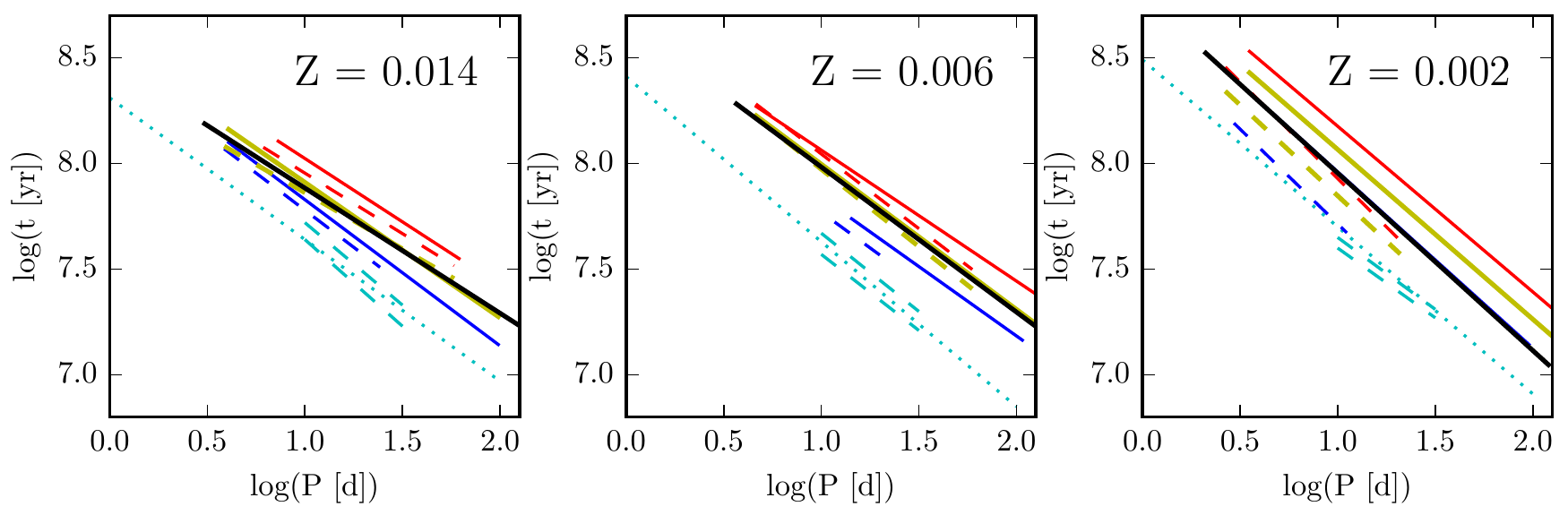}
\includegraphics[]{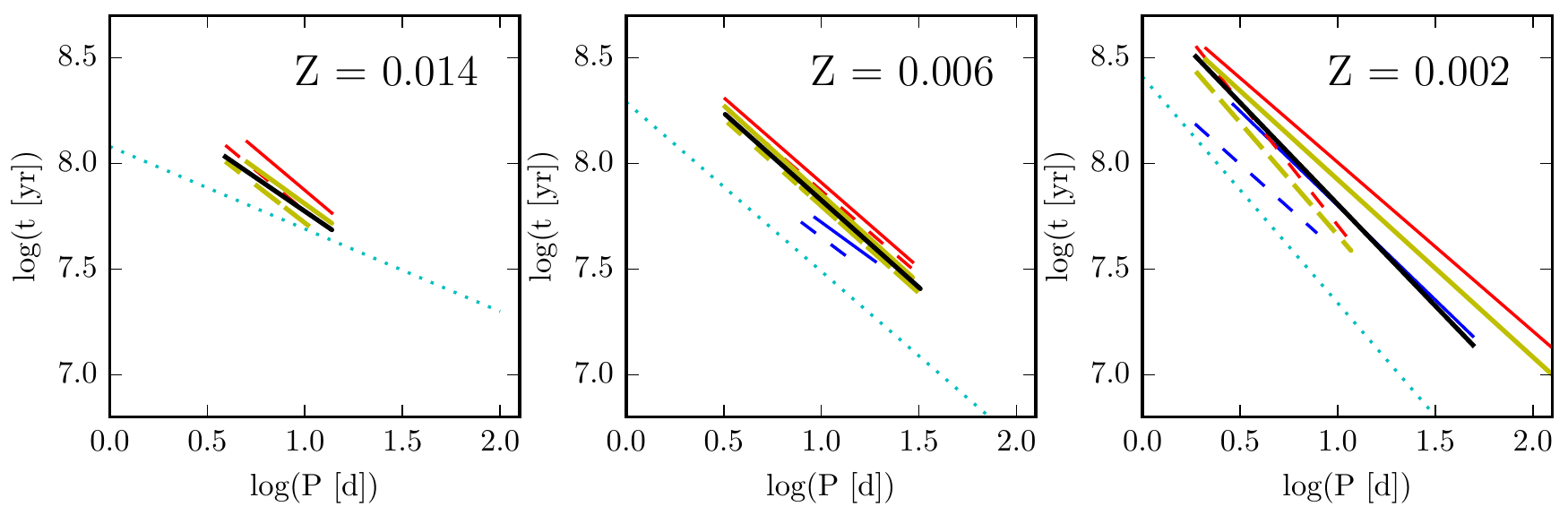}
\caption{Period-age relations for fundamental mode (top panels) and first overtone (bottom panels) Cepheids with $\omega=0.5$ (average rotation), ordered by metallicity as labeled, cf. Tab.\,\ref{tab:res:PeriodAge}. Relations are fitted separately on second (dashed lines) and third (solid lines) crossings near the hot (blue) and cool (red) IS edge over the range of periods predicted (cf. Sec.\,\ref{sec:periodrange}). Yellow lines represent IS-averaged relations, and black solid lines are averaged over IS position and crossing. Dotted cyan line shows the relations by \citet{2005ApJ...621..966B}  that do not account for rotation, plotted for the arbitrary period range of $1-100$\d. Shorter cyan dashed lines show the impact of the IS width \citep[table 7]{2005ApJ...621..966B} for fundamental mode Cepheids. Ages based on our {\it crossing average} relations (black solid lines) are older than ages based on these literature relations  by $\Delta\log{t} \sim 0.2$ at a period of 10\,days.}
\label{fig:PeriodAgeFit}
\end{figure*}

\begin{figure*}
\centering
\includegraphics[]{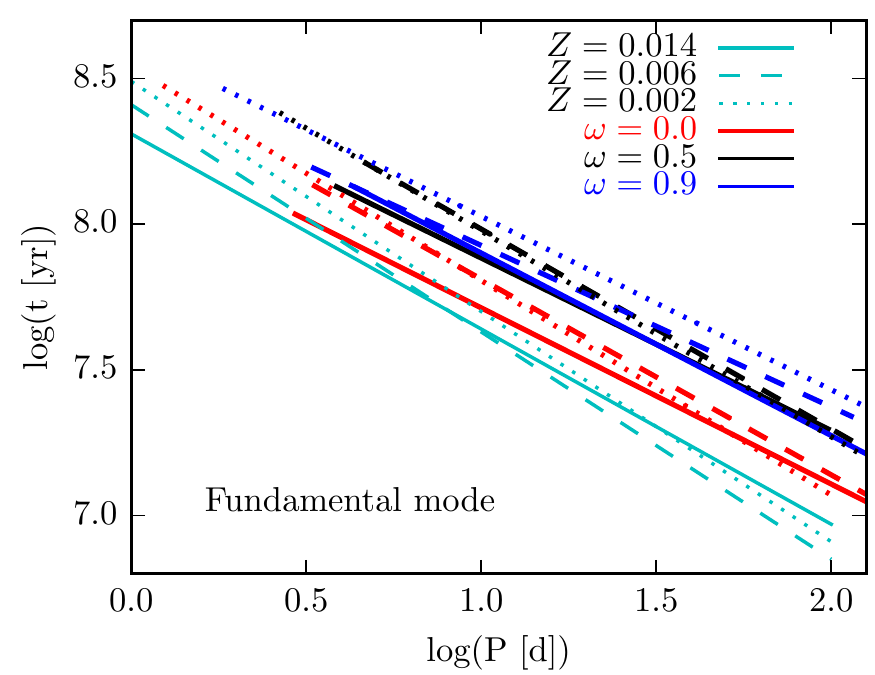}
\includegraphics[]{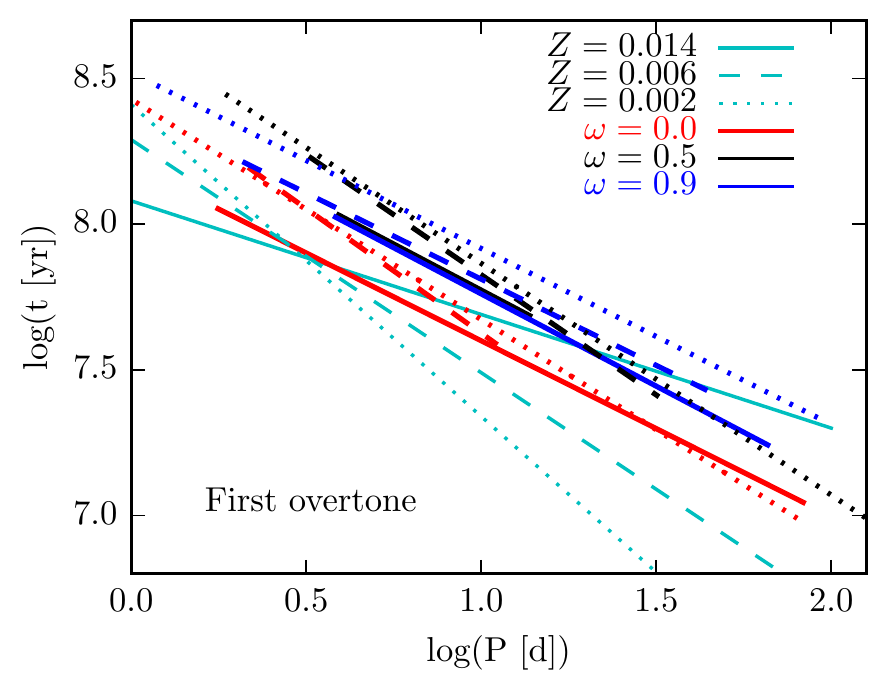}
\caption{Period age relations as function of different initial rotation rate and metallicity, averaged over 2nd and 3rd crossing and the width of the instability strip. The cyan lines show PA-relations from \citet{2005ApJ...621..966B} for similar metallicities ($Z=0.02, 0.01, 0.004$) plotted for a fixed $\log{P}$ range. Our models are shown only for the period-range accessible from the computed models. Non-rotating models are shown in red, average rotation in black, and fast in blue.}
\label{fig:PArelation_RotationDiff}
\end{figure*}

We show our period-age relations in Fig.\,\ref{fig:PeriodAgeFit}, which clearly demonstrates that all the mentioned effects have significant impact on age estimates based on Cepheid periods. Figure\,\ref{fig:PArelation_RotationDiff} further illustrates the effect of metallicity and rotation on period-age relations for both fundamental and first overtone Cepheids. The period-age relation thus depends on:
\begin{enumerate}
\item Crossing number ($\Delta\log{t} \lesssim 0.2$), larger differences with lower metallicity. Since the three crossings occur in sequence, 3rd crossing Cepheids are older than those on a second crossing. The largest difference in age at fixed period is observed between the first and second crossing, since the majority of core He burning occurs during the red giant phase before blue loop evolution \citep{2014A&A...564A.100A}. However, comparing the predictions for different metallicities reveals that age estimates during second and third crossings differ more for lower metallicity models. This can be understood by the larger extent of blue loops at lower metallicity, since the evolutionary time scale along the loop is slowest near the turning point of the blue loop.
\item Position inside IS ($\Delta\log{t} \lesssim 0.25$). This is a consequence of the period-age-color relation \citep{2005ApJ...621..966B}. As Cepheids cross the IS, their pulsation periods change due to overall expansion or contraction of the outer envelopes. Since the position of the IS does not change much with metallicity, the age effect of the position inside the IS is relatively constant. 
\item Rotation ($\Delta\log{t} \lesssim 0.25$). Rotating stars have longer MS lifetimes than non-rotating stars due to an increased availability of hydrogen in the core. Comparing ages for 2nd and 3rd crossing Cepheids, we find age differences of up to $\Delta\log{t} \sim 0.25$ between non-rotating and $\omega = 0.5$ as well as $\omega = 0.9$ models (not shown in Fig.\,\ref{fig:PeriodAgeFit} for clarity).
 
\end{enumerate}

Based on these considerations, it is clear that individual Cepheid ages are uncertain to approximately $50\%$ if crossing numbers, IS position, and, importantly, the rotational histories are not known. On the other hand, these parameters can in principle be constrained by measured rates of period change (crossing number), color or effective temperature, and CNO surface abundances \citep{2014A&A...564A.100A}. As Fig.\,\ref{fig:PArelation_RotationDiff} shows, the effect of rotation on the period-age relation is similar to or greater than the effect of metallicity. Therefore, we recommend using period-age relations derived for models with average rotation ($\omega=0.5$), averaged over the 2nd and 3rd crossing and IS width in cases where a determination of these parameters is not feasible. Ages inferred from such mean relations tend to be approx. $\Delta \log{t} \sim 0.2 - 0.3$, i.e., $50 - 100\%$ higher than those predicted by literature relations without rotation  \citep{2005ApJ...621..966B,2003ARep...47.1000E}, depending on the period (more discrepant for longer periods, cf. Fig.\,\ref{fig:PArelation_RotationDiff}). 

\begin{table*}
\centering
\begin{tabular}{@{}ll@{\hspace{2mm}}rrrrrr@{\hspace{6mm}}rrrrrr@{}}
\hline
 & & \multicolumn{6}{c}{Fundamental Modes} & \multicolumn{6}{c}{First Overtones} \\
Z & Xing & $\alpha_b$ & $\beta_b$ & $\alpha_r$ & $\beta_r$ & $\alpha$ & $\beta$ & $\alpha_b$ & $\beta_b$ & $\alpha_r$ & $\beta_r$ & $\alpha$ & $\beta$\\
\hline
 & & \multicolumn{2}{c}{FU Blue Edge} & \multicolumn{2}{c}{FU Red Edge}  & \multicolumn{2}{c}{FU IS avg} & \multicolumn{2}{c}{1O Blue Edge} & \multicolumn{2}{c}{1O Red Edge} & \multicolumn{2}{c}{1O IS avg}\\
0.014 & 2nd & -0.702 & 8.481 & -0.573 & 8.527 & -0.532 & 8.393 & --  &  --  & -0.764 & 8.538 & -0.713 & 8.432 \\
          & 3rd  & -0.692 & 8.520 & -0.599 & 8.623 & -0.641 & 8.551 &  --  &  -- & -0.778 & 8.65 & -0.666 & 8.475 \\
0.006 & 2nd & -0.675 & 8.444 & -0.696 & 8.727 & -0.706 & 8.654 & -0.695 & 8.346 & -0.793 & 8.665 & -0.823 & 8.622 \\
          & 3rd & -0.656 & 8.497 & -0.620 & 8.685 & -0.671 & 8.653 & -0.671 & 8.394 & -0.804 & 8.713 & -0.840 & 8.694 \\
0.002 & 2nd & -0.896 & 8.610 & -0.916 & 8.844 & -0.827  & -0.859 & 8.706 & 8.41 & -1.166 & 8.873 & -1.065 & 8.726 \\ 
          & 3rd & -0.833 & 8.793 & -0.784 & 8.960 & -0.803 & 8.869 & -0.892 & 8.694 & -0.798 & 8.803 & -0.839 & 8.763 \\
\hline
 & & & & & & \multicolumn{2}{c}{FU Xing avg} & & & & & \multicolumn{2}{c}{1O Xing avg} \\
0.014 & avg & & & & & -0.592 & 8.476 & & & & & -0.633 & 8.406 \\
0.006 & avg & & && &  -0.665 & 8.628 & & & & & -0.825 & 8.651\\
0.002 & avg & & & & & -0.840 & 8.794 & & & & & -0.961 & 8.768\\   
\hline
\end{tabular}
\caption{Period-age relations ($\log{t} = \alpha \cdot \log{P} + \beta$) for $\omega=0.5$, i.e., {\it average} initial rotation, for second and third IS crossings, cf. Fig.\,\ref{fig:PeriodAgeFit}. Relations for unknown location within the IS, and also for unknown crossings are also provided. See text for important caveats regarding the use of period-age relations, which depend significantly on initial rotation rate $\omega$, as well as on IS crossing and IS position.}
\label{tab:res:PeriodAge}
\end{table*}

\subsection{Period-Radius Relations from Rotating Models}
\label{sec:res:radius}

\begin{figure*}
\centering
\includegraphics[]{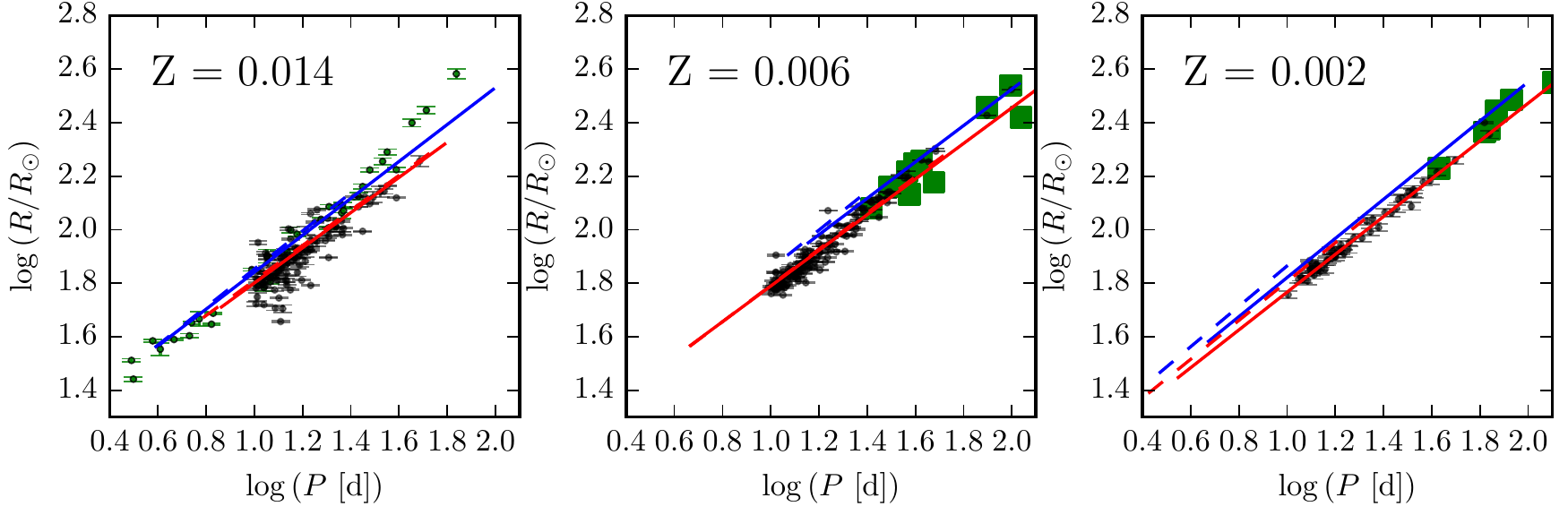}
\caption{Period-radius relations for fundamental-mode pulsation at the hot and cool IS boundaries (blue and red lines) for models with average rotation ($\omega=0.5$), for different metallicities. Second crossings are shown as dashed lines, third crossing as solid lines. We compare our predictions to radii empirically-determined by \citet[MW only, blue errorbars]{1998ApJ...496...17G}, \citet[black errorbars]{2012ApJ...748..107P}, and \citet[green squares, LMC \& SMC]{1999ApJ...512..553G}.}
\label{fig:res:period-radius_FU}
\end{figure*}

Since pulsation periods are approximately proportional to $M^{-0.5}R^{1.5}$, the effect of the radius on pulsation period significantly outweighs the effect of mass. As a result, there exists a nearly  one-to-one correspondence between radius and period \citep[as noted by][]{1998ApJ...496...17G}. Moreover, the PR relations at the blue and red IS edges are very close to each other and indistinguishable at the current level of empirical accuracy.
As seen in Fig.\,\ref{fig:res:period-radius_FU}, PR relations are essentially linear in the $\log{P} - \log{R}$ plane. 
We tabulate our results in Tab.\,\ref{tab:res:PeriodRadius}, where we fitted separately the relations at the blue and red edges of the IS for different metallicities and rotation rates. 

Figure\,\ref{fig:res:period-radius_FU} shows our period-radius relations for fundamental mode pulsators at the blue and red IS boundaries,  compared to radii of Milky Way Cepheids by \citet{1998ApJ...496...17G}, LMC and SMC Cepheids by \citet{1999ApJ...512..553G}, as well as MW, LMC, and SMC Cepheids by \citet{2012ApJ...748..107P}. Rotation and metallicity affect the PR-relation relatively weakly, despite the significant impact of either effect on radius and period predicted for a model of a given mass (lower metallicity: smaller radius; faster rotation: larger radius). 

Overall, the predicted relations agree very well with observations, despite large scatter among empirical results. This scatter may be partially explained by difficulties related to the projection factors needed to translate observed radial velocities into pulsational velocities \citep[e.g.][]{2007A&A...471..661N, 2016MNRAS.455.4231A,2016arXiv160104727B}. The short period end matches particularly well for solar metallicity, although the three longest period Cepheids by \citet{1998ApJ...496...17G} are systematically offset to larger radii. We further find that the empirical radii determined via a global modeling effort of many Cepheids  \citep{2012ApJ...748..107P} tend to cluster near our predicted red edge PR-relations. However, period-$T_{\rm{eff}}$ distributions by the same authors show the opposite trend, i.e., they cluster blueward of our hot edge (see Sec.\,\ref{sec:res:pTeff}). Unfortunately, radius and $T_{\rm{eff}}$ are not independent parameters in this study, and more empirical data is needed to investigate this point in more detail. Interferometric observations together with {\it Gaia} parallaxes will soon provide a much more detailed view of the period-radius relation.

\begin{table*}
\centering
\begin{tabular}{@{}ll@{\hspace{2mm}}rrrrrr@{\hspace{8mm}}rrrrrr@{}}
\hline
 & & \multicolumn{6}{c}{Fundamental Modes} & \multicolumn{6}{c}{First Overtones} \\
Z & Xing & $\alpha_b$ & $\beta_b$ & $\alpha_r$ & $\beta_r$ & $\alpha$ & $\beta$ & $\alpha_b$ & $\beta_b$ & $\alpha_r$ & $\beta_r$  & $\alpha$ & $\beta$ \\
\hline
 & & \multicolumn{2}{c}{FU Blue Edge} & \multicolumn{2}{c}{FU Red Edge} & \multicolumn{2}{c}{FU IS avg} & \multicolumn{2}{c}{1O Blue Edge} & \multicolumn{2}{c}{1O Red Edge} & \multicolumn{2}{c}{1O IS avg} \\
0.014 & 2nd &  0.710 & 1.144 & 0.654 & 1.157 & 0.652 & 1.180 &  --  &  --  & 0.719 & 1.225 & 0.708 & 1.246 \\
          & 3rd  & 0.689 & 1.152 & 0.657 & 1.144 & 0.674 & 1.150 &  --  &  --  & 0.720 & 1.208 & 0.699 & 1.241 \\
0.006 & 2nd & 0.722 & 1.133 & 0.677 & 1.115 & 0.687 & 1.124 & 0.754 & 1.217 & 0.743 & 1.185 & 0.751 & 1.191 \\
          & 3rd & 0.679 & 1.166 & 0.666 & 1.124 & 0.684 & 1.123 & 0.720 & 1.236 & 0.741 & 1.180 & 0.749 & 1.183 \\
0.002 & 2nd & 0.755 & 1.110 & 0.725 & 1.081 & 0.721 & 1.106 & 0.746 & 1.230 & 0.795 & 1.152 & 0.781 & 1.177 \\ 
          & 3rd & 0.730 & 1.091 & 0.706 & 1.061 & 0.716 & 1.079 & 0.788 & 1.164 & 0.756 & 1.154 & 0.768 & 1.160 \\
\hline
 & &  & & & & \multicolumn{2}{c}{FU Xing avg} & & & & & \multicolumn{2}{c}{1O Xing avg} \\
0.014 & avg   & & & & & 0.665 & 1.164 & & & & & 0.695 & 1.251 \\
0.006 & avg  & & & & & 0.683 & 1.126 & & & & & 0.749 & 1.188\\
0.002 & avg  & & & & & 0.737 & 1.078 & && &  & 0.795 & 1.154\\   
\hline
\end{tabular}
\caption{Period-radius relations ($\log{R/R_\odot} = \alpha \cdot \log{P} + \beta$) of fundamental mode and first overtone Cepheids based on models with {\it average} initial rotation, for second and third IS crossings along the blue and red edge, as well as for relations averaged over IS position alone (crossing separate), and IS position plus crossing number together, cf. Fig.\,\ref{fig:res:period-radius_FU}.}
\label{tab:res:PeriodRadius}
\end{table*}

\subsection{Period-Temperature Relation}\label{sec:res:pTeff}

The finite width of the IS naturally leads to a period-temperature relation, which has been a subject of extensive research \citep[e.g.][]{1958ApJ...127..513S,2009A&A...493..471S}  in the context of discussing the intrinsic dispersion of the PLR. While the observed dispersion tends to be dominated by reddening and extinction, the theoretical arguments for a significant period-color (or temperature) relation remain valid \citep[e.g.][]{2009pfer.book.....M}. 

\begin{figure*}
\centering
\includegraphics[]{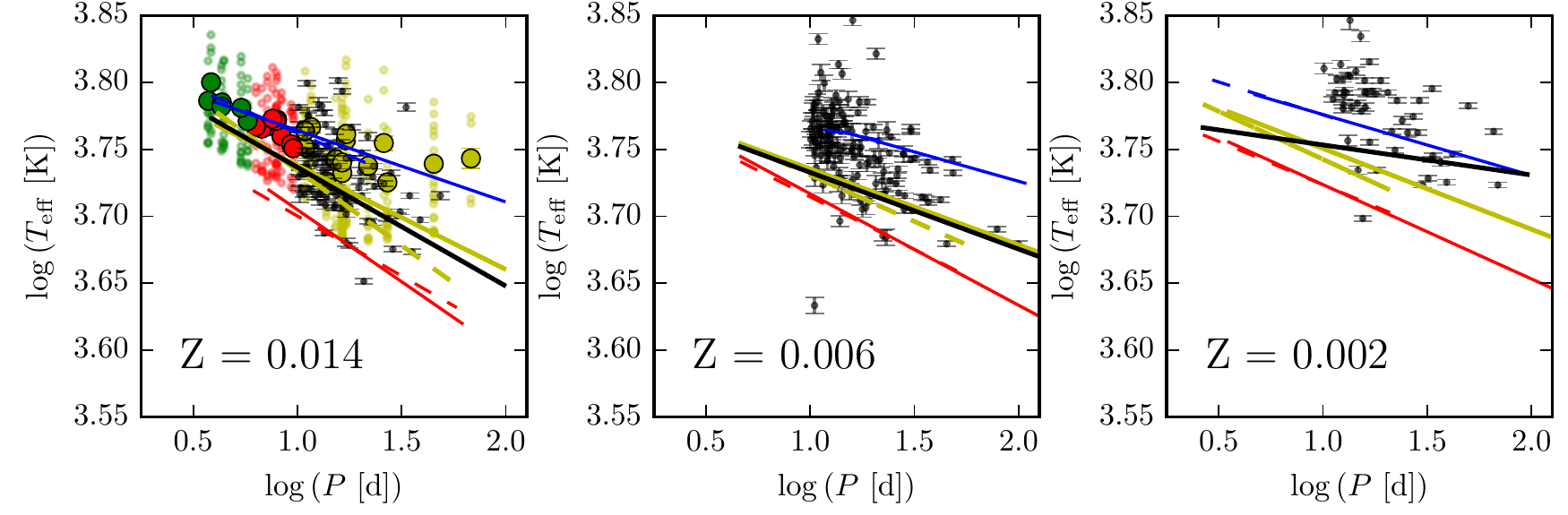}
\caption{Period-(average) effective temperature relations for fundamental mode Cepheids based on models with average rotation ($\omega=0.5$) and different metallicities. We show predictions at the blue and red IS edges in same colors, IS-averaged relations in yellow, and averages over crossing and IS position in black. Dashed lines represent second crossings, solid lines third crossings. We also show spectroscopically-determined $T_{\rm{eff}}$ of Galactic Cepheids \citep[left panel, where red, yellow, and green errorbars correspond to data from][]{2004AJ....128..343L,2005AJ....129..433K,2005AJ....130.1880A}. We plot their phase-averaged values as larger symbols with standard errors and the individual (phase-resolved) measurements as smaller circles of the same color. Average effective temperatures inferred by global modeling for Galactic, LMC and SMC Cepheids \citep{2012ApJ...748..107P} are shown as small black error bars. See text in Sec.\,\ref{sec:res:pTeff} for a discussion of the mismatch at lower metallicities.}
\label{fig:pTeff}
\end{figure*}

As Fig.\,\ref{fig:PeriodTeffMetallicity} shows, low metallicity models are hotter at fixed period than their high-metallicity counterparts. In Figure\,\ref{fig:pTeff}, we  thus compare our predicted $\log{P}$ and $\log{T_{\rm{eff}}}$ relations with empirical results from (1) spectroscopically-determined $T_{\rm{eff}}$ of Galactic Cepheids \citep{2004AJ....128..343L,2005AJ....129..433K,2005AJ....130.1880A} and (2) empirically inferred (by means of global pulsation modeling) mean temperatures by \citet{2012ApJ...748..107P}. 
While this provides a useful comparison, it is important to remember that there are conceptual differences between $T_{\rm{eff}}$ derived from spectral lines and $T_{\rm{eff}}$ predicted by the models. To wit, the temperatures of our models are computed from the Stefan-Boltzmann law, i.e., $T_{\rm{eff}}$ corresponds to the total power emitted per surface area, whereas spectroscopic temperatures depend on the temperature stratification over spectral line forming regions. 

For Galactic metallicity, the comparison between theory and data is very good. On the other hand,  the data for LMC and SMC Cepheids \citep{2012ApJ...748..107P} become increasingly discrepant with decreasing metallicity. Since $R$ and $T_{\rm{eff}}$ are not independently determined in the study by \citet{2012ApJ...748..107P} this apparent discrepancy is likely related to the apparent discrepancy in the period-radius relation discussed in Sec.\,\ref{sec:res:radius}.
For the SMC, where the discrepancy is largest, we find that an offset of $\Delta\log{T_{\rm eff}} \sim 0.025$ would be required to match predictions and empirical results. Such an offset is easily compensated by a change in radius of $\Delta \log{R} \sim 0.05$, which would preserve the same luminosity and is certainly consistent with the apparent discrepancy of radii shown in Fig.\,\ref{fig:res:period-radius_FU}. Unfortunately, we are at this point unable to conclude whether the empirical results by \citet{2012ApJ...748..107P} or our predictions require adjusting to resolve the matter.

\subsection{Rates of Period Change}\label{sec:res:pdot}

Here we take a look at the rate with which Cepheid pulsation periods are predicted to change due to secular evolution based on our results. 

The predicted rates of period change depend quite significantly on rotation, although no clear pattern emerges. Non-rotating models tend to indicate faster \Pdot\ than rotating models, although \Pdot\ does not monotonously depend on $\omega$. First overtone models exhibit much faster (a factor of a few up to 10) higher \Pdot\ than fundamental mode models. This is consistent with first overtone pulsators exhibiting faster and more erratic period changes than fundamental mode Cepheids \citep{2008AcA....58..313P}. Very long-period Cepheids, i.e., those which cross the IS only once (cf. Sec.\,\ref{sec:evol}) are nicely consistent with an extension of \Pdot\ observed for shorter periods. The rotation dependence of \Pdot\ provides a simple explanation for the observed scatter in the \Pdot\ - P relation. 

Our predicted \Pdot\ values are in general agreement with predictions by \citet{2014AstL...40..301F}, although our rotating models appear to better reproduce the slope (and dispersion) of the observed \Pdot. Moreover, our models start to deviate significantly from \citet{2014AstL...40..301F} for very long-period Cepheids ($\log{P} \gtrsim 1.7$). 

We compare predicted \Pdot\ to empirical values from the following sources. David Turner (private communication) kindly provided a recent update of his observational data for Galactic Cepheids \citep{2006PASP..118..410T}, comprising 66 Cepheids with negative recorded period changes (2nd crossing Cepheids) and 124 Cepheids with positive period changes (1st or 3rd crossing). For the LMC, we use the sample of fundamental-mode Cepheids ($160$ Cepheids with positive period changes, $184$ negative) by \citet{2001AcA....51..247P} based on OGLE-II data \citep{1997AcA....47..319U}. For the SMC, we use $213$ Cepheids with positive and $285$ Cepheids with negative \Pdot\ as listed in \citet{2002AcA....52..177P}.
\citet{2008AcA....58..313P} cautions that \Pdot\ values derived from insufficiently long temporal baselines are not necessarily suitable for comparisons with stellar evolution models, since such \Pdot\ values can be dominated by short-term fluctuations rather than secular evolution \citep[see also][]{2000NewA....4..625B,2009AstL...35..406B}. Studies of \Pdot\ over longer temporal baselines would be very useful for more accurate tests of stellar models.

\begin{figure*}
\centering
\begin{tabular}{cc}
\includegraphics{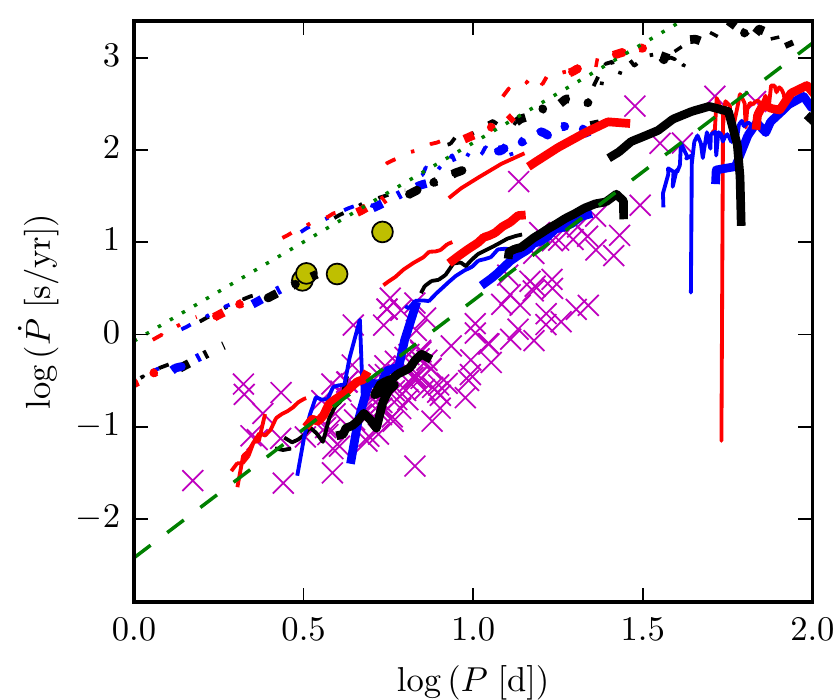} &
\includegraphics{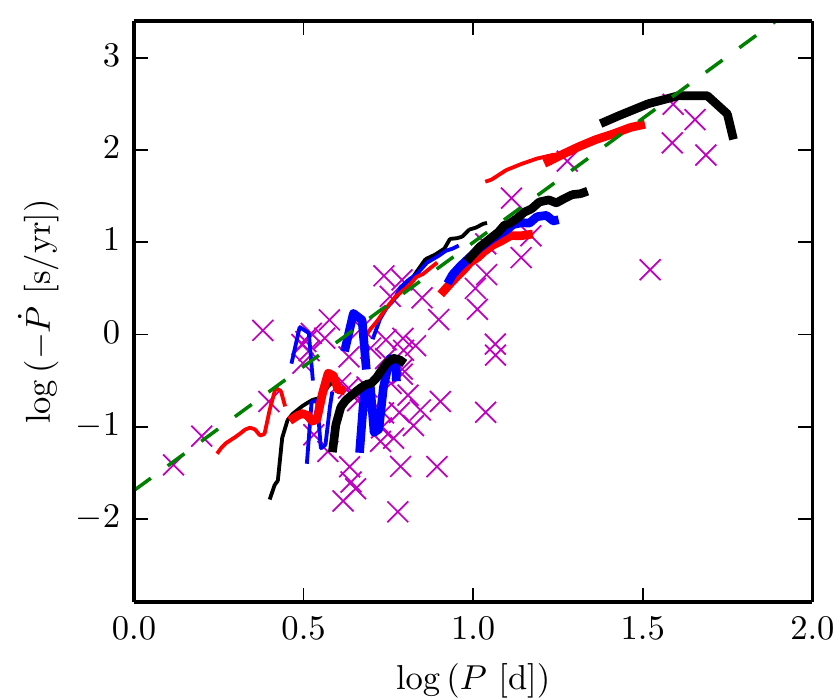}\\
\includegraphics{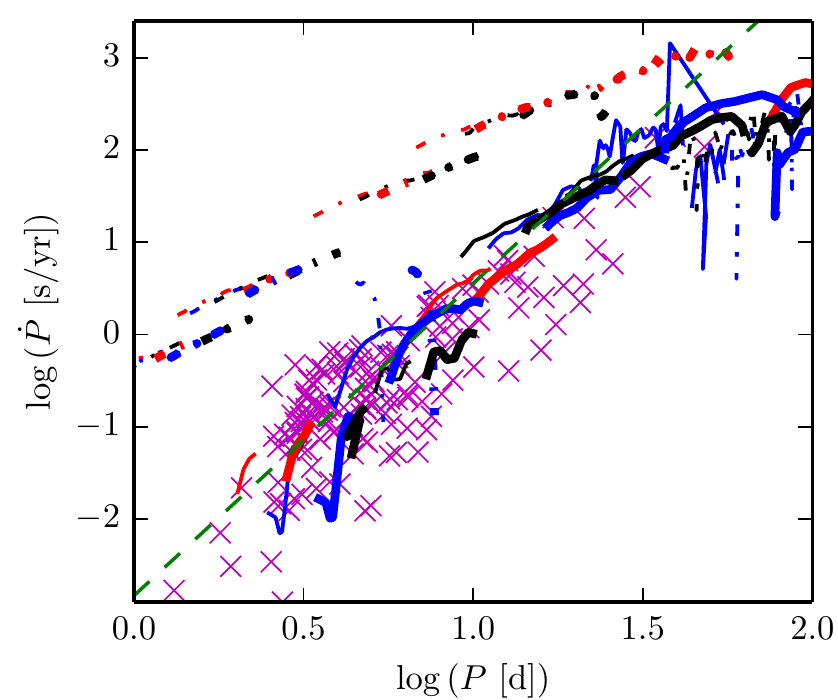} &
\includegraphics{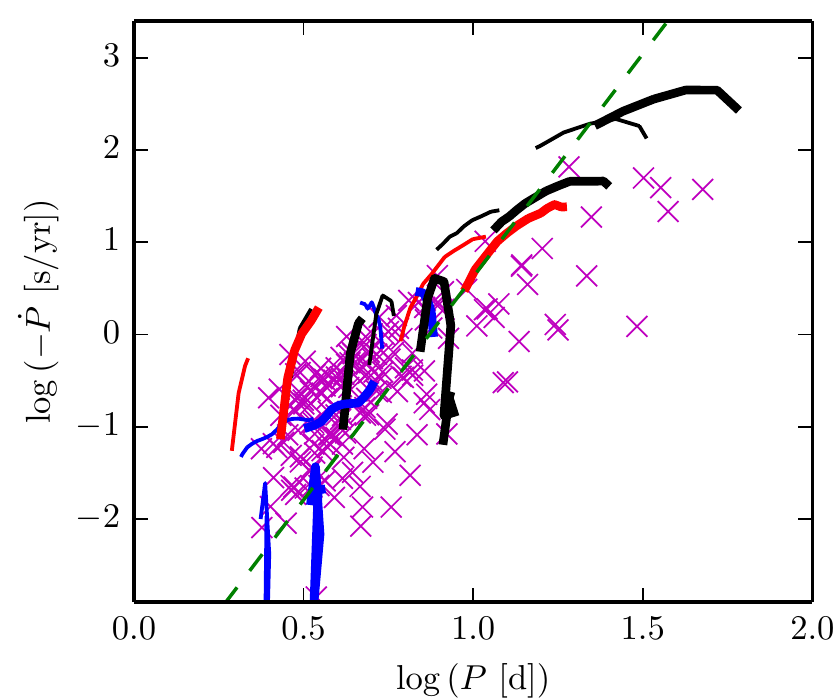} \\
\includegraphics{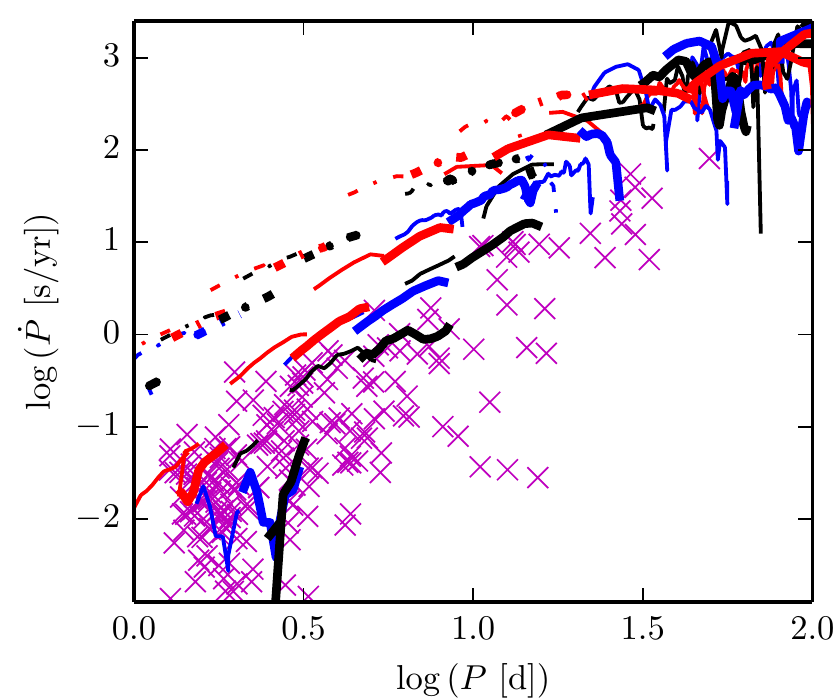} &
\includegraphics{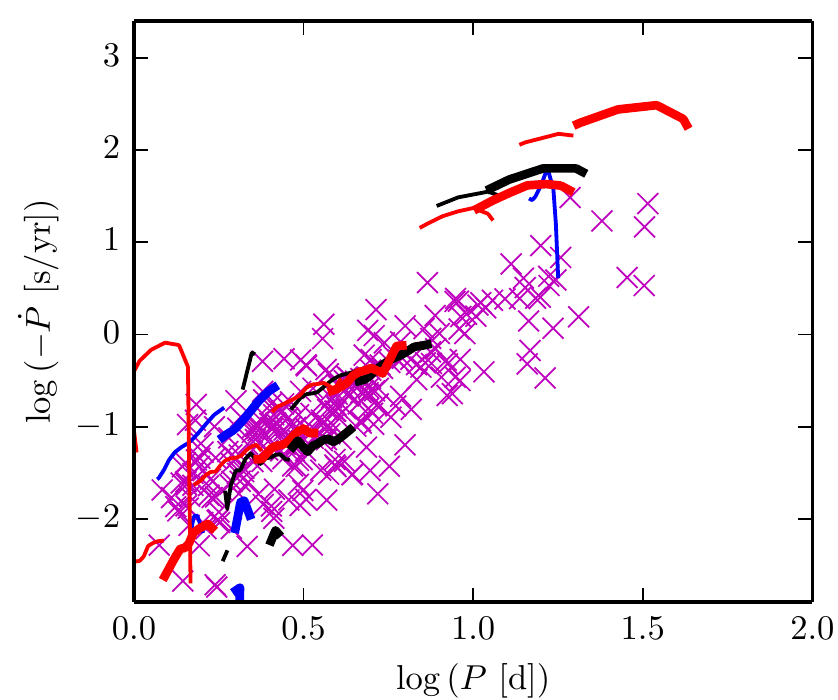}
\end{tabular}
\caption{Comparison between predicted and empirical rates of period change, \Pdot. Fundamental mode models are shown as thicker lines, first overtones as thinner lines. Initial rotation velocity is color coded: non-rotating models are red, average rotation black, fast rotation blue. Left hand panels show increasing periods (first, dashed-dotted lines, and third crossings, solid lines), right hand panels decreasing periods (second crossings). Metallicity decreases from top to bottom: $Z=0.014$ and data for Galactic Cepheids \citep{2006PASP..118..410T} with first crossing candidates Polaris, BY\,Cas, DX\,Gem, and HD\,344787 highlighted as yellow filled circles \citep{2013ApJ...772L..10T}; $Z=0.006$ and data from LMC Cepheids by \citet{2001AcA....51..247P}; $Z=0.002$ and data from SMC Cepheids by \citet{2002AcA....52..177P}. Theoretical $\dot{P}-P$ relations by \citet{2014AstL...40..301F} and \citet{2013AstL...39..746F} are shown as green dotted lines for 1st crossings and green dashed lines for 3rd crossings, where available.}
\label{fig:res:pdot}
\end{figure*}

We show this comparison for first overtone and fundamental mode pulsators with the above empirical data in Figure\,\ref{fig:res:pdot}, ordered by metallicity. 
Qualitatively, our predictions indicate that virtually all Cepheids are observed during the second or third crossing as expected by virtue of the fast evolution along the Hertzsprung gap. The best candidates for first crossing Cepheids are Polaris, DX\,Gem, BY\,Cas, and HD\,344787 (= BD+22\,3786) \citep{2013ApJ...772L..10T}. Two Cepheids (SZ\,Cas and AQ\,Pup) with periods longer than $10$\,d are found relatively close to the predicted first crossing range, although their location in the diagram also agrees with a third crossing and/or overtone pulsation. As argued in \citet{2014A&A...564A.100A}, high-mass Cepheids are more likely to be observed on first crossings than low-mass Cepheids, since the first crossing timescale is fractionally larger compared to other crossings in high-mass Cepheids. A more detailed investigation of SZ\,Cas and AQ\,Pup could be of interest to clarify their evolutionary status and pulsation mode.

Our predictions are in excellent agreement especially for short-period Cepheids ($\log{P} < 0.8$) on the third crossing and most Cepheids on the second crossing. However, for the low-metallicity (SMC) long-period Cepheids, our results predict apparently excessive \Pdot\ on the third crossing.

Overall, Fig.\,\ref{fig:res:pdot} indicates very good agreement between our results and observations over the full range of metallicities, with some obvious shortcomings for long-period SMC Cepheids. The effect of rotation on \Pdot\ is discernible, albeit not expressed linearly. As is the case for luminosity \citep{2014A&A...564A.100A}, rotation can both increase (up to a point) and decrease (beyond a maximal increase) the predicted rate of period change at a fixed period. This pattern is most likely due to the competing mixing and hydrostatic effects in rotating stars. 

Since rotation affects both luminosity and \Pdot, it should be possible to infer the initial (i.e., ZAMS) rotation rate of Cepheid progenitors. A first such test was recently done for the prototype $\delta$\,Cephei \citep{2015ApJ...804..144A} based on the evolution of the average density, rather than the pulsation period. With the present model predictions, it will be possible to make such comparisons more accurately in the near future. 

\section{Discussion}\label{sec:discussion}

In the preceding sections, we have compared our predictions of the IS borders in the HRD, as well as relations between period and luminosity, radius, effective temperature, and period changes with observations based on Galactic, LMC, and SMC Cepheids, finding generally very good agreement. These observational tests of our models are important for improving the understanding of mixing processes in intermediate-mass stars and their impact on the evolutionary paths. Likewise, demonstrating agreement between predictions and observations corroborates other predictions that are more difficult to test directly, or with sufficient accuracy. In the following, we discuss the implications of our work in terms of the Cepheid mass discrepancy, investigate various effects impacting the PLR, and show for the first time that the flux-weighted-gravity-luminosity relation (FWGLR) provides an additional means for determining precise distances to Cepheids using spectroscopy.

\subsection{The Cepheid mass discrepancy}\label{sec:MassDiscrepancy}
A long-standing issue in stellar evolution are mass discrepancies that have been identified in particular in intermediate- and high-mass stars. The ``symptom'' of a mass discrepancy is that evolutionary models yield systematically higher masses than other means for determining stellar masses \citep{1968QJRAS...9...13C,1969MNRAS.144..461S,1969MNRAS.144..485S,1969MNRAS.144..511S,2008ApJ...677..483K,2010Natur.468..542P,2012ApJ...749..108P}. From a stellar evolution modeling perspective, mass discrepancies can be ``solved'' either by increasing luminosity at fixed mass, or allowing for strong mass-loss in order to reduce the present-day mass at fixed luminosity. To achieve the former type of ``solution'' generally requires to increase the effective size of the convective core. Two main mechanisms are available to this end: convective overshooting and rotational mixing. 

In \citet{2014A&A...564A.100A} we argued that a combination of rotation and crossing number indeed offers a viable explanation for the currently accepted level to which masses are discrepant, i.e., $10-20\%$ \citep[e.g.][]{2008ApJ...677..483K}. 

\begin{figure}
\centering 
\includegraphics[]{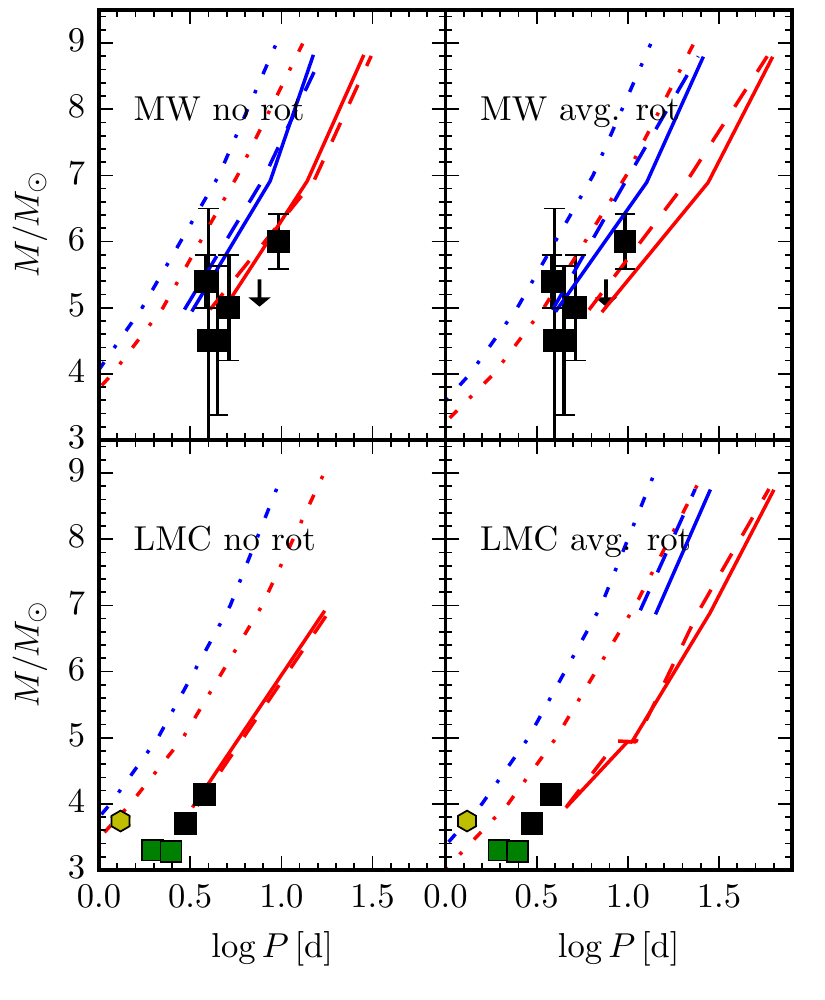}
\caption{
Predicted Cepheid masses against fundamental mode pulsation period compared to model-independent Cepheid masses from \citep[and references therein]{2006ApJ...647.1387E,2008AJ....136.1137E,2009AJ....137.3700E,2011AJ....142...87E} for Galactic Cepheids (upper panel), and masses for LMC Cepheids (lower panel) in eclipsing binary systems from the Araucaria project \citep{2010Natur.468..542P,2011ApJ...742L..20P,2014ApJ...786...80G,2015ApJ...815...28G}. Upper panels are for solar metallicity, and lower panels for LMC metallicity, whereas left panels show models without rotation and right panels those with average rotation ($\omega=0.5$). Relations are shown separately for first (dash-dotted), second (dashed) and third crossing (solid). Red lines represent the cool edge, blue lines the blue IS edge.
Observational data for fundamental mode Cepheids is plotted as black squares with errorbars, overtone pulsators as green squares with errorbars. OGLE-LMC-CEP-1812 (yellow hexagon, LMC) is an expected outlier as there is evidence it may have undergone a merger and does not appear to be a classical Cepheid \citep{2015A&A...581L...1N}. Note that blue loops of low-mass LMC models do not cross the blue IS edge, hence only red boundary predictions can be shown.}
\label{fig:NoMassDiscrepancy}
\end{figure}

In Fig.\,\ref{fig:NoMassDiscrepancy}, we provide an updated view of \citet[Fig.7]{2014A&A...564A.100A}, plotting directly-measured masses of Galactic and LMC as a function of period and comparing them to our model predictions for fundamental mode pulsation. The figure clearly shows the superior agreement of models with average rotation compared to models with no rotation. The finite mass-resolution of our model grid unfortunately limits our ability to predict pulsation periods of the lowest-mass models (see discussion in Sec.\,\ref{sec:periodrange}). The mass-period relations on the blue edge do not extend to low masses for LMC metallicity, since the blue loops of these models turn back towards the red giant branch before crossing the blue IS boundary. Additional directly measured Cepheid masses, especially of long-period Cepheids, in addition to a finer model grid towards lower masses are required to improve this comparison. Projects involving long-baseline interferometry \citep[e.g.][]{2014A&A...561L...3G} and astrometric orbits with {\it Gaia} will soon significantly improve this situation.

\subsection{Dependence of the PL-relation on IS position, crossing, metallicity, and rotation}
\label{sec:PLRdependence}

In Sect.\,\ref{sec:res:PLRs}, we have compared the predicted PLRs in four different photometric passbands (or combinations thereof) and found excellent agreement with observations over the full range of metallicities. Here, we discuss the relative impact of the four most relevant effects that introduce scatter in observed period-luminosity distributions of Cepheids in complex stellar populations. 

Besides initial mass, there are four main quantities that impact the predicted pulsation period and luminosity of a given classical Cepheid: (i) the position within the IS; (ii) the crossing number; (iii) metallicity; (iv) the initial (ZAMS) rotation rate, $\omega$. Each of these effects act on Cepheids at an individual level, thus causing scatter in observed P-L distributions. In the following, we investigate these effects separately using PLRs averaged over the effects not to be considered at this point. For instance, when considering the effect of the IS width on the PLR, we determine the PLRs for models on any crossing and with any initial rotation rate. In general, we use Solar metallicity models for these comparison. To study the effect of metallicity, we consider LMC and SMC relations relative to Solar, averaged over all $\omega$, crossings, and the IS width. To further facilitate comparison with experimental work, we carry out this comparison in four different photometric pass-bands or filter combinations: V-band, H-band, and Wesenheit indices $W_{VI}$ and $W_{H,VI}$. Of course, extinction, which provides an additional motivation for using Wesenheit indices in empirical studies, is not included in our models.

\begin{figure*}
\centering 
\includegraphics[]{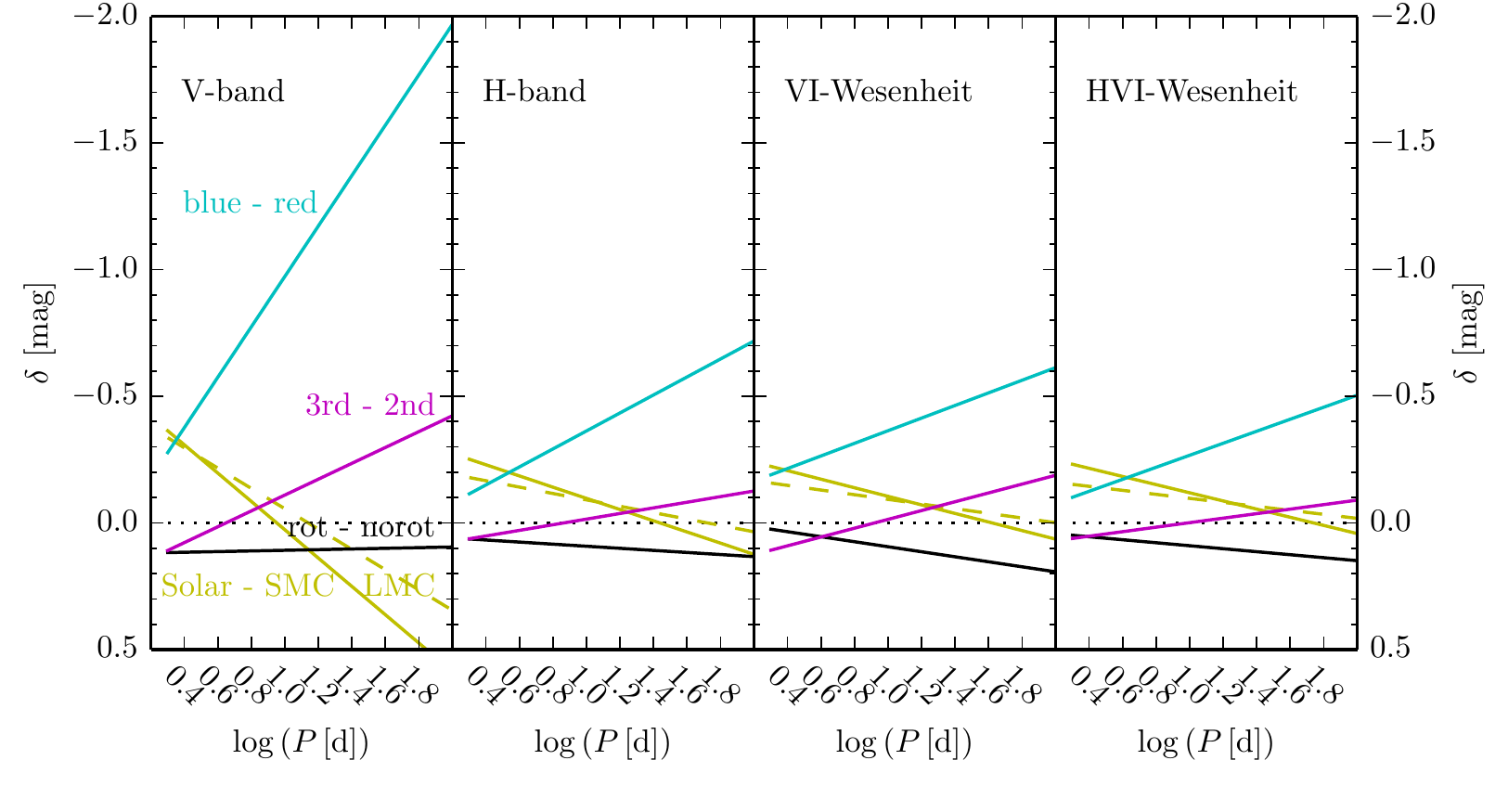}
\caption{Effects of IS position, IS crossing, metallicity, and rotation on the PLR. The abscissa shows $\delta = M_{p1} - M_{p2}$, where p1 and p2 correspond to the effects as labeled: blue minus red edge, 3rd minus 2nd crossing, etc. Panels from left to right for $V$-band, $H$-band, and the Wesenheit indices $W_{VI}$ and $W_{H,VI}$. Cyan solid lines show the effect of IS position (blue edge minus red edge), yellow lines show metallicity effects for Z=0.014 minus Z=0.006 (dashed) and Z=0.002 (solid). Magenta solid lines show the effect of Cepheids on a 3rd minus 2nd crossing. Black solid line shows the effect of rotation for average rotation relations minus non-rotating Cepheids. Interestingly, the effects of rotation and crossing number are of similar size with opposite sign. Note also the period-dependence of all the effects, in particular of metallicity and crossing number.}
\label{fig:PLReffects}
\end{figure*}

We compute magnitude differences as function of pulsation period for each of the four effects mentioned above. For instance, to understand the (maximal) impact of the IS width, we compute the difference in magnitudes for blue and red IS edge average PLRs, and so on. We thus illustrate the relative importance of neglecting IS width, crossing, metallicity, and rotation as a function of the filter combination used in Figure\,\ref{fig:PLReffects}. It should be noted that these magnitude differences are much larger than what is to be expected for observed PLR scatter of real Cepheid populations, since (1) observed PLR scatter represent a dispersion around a mean, whereas this comparison explores the {\it maximal} difference, e.g. between the two edges of the IS; (2) no assumption on the true distribution, e.g. of metallicity or initial rotation rates, is made, although initial rotation rates are not uniformly distributed and Cepheids in real populations will exhibit some differences in chemical composition. Population synthesis is required to predict the dispersion of realistic stellar populations, and we plan to carry out such simulations in the future using {\tt SYCLIST} \citep{2014A&A...566A..21G}. Similarly, population synthesis is also required to investigate whether our models indicate any non-linearities (breaks) in PLRs \citep[see][and references therein]{2013ApJ...764...84I,2016MNRAS.457.1644B}.

In most filters, the {\it width of the IS} is the dominating contribution to uncertainties in the PLR (Fig.\,\ref{fig:PLReffects}), leading to brighter Cepheids for hotter temperatures. At a period of $\sim 10$\,days, the difference of the PLR near the blue and red edges is approximately $1\,$mag in $V$-band, $0.4$\,mag in both $H$-band and $W_{VI}$, and $0.3$\,mag in $W_{H,VI}$. Reducing the width of the IS by using Wesenheit indices or $H-$band photometry is thus an elegant way of reducing the intrinsic scatter of the PLR, in addition to rendering observations less sensitive to reddening \citep[see also][]{2010ApJ...715..277B}.

The {\it metallicity} effect of the PLR is opposite to the effect due to the IS boundaries, yielding fainter Cepheids for lower metallicity models as period increases. Interestingly, the PLR difference due to metallicity is roughly equal to the difference due to the width of the IS at short periods ($\log{P}\sim 0.4$). In $H$-band as well as both Wesenheit indices, this difference becomes noticeably smaller than in $V$-band and develops a nearly opposite slope to the IS width effect. Specifically, it is noteworthy that the metallicity effect on the PLR vanishes for certain periods, e.g. near $\log{P} = 1$ in $V-$band, $1.2$ (16\,d) in $H$ and $W_{VI}$, and $1.4$ (25\,d) in $W_{H,VI}$. In essence, this effect is much weaker than the IS width for all periods longer than $10$\,d. We note that the period-dependence of the PLR's metallicity dependence implies that the observed distance bias to a given sample of Cepheids (e.g. in distant galaxies) depends on the sampled period distribution as well as the filter set used. This is important when aiming to quantify the metallicity effect on distance moduli derived from Cepheids  \citep[e.g.][and references therein]{1998ApJ...498..181K,2004ApJ...608...42S,2008A&A...488..731R}.

The {\it crossing number} affects the PLR most strongly in optical passbands. While the luminosity difference between 2nd and 3rd crossing is reduced in $W_{VI}$, the $H-$band appears to be an even better choice in this regard. Combining $H-$band and $V-I$ color in the $W_{H,VI}$ index leads to the weakest impact of crossing differences.

Figure\,\ref{fig:PLReffects} also shows for the first time the {\it effect of rotation} on the PLR, to which we alluded in prior work \citep{2014A&A...564A.100A}. As discussed there and in Sec.\,\ref{sec:evol} above, rotating models predict more luminous Cepheids for a given initial mass than non-rotating models. However, this relationship is not monotonous due to competing mixing (luminosity-increasing) and hydrostatic (luminosity decreasing) effects, with a maximal luminosity for $\omega \sim 0.5$. Similarly, rotating models have lower luminosity than non-rotating models at a given period. This is markedly different from the effect of rotation on the mass-luminosity relation, where rotating models have higher luminosity for a given mass than models without rotation. This behavior is to be expected since the PLR is a projection of the PL-$T_{\rm{eff}}$ relation, which results from the mass-luminosity relation and the pulsation equation $P \propto 1/\sqrt{\bar{\rho}}$ \citep{1879Ritter}, see also \citet{2009pfer.book.....M}. 

While rotation consistently has the weakest effect on the PLR of all the effects considered, it is the only effect that grows towards longer wavelengths. Similarly to the opposite effects of IS width and metallicity, rotation acts in opposition to the crossing effects ($V-$band is an exception here). 

While it has previously been known that Wesenheit indices collapse the width of the IS, our work shows that such filter combinations also reduce the impact of other effects such as the luminosity span of blue loops (difference between crossings) and even rotation. Among the filter combinations explored here, the Wesenheit index $W_{H,VI}$ is an ideal choice to minimize the intrinsic scatter of the PLR \citep[see also][]{2013MNRAS.434.2866F}. From the point of view of limiting the intrinsic scatter of observed PLRs, we recommend the period range $0.8 \lesssim \log{P} \lesssim 1.4$. Observations in $W_{H,VI}$ in this period range are predicted to show the least scatter due to the physical properties of the Cepheids, and are ideal due to the $H-$band's natural insensitivity to extinction, and the Wesenheit index's ``reddening-free'' nature.

\begin{table}
\centering 
\begin{tabular}{@{}lrrrrrr@{}}
& \multicolumn{2}{c}{$Z = 0.014$} & \multicolumn{2}{c}{$Z = 0.006$} & \multicolumn{2}{c}{Metallicity} \\
$\log{P}$ & \multicolumn{2}{c}{blue - red} & \multicolumn{2}{c}{blue - red} & \multicolumn{2}{c}{0.014 - 0.006} \\
 & $W_{VI}$ & $W_{H,VI}$ & $W_{VI}$ & $W_{H,VI}$ & $W_{VI}$ & $W_{H,VI}$ \smallskip \\
\hline
0.301 & -0.190 & -0.101 & -0.342 & -0.299 & -0.157 & -0.152  \\
0.390 & -0.212 & -0.123 & -0.356 & -0.308 & -0.149 & -0.145  \\
0.480 & -0.234 & -0.144 & -0.369 & -0.317 & -0.141 & -0.138  \\
0.569 & -0.257 & -0.165 & -0.383 & -0.326 & -0.133 & -0.131  \\
0.659 & -0.279 & -0.186 & -0.397 & -0.335 & -0.124 & -0.124  \\
0.748 & -0.301 & -0.207 & -0.411 & -0.344 & -0.116 & -0.117  \\
0.838 & -0.324 & -0.228 & -0.425 & -0.353 & -0.108 & -0.110  \\
0.927 & -0.346 & -0.250 & -0.439 & -0.362 & -0.100 & -0.103  \\
1.016 & -0.368 & -0.271 & -0.453 & -0.371 & -0.091 & -0.096  \\
1.106 & -0.390 & -0.292 & -0.466 & -0.380 & -0.083 & -0.089  \\
1.195 & -0.413 & -0.313 & -0.480 & -0.389 & -0.075 & -0.082  \\
1.285 & -0.435 & -0.334 & -0.494 & -0.397 & -0.067 & -0.075  \\
1.374 & -0.457 & -0.356 & -0.508 & -0.406 & -0.059 & -0.067  \\
1.463 & -0.479 & -0.377 & -0.522 & -0.415 & -0.050 & -0.060  \\
1.553 & -0.502 & -0.398 & -0.536 & -0.424 & -0.042 & -0.053  \\
1.642 & -0.524 & -0.419 & -0.550 & -0.433 & -0.034 & -0.046  \\
1.732 & -0.546 & -0.440 & -0.563 & -0.442 & -0.026 & -0.039  \\
1.821 & -0.568 & -0.462 & -0.577 & -0.451 & -0.017 & -0.032  \\
1.911 & -0.591 & -0.483 & -0.591 & -0.460 & -0.009 & -0.025  \\
2.000 & -0.613 & -0.504 & -0.605 & -0.469 & -0.001 & -0.018  \\
\hline
\end{tabular}
\caption{Tabulated PLR differences due to the finite width of the IS, calculated as blue edge minus red edge, and due to metallicity differences as Solar minus LMC metallicity for Wesenheit indices $W_{VI}$ and $W_{H,VI}$ as shown in Fig.\,\ref{fig:PLReffects}. For a fixed period, Cepheids near the blue edge are brighter than those near the red edge, and Solar metallicity Cepheids brighter than LMC metallicity Cepheids, hence the tabulated magnitude differences are negative.}
\label{tab:PLRdifferences}
\end{table}

\subsection{The Flux-Weighted Gravity-Luminosity Relation for classical Cepheids}\label{sec:FWGLR}

\citet{Kudritzki2003} \citep[see also][]{Kud2008}  proposed the use of a new empirical relation for distance determinations of blue supergiants, the flux-weighted-gravity-luminosity relation (FWGLR). This relation links the bolometric luminosity of a star with a factor defined as $g/T_{\rm eff,4}^4$, where $g$ is the surface gravity and $T_{\rm eff,4}$ the effective temperature expressed in units of $10\,000$ K. The spectroscopic determination of the quantity $g/T_{\rm eff,4}^4$ provides access to the intrinsic luminosity of the star using a well-calibrated FWGLR and thus to its distance. As shown by \citet{Kudritzki2003}, blue supergiants obey a very tight relation between luminosity and the quantity $g/T_{\rm eff,4}^4$.

For a class of stars to obey such a relation, two conditions must be met. First, within the class, stars of a given initial mass should exhibit little variation in both luminosity and (current) mass, so that a given initial mass star in the class can be associated to a well defined luminosity and vice-versa. Second, stars of the class considered should follow a well-defined luminosity-to-mass relation\footnote{This is because $g/T_{\rm eff,4}^4 \propto M/L$.}. 

Recently, \citet{MeynetKud2015} studied the possible sources of scatter in the FWGLR of blue supergiants from a theoretical point of view 
and reached two conclusions: (i) the FWGLR for blue supergiants is a robust relation that exhibits little dependence on many intricacies of stellar models;  
(ii) the very small observed scatter provides interesting constraints on the post Main-Sequence evolution of massive stars. Specifically, they showed that the observed small scatter is incompatible with many blue supergiants having evolved to their blue location as a result of strong mass loss during a previous red supergiant phase. Here we investigate the FWGLR in the context of classical Cepheids to address the following questions: Do Cepheids follow such a relation? Is the FWGLR useful for determining distances, and/or for constraining stellar models?

\begin{figure*}
\centering
\includegraphics[scale=0.60, angle=-90]{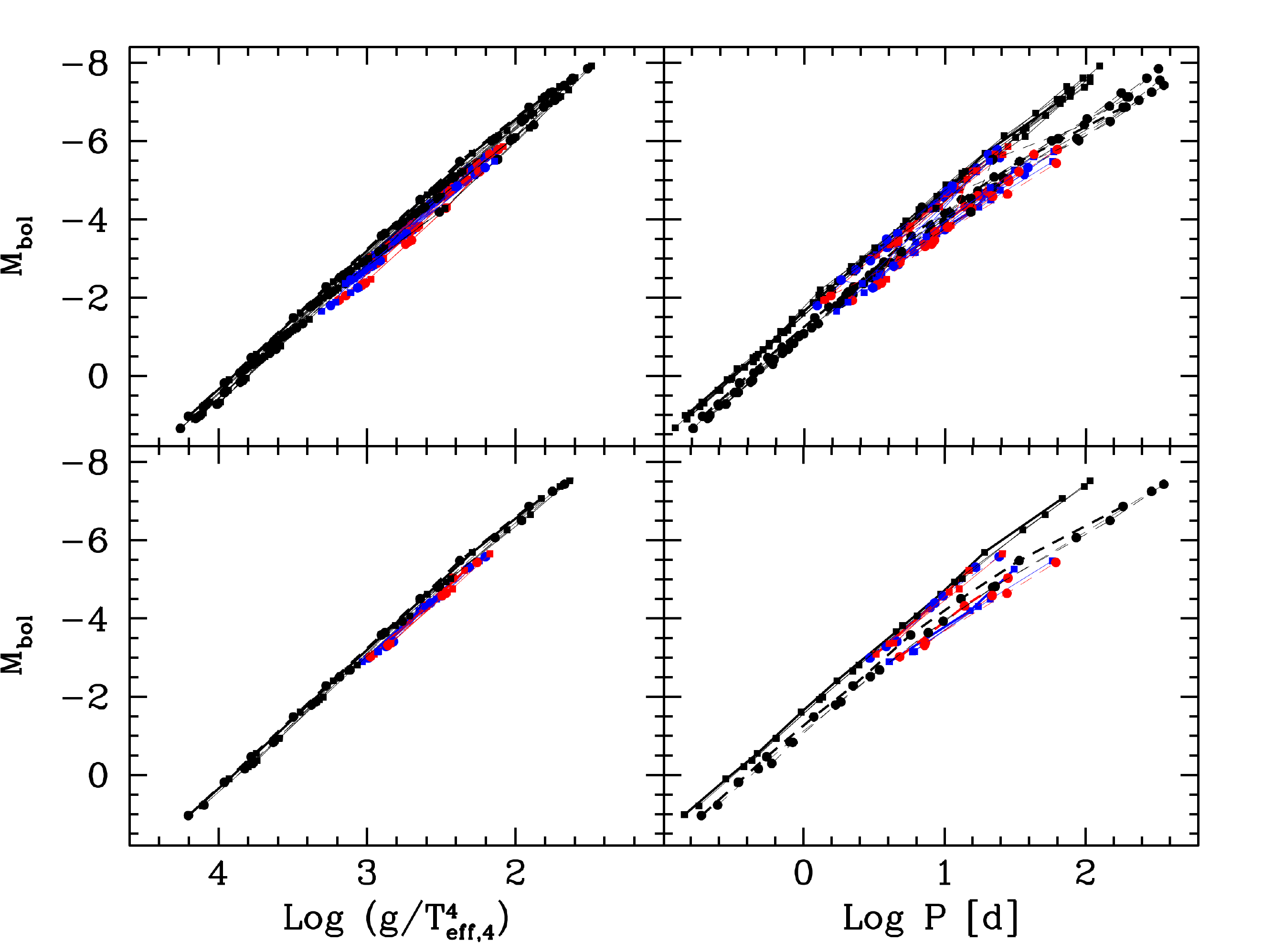}
\caption{Comparison between the FWGLR (left panels) and the PLR (right panels). We show bolometric magnitudes of Cepheids of different initial masses  as a function of $g/T_{\rm eff,4}^4$ (left panels) and as a function of the period (right panels). Differently colored lines correspond to first (black), second (blue), and third (red) crossing. The points along the lines correspond to the position of various initial mass stellar models (increasing from left to right). The filled squares correspond to the entry point in the IS and the filled disks to the exit point. The upper panels show the situation when all the stellar models for different initial masses, metallicities and rotations are plotted, that means all the models shown in Tables \ref{tab:models_z14} to \ref{tab:models_z02}. The lower panels show only solar metallicity models.}
\label{fig:fwglr-cep-georges}
\end{figure*}

In Fig.~\ref{fig:fwglr-cep-georges}, we compare our predicted FWGLRs and PLRs based on our analysis. The upper panel shows all predictions listed in Tables~\ref{tab:models_z14} through \ref{tab:models_z02}. 
We thus find the scatter of the FWGLR to be much smaller than the expected scatter of the PLR. In particular the shift in M$_{\rm bol}$ at a fixed period due to the width of the IS is greatly reduced in the FWGLR, and independent of the crossing considered. 

The fact that the FWGLR can be constructed from spectroscopy means that information on the current surface abundances (as a proxy for initial metallicity) can be obtained at the same time. Hence, it is possible to construct FWGLRs for specific metallicities. As the lower left panel of Fig.~\ref{fig:fwglr-cep-georges} shows, the scatter in the FWGLR is further reduced when a single metallicity is considered. We thus find that Cepheid models do follow a very well defined FWGLR, answering our first question. 

To answer the second question whether the FWGLR is useful in terms of distance determinations of Cepheids, we consider the benefits and shortcomings of such an approach.
The primary advantages are that the FWGLR is intrinsically much tighter than the PLR and that the (weak) dependence on metallicity can be dealt with explicitly and simultaneously, see above. However, the main drawback of the method is its need for time-resolved high-resolution spectroscopy, which makes it observationally more expensive than photometric methods. Future extremely large telescopes (ELTs) with their advanced capabilities such as AO-assisted spectroscopy and large collecting surface may enable applications of this method to extragalactic Cepheids well beyond the Magellanic Clouds.

Apart from its possible distance applications, the FWGLR provides interesting constraints for stellar models, in particular with respect to the mass discrepancy (cf. Sec.\,\ref{sec:MassDiscrepancy}), since $g/T_{\rm eff,4}^4 \propto M/L$. Hence, spectroscopically determined  $\log{g}$ and $T_{\rm{eff}}$ can be compared to evolutionary $T_{\rm{eff}}$ and gravity (as derived from the predicted mass and radius, assuming sphericity).  

\begin{figure}
\centering
\includegraphics[]{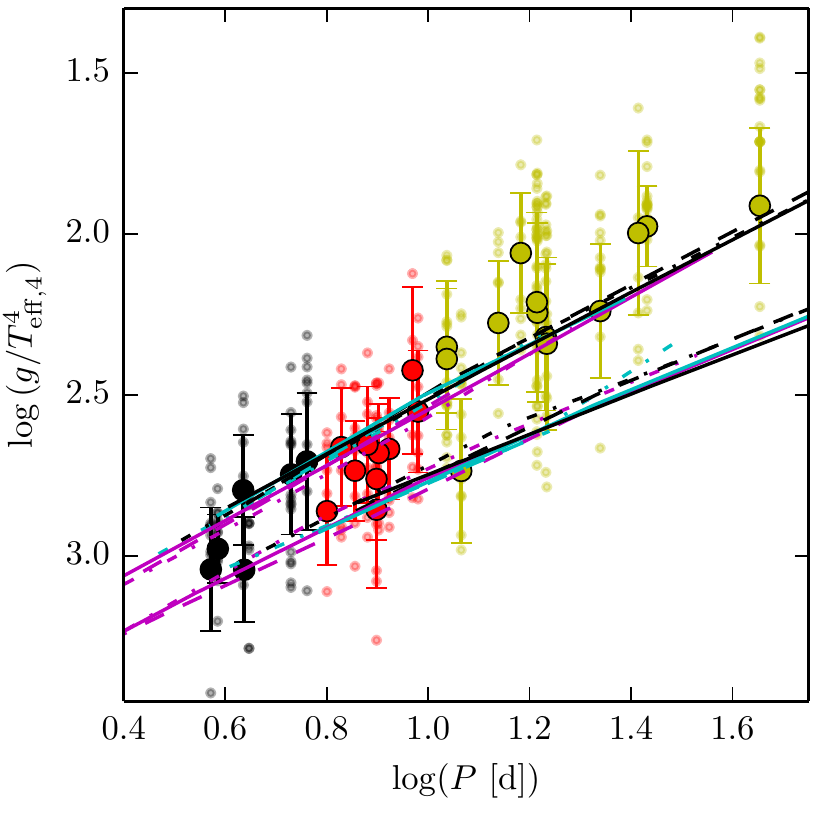}
\caption{FWGLR for classical fundamental mode Cepheids: predictions (lines, $Z=0.014$) and observations (circles). Line styles (dash-dotted, solid, and dashed) represent different initial rotation rates (non-rotating, average, fast), and crossing numbers (first, second, and third) are shown with different colors (magenta, cyan, and black). Data shown are from: \citet{2004AJ....128..343L,2005AJ....129..433K,2005AJ....130.1880A}. Phase-dependent data are drawn as transparent circles. Phase-averaged  values are shown as errorbars with rough estimates of uncertainties (standard deviation of individual values, centered on center value of observed range). Agreement between predictions and observations is excellent for periods below $10$\,d. Periods longer than $10$\,d \citep[yellow errorbars, data from][]{2005AJ....129..433K} appear systematically offset to lower values of $\log{(g/T_{\rm{eff},4}^4)}$.}
\label{fig:FWGLR}
\end{figure}

Figure\,\ref{fig:FWGLR} shows this comparison using data from \citet{2004AJ....128..343L}, \citet{2005AJ....129..433K}, and \citet{2005AJ....130.1880A} using both their phase-dependent and phase-averaged values. The identical data treatment and phase-resolved measurements provided by these authors renders their measurements ideal for this comparison. We thus find very good agreement between predicted and empirical FWGLRs for periods shorter than approx. 10 days. For longer periods, however, the predicted FWGLR tends to yield systematically higher values. We have not included here the measurements by \citet{2013MNRAS.432..769T}, since these authors report a possible systematic uncertainty in their spectroscopic $\log{g}$ values, which leads to $g/T_{\rm eff,4}^4$ values in conflict with the observations by Luck, Andrievsky, Kovtyukh, et al.

One possible explanation for this apparent mismatch is that higher-mass Cepheids ($> \sim 8$ M$_\odot$) have higher spectroscopic masses than evolutionary masses, i.e., that an inverted mass discrepancy exists in this mass range. However, this possibility can almost certainly be readily excluded, since it would require an {\it increase} of the mass deduced from the stellar models by approximately a factor of 2 to reconcile the evolutionary with the spectroscopic masses, which seems unreasonable. 

A more likely explanation for the mismatch at longer periods could be related to systematics involved in the spectroscopic measurements. Long-period Cepheids have larger radii and increasingly diluted atmospheres, where non-LTE effects and sphericity become increasingly important. More generally, there is also a conceptual difference between the optical-depth dependent temperature distribution  responsible for the spectral characteristics and the effective temperature defined by the total power emitted by surface area via the Stefan-Boltzmann law \citep[see e.g.][Chap. 14]{2005oasp.book.....G}. Further research into these areas would provide a welcome opportunity to directly test the M-L relations of Cepheids, thereby providing important tests of the mixing and transport mechanisms acting in intermediate and high-mass stars.

\section{Conclusions}\label{sec:conclusions}
We here present the first pulsational instability analysis of stellar evolution models that include a treatment of rotation for the mass range relevant for classical Cepheid variable stars. Our models \citep{2013A&A...553A..24G} cover a range of metallicities corresponding to the values of the Sun, the LMC, and the SMC, and three different initial rotation rates (non-rotating, average rotation, fast rotation). Our analysis includes both fundamental mode and first overtone pulsation. We determine hot and cool (blue and red) boundaries of the classical IS via a linear non-adiabatic radial pulsation code that takes into account the perturbation of convective flux due to pulsations. We find that this perturbation, while required to model the red IS edge, is also required to accurately reproduce the location of the blue IS edge. We do not find a strong dependence of the blue IS edge on rotation, and only a weak dependence on metallicity. The red edge, however, is slightly more strongly affected by both effects.

We treat rotation as a purely evolutionary effect, i.e., mechanical effects are not included as additional perturbations in the computation of pulsation periods. Within the derived IS boundaries, we predict pulsation periods and rates of period change. Using the remaining parameters from our models, we construct relations between period and luminosity, age, radius, and temperature. Therefore, all relations presented here are based directly on the stellar structures computed, and not on adopted relations, e.g. for mass and luminosity. 

We confront each of these predictions with observations and find generally excellent agreement, in particular for PLRs. Notably, we find that our models with rotation tend to better reproduce observed trends than our models with zero initial rotation.

We discuss our results with regards to the much discussed Cepheid mass discrepancy, of which we find no indication when using models including the effects of rotation. We confirm that rotation indeed contributes to the intrinsic scatter of the PLR \citep{2014A&A...564A.100A}, albeit at a low level.   We discuss the impact of several effects on the PLR, including the intrinsic width of the IS, the crossing number, metallicity, and rotation for each of four photometric bandpasses or filter combinations. Based on this discussion, we find that all these effects are significantly period-dependent and that the intrinsic scatter of the PLR increases with period. We determine that the Wesenheit index $W_{H,VI}$ is an ideal choice for limiting these dispersing effects in addition to being insensitive to extinction. Finally, we find that the period range between $0.8 \lesssim \log{P} \lesssim 1.4$ is ideal in terms of the various sources contributing to the scatter of the PLR.

We further take a first look at the FWGLR of classical Cepheids, which we show to form an intrinsically tighter relation than the PLR. The FWGLR of Cepheid deserves observational follow-up, since it provides direct tests of the mass-luminosity relation of Cepheids and provides a potentially interesting, complementary alternative to the PLR for extragalactic distance estimation. 

We find that period-age relations of Cepheids are expected to exhibit a large scatter due to differences in initial rotation, crossing number, metallicity, and the width of the IS. Depending on period, rotation increases Cepheid ages by approximately $\Delta\log{t} \sim 0.2 - 0.3$ compared to period-age relations that do not take into account rotational mixing effects. This should be taken into account when using Cepheid periods for dating star formation events, e.g. in the Galactic Bulge or in an extragalactic context. 

Our predicted rates of period change are in excellent agreement with observations. In particular, rotation helps to better reproduce the observed distribution and explains its intrinsic scatter at fixed $\log{P}$. We confirm that the Cepheids Polaris, BY\,Cas, DX\,Gem, and HD\,344787 are likely undergoing a first crossing of the IS based on their observed rates of period change \citep{2006PASP..118..410T,2013ApJ...772L..10T}. In general, rotating Cepheids change pulsation more slowly than non-rotating ones, although this trend does not monotonously depend on initial rotation rate. This effect may be understood as being due to a generally slower secular evolution for rotating stars (rotating stars live longer and spend more time in the IS).

For long-period Cepheids, additional information besides rates of pulsation period change is required to conclude on first or third crossings of the IS. Our models show that the predicted \Pdot\ values for both crossings approach each other, before blue loop evolution is eventually suppressed for higher mass models, i.e., also at longer periods. Crucial tests for telling crossing numbers include absolute magnitudes, e.g. via {\it Gaia} parallaxes, and CNO abundances \citep{2014A&A...564A.100A}.

In summary, we find that the evolutionary effect of rotation has important consequences for the pulsational properties of classical Cepheid variable stars, increasing (slightly) the intrinsic scatter of the crucial PLR, increasing (significantly) inferred ages, and altering the way with which Cepheids change pulsation period during their secular evolution. 

We project that initial mass and initial rotation rate can be inferred by measuring highly accurate absolute magnitude (requires high accuracy parallax), color, and rate of period change \citep[e.g.][]{2015ApJ...804..144A}. The ESA space mission \textit{Gaia} will soon provide highly accurate parallaxes of thousands of Cepheids \citep{2012Ap&SS.341..207E}, thus enabling even more accurate tests of these models. 

\section*{Acknowledgments}
We thank David G. Turner for communicating updated information on the observed rates of period changes for Galactic Cepheids. We thank the anonymous referee for a detailed and constructive report that helped to improve the quality of this manuscript. RIA acknowledges useful discussions with S. Casertano and A. Riess. 

RIA acknowledges funding from the Swiss National Science Foundation (SNSF) through an Early.Postdoc Mobility Fellowship. This work was partly supported by the SNSF (project number 200020-146401) to GM. SE and CG also acknowledge funding from the SNSF.

This research has made use of the following services: the CDS Astronomical Databases
SIMBAD and VizieR catalogue access tool\footnote{Available at
\url{http://cdsweb.u-strasbg.fr/}} and NASA's Astrophysics Data System.

\bibliographystyle{aa}
\bibliography{Bib_mine,Bib_Cepheids,Bib_StellarEvolution,Bib_StellarRotation,Bib_Pulsation,Bib_general} 



\Online

\begin{appendix}
\section{Physical quantities of evolutionary models at the boundaries of the instability strip}\label{app:tables}

Tables\,\ref{tab:models_z14}, \ref{tab:models_z06}, and \ref{tab:models_z02} provide physical quantities of models entering and exiting the IS for fundamental mode pulsation for metallicities $Z=0.014$, $0.006$, and $0.002$. Tables\,\ref{tab:models_z14_FO} through \ref{tab:models_z02_FO} provide the same information for first overtone pulsation.
The first integer of each row stands for the 'n-th' entering into the IS, where $^*$ indicates that the evolution track returns to the same side of the the IS, e.g., where blue loops do not extend all the way to the position where the hot IS boundary is to be expected based on models on first crossings or extrapolation from higher-mass models. Models labeled as $4^{\rm{th}}$ crossings indicate cases where numerical instabilities introduce additional helium into the core, leading to additional small (likely unphysical) loop patterns near the blue loop turning point, cf. He-spikes in Sect.\,\ref{sec:evol}.

\begin{table*}
\centering
\caption{Parameters of fundamental mode pulsation for models entering and exiting IS for $Z=0.014$ }
\begin{tabular}{@{}c@{\hspace{2mm}}r@{\hspace{6mm}}r@{\hspace{2mm}}r@{\hspace{2mm}}r@{\hspace{2mm}}r@{\hspace{2mm}}r@{\hspace{2mm}}r@{\hspace{2mm}}r@{\hspace{6mm}}r@{\hspace{2mm}}r@{\hspace{2mm}}r@{\hspace{2mm}}r@{\hspace{2mm}}r@{\hspace{2mm}}r@{\hspace{2mm}}r@{\hspace{2mm}}r@{}} 
\hline
 & & \multicolumn{7}{c}{{\bf Entering stage}} & \multicolumn{7}{c}{{\bf Exiting stage}}  \\
Xing & $M_{\rm{ini}}$ & $\log t$ & P~~ & $M$~~ & $\log T_{\rm eff}$ & $\log L$ & $Y_{\rm c}$ & $V_{\rm e}$~~~~ & N/C & N/O & $\log \Delta t$ & P~~  & $\log T_{\rm eff}$ & $\log L$ & $Y_{\rm c}$ & $V_{\rm e}$~~~~\\
 & [$M_\odot$] & [yr] & [d] & [$M_\odot$] & $[K]$ & [$L_\odot$] & & [km/s] & num & num & [yr] & [d] & [K] & [$L_\odot$] & & [km/s]\\
\hline
& & \multicolumn{14}{c}{$\omega_{\rm ini} = 0.0$} \smallskip \\
 1 & 2.0 & 9.012 &  0.143 &  2.00 & 3.872 & 1.494 & 0.986 & 0.0 & 0.247 & 0.132 & 7.096 &  0.189 &  3.831 & 1.486 & 0.986 & 0.0 \\
 1 & 2.5 & 8.743 & 0.282 & 2.50 & 3.851 & 1.859 & 0.986 & 0.0 & 0.247 & 0.132 & 6.017 &  0.346 &  3.816 &  1.826 & 0.986 & 0.0\\
 1 & 3.0 & 8.518 & 0.470 & 3.00 & 3.833 & 2.120 & 0.986 & 0.0 &0.247 & 0.132 & 5.341 &  0.549  & 3.804 &  2.086 & 0.986 & 0.0\\
 1 & 4.0 & 8.193 & 0.964 &  4.00 & 3.819 & 2.542 & 0.986 & 0.0 & 0.247 & 0.132 & 4.754  & 1.185 &  3.781 & 2.495 & 0.986 & 0.0\\
 1 & 5.0 & 7.955 & 1.725 & 5.00 & 3.804 & 2.863 & 0.986 & 0.0 & 0.247 & 0.132 & 4.356 &  2.244  & 3.758 & 2.812 & 0.986 & 0.0\\  
 2 & 5.0 & 8.010 & 4.066 & 4.97 & 3.743 & 3.055 & 0.329 & 0.0 & 1.27 & 0.442 & 5.758 & 2.950  & 3.793 & 3.094 & 0.338 & 0.0\\
 3 & 5.0 & 8.030 & 3.269 & 4.97 & 3.790 & 3.136 & 0.168 & 0.0 & 1.27 & 0.442 & 5.859 & 4.814  & 3.736 & 3.109 & 0.092 & 0.0\\ 
 1 & 7.0 & 7.628 & 4.546 & 7.00 & 3.776 & 3.365 & 0.986 & 0.0 & 0.247 & 0.132 & 3.602 & 5.752  & 3.739 & 3.331 & 0.986 & 0.0 \\
 2 & 7.0 & 7.674 & 15.20 & 6.93 & 3.684 & 3.580 & 0.314 & 0.0 & 1.30 & 0.453 & 4.951 & 7.900  & 3.769 & 3.611 & 0.294 & 0.0\\  
 3 & 7.0 & 7.697 & 8.686 & 6.91 & 3.769 & 3.660 & 0.019 & 0.0 & 1.30 & 0.453 & 4.598 & 13.87  & 3.704 & 3.627 & 0.012 & 0.0\\ 
 1 & 9.0 & 7.426 & 9.413 & 8.99 & 3.759 & 3.749 & 0.986 & 0.0 & 0.247 & 0.132 & 3.269 & 13.09  & 3.707 & 3.701 & 0.986 & 0.0\\ 
 2 & 9.0 & 7.465 & 31.19 & 8.81 & 3.680 & 4.004 & 0.233 & 0.0 & 1.38 & 0.482 & 3.994 & 16.65  & 3.758 & 4.023 & 0.229 & 0.0\\ 
 3 & 9.0 & 7.479 & 14.91 & 8.80 & 3.764 & 3.991 & 0.001 & 0.0 & 1.38 & 0.482 & 3.918 & 28.18  & 3.670 & 3.913 & 0.000 & 0.0\\ 
 1 & 12.0 & 7.192 & 19.24 & 11.94 & 3.752 & 4.176 & 0.984 & 0.0 & 0.247 & 0.132 & 3.093 & 33.88  & 3.665  & 4.091 & 0.984 & 0.0\\ 
 1 & 15.0 & 7.065 & 68.42 & 14.63 & 3.722 & 4.728 & 0.640 & 0.0 & 0.247 & 0.132 &  4.340 & 185.0  & 3.589 &  4.646 & 0.621 & 0.0\\ 
& & \multicolumn{14}{c}{$\omega_{\rm ini} = 0.5$} \smallskip \\
 1 & 2.0 & 9.108 &  0.181 & 2.00 & 3.858 & 1.584 & 0.986 & 95.9 & 0.287 & 0.148 & 7.188 & 0.246 & 3.819 & 1.593 & 0.986 &79.3\\ 
 1 & 2.5 & 8.833 &  0.377 & 2.50 & 3.842 & 1.987 & 0.986 & 65.5 & 0.300 & 0.153 & 6.068 &  0.480 & 3.805 & 1.963 & 0.986 & 55.4\\ 
 1 & 3.0 & 8.614 &  0.639 & 3.00 & 3.828 & 2.271 & 0.986 & 49.8 & 0.319 & 0.161 & 5.440 & 0.799 & 3.792 & 2.239 & 0.986 & 42.3\\
 1 & 4.0 & 8.283 & 1.364 & 4.00 & 3.811 & 2.696 & 0.986 & 34.5 & 0.364 & 0.179 & 4.854 & 1.841 & 3.761 & 2.646 & 0.986 & 27.4\\  
 1 & 5.0 & 8.046 & 2.478 & 5.00 & 3.796 & 3.023 & 0.986 &25.8& 0.435 & 0.208 & 4.401 & 3.462 & 3.743 & 2.973 & 0.986 &19.9\\ 
 2 & 5.0 & 8.085 & 6.135 & 4.97 & 3.719 & 3.163 & 0.367 &21.7& 2.92 & 0.765 & 5.878 & 3.859 & 3.787 & 3.214 & 0.322 &30.4\\
 3 & 5.0 & 8.094 & 4.064 & 4.96 & 3.788 & 3.242 & 0.226 &26.5& 2.92 & 0.765 & 6.409 & 7.324 & 3.720 & 3.253 & 0.123 &11.4\\
 1 & 7.0 & 7.715 & 6.440 & 7.00 & 3.772 & 3.524 & 0.986 &15.3& 0.626 & 0.271 & 3.789 & 9.797 & 3.708 & 3.472  & 0.986 &11.1\\ 
 2 & 7.0 & 7.747 & 21.15 & 6.92 & 3.672 & 3.698 & 0.396 &20.5& 3.46 & 0.825 & 4.718 & 9.766 & 3.768 & 3.728 & 0.384 &41.5\\ 
 3 & 7.0 & 7.765 & 12.72 & 6.89 & 3.756 & 3.803 & 0.034 &26.8& 3.46 & 0.825 & 4.900 & 27.76 & 3.658 & 3.757 &  0.022 &11.2\\
 1 & 9.0 & 7.503 & 13.40 & 8.99 & 3.754 & 3.906 & 0.986 &11.8& 0.893 & 0.350 & 3.461 & 22.91 & 3.674 & 3.828  & 0.986 & 7.5\\ 
 2 & 9.0 & 7.528 & 58.13 & 8.80 & 3.631 & 4.089 & 0.454 &12.1& 4.08 & 0.902 & 3.981 & 24.41 & 3.734 & 4.131  & 0.450 &46.6\\   
 3 & 9.0 & 7.546 & 25.72 & 8.77 & 3.739 & 4.161 & 0.013 &23.4& 4.08 & 0.902 & 4.268 & 61.66 & 3.620 & 4.073  & 0.007 &4.1\\ 
 1 & 12.0 & 7.273 & 36.05 & 11.90 & 3.734 & 4.407 & 0.982 &14.5& 1.34 & 0.461 & 3.404 & 86.11 & 3.619 &  4.327 & 0.981 & 5.4\\
 1 & 15.0 & 7.140 & 98.39 & 14.47 & 3.714 & 4.850 & 0.784 &3.8& 1.87 & 0.597 & 4.694 & 294.0 & 3.572 & 4.799 &  0.747 & 1.\\  
& & \multicolumn{14}{c}{$\omega_{\rm ini} = 0.9$} \smallskip \\
 1 & 2.5 & 8.875 & 0.434 & 2.50 & 3.830 & 2.046 & 0.986 &119.4& 0.702 & 0.283 & 6.125 & 0.596 & 3.786 & 2.018  & 0.986 &92.1\\ 
 1 & 3.0 & 8.636 & 0.645 & 3.00 & 3.823 & 2.277 & 0.986 &94.9& 0.841 & 0.323 & 5.523 & 0.841 & 3.781 & 2.234 & 0.986 &77.2\\ 
 1 & 4.0 & 8.302 & 1.295 & 4.00 & 3.809 & 2.671 & 0.986 &66.9& 1.15 & 0.400 & 4.821 & 1.693 & 3.763 & 2.617 & 0.986 &54.1\\ 
 1 & 5.0 & 8.062 & 2.236 & 5.00 & 3.793 & 2.961 & 0.986 &50.2& 1.64 & 0.490 & 4.354 & 2.976 & 3.744 & 2.905  & 0.986 &40.0\\ 
 2 & 5.0 & 8.117 & 5.935 & 4.97 & 3.721 & 3.158 & 0.248 &33.5& 6.21 & 0.939 & 6.495 & 4.608 & 3.778 & 3.264 & 0.264 &27.4\\
 3 & 5.0 & 8.128 & 4.380 & 4.96 & 3.780 & 3.247 & 0.225 &24.8& 6.21 & 0.939 & 5.925 & 7.214 & 3.714 & 3.226 & 0.119 &20.8\\
 1 & 7.0 & 7.727 & 5.059 & 7.00 & 3.777 & 3.426 & 0.986 &33.3& 3.01 & 0.659 & 3.824 & 7.681 & 3.710 & 3.355 & 0.986 &23.7\\ 
 2 & 7.0 & 7.761 & 17.35 & 6.95 & 3.675 & 3.621 & 0.492 &31.9& 9.57 & 1.03 & 4.842 & 8.527 & 3.763 & 3.657 &  0.481 &65.1\\
 3 & 7.0 & 7.787 & 10.83 & 6.91 & 3.767 & 3.770 & 0.055 &36.6& 9.57 & 1.03 & 5.013 & 21.78 & 3.677 & 3.734 &  0.036 &19.2\\ 
 1 & 9.0 & 7.515 & 11.79 &  8.99 & 3.762 & 3.875 & 0.982 &20.6& 4.82 & 0.797 & 3.792 & 22.02 & 3.675 & 3.819 & 0.981 &12.9\\ 
 1 & 12.0 & 7.312 & 51.88 & 11.76 & 3.731 & 4.562 & 0.512 &14.6& 7.33 & 0.945 & 4.594 & 149.3 & 3.601 & 4.501 & 0.492 &3.6\\ 
 1 & 15.0 & 7.183 & 107.6 & 14.21 & 3.719 & 4.907 & 0.268 &3.8& 6.84 & 1.04 & 4.946 & 359.2 & 3.571 & 4.871 & 0.258 &2.\\ 
\hline
\end{tabular}
\label{tab:models_z14}
\end{table*}

\begin{table*}
\centering
\caption{Parameters of fundamental mode pulsation for models entering and exiting IS for $Z=0.006$ }
\begin{tabular}{@{}c@{\hspace{2mm}}r@{\hspace{6mm}}r@{\hspace{2mm}}r@{\hspace{2mm}}r@{\hspace{2mm}}r@{\hspace{2mm}}r@{\hspace{2mm}}r@{\hspace{2mm}}r@{\hspace{6mm}}r@{\hspace{2mm}}r@{\hspace{2mm}}r@{\hspace{2mm}}r@{\hspace{2mm}}r@{\hspace{2mm}}r@{\hspace{2mm}}r@{\hspace{2mm}}r@{}} 
\hline
Xing & & \multicolumn{7}{c}{{\bf Entering stage}} & \multicolumn{7}{c}{{\bf Exiting stage}}  \\
 & $M_{\rm{ini}}$ & $\log t$ & P~~ & $M$~~ & $\log T_{\rm eff}$ & $\log L$ & $Y_{\rm c}$ & $V_{\rm e}$~~~~ & N/C & N/O & $\log \Delta t$ & P~~  & $\log T_{\rm eff}$ & $\log L$ & $Y_{\rm c}$ & $V_{\rm e}$~~~~\\
 & [$M_\odot$] & [yr] & [d] & [$M_\odot$] & $[K]$ & [$L_\odot$] & & [km/s] & num & num & [yr] & [d] & [K] & [$L_\odot$] & & [km/s]\\
\hline
& & \multicolumn{14}{c}{$\omega_{\rm ini} = 0.0$} \smallskip\\
 1 & 1.7 & 9.125 & 0.121 & 1.70 & 3.875 & 1.370 & 0.994 &0.0& 0.247 & 0.132 & 7.135 & 0.162 & 3.833 & 1.362 & 0.994 &0.0 \\
 1 & 2.0 & 8.922 & 0.192 & 2.00 & 3.864 & 1.634 & 0.994 &0.0& 0.247 & 0.132 & 6.385 & 0.244 & 3.825 & 1.609 & 0.994 &0.0 \\ 
 1 & 2.5 & 8.658 & 0.332 & 2.50 & 3.856 & 1.974 & 0.994 &0.0& 0.247 & 0.132 & 5.774 & 0.438 & 3.808 & 1.929 & 0.994 &0.0\\ 
 1 & 3.0 & 8.454 & 0.565 & 3.00 & 3.833 & 2.230 & 0.994 &0.0& 0.247 & 0.132 & 5.172 & 0.694 & 3.797 & 2.192 & 0.994 &0.0\\ 
 1 & 4.0 & 8.150 & 1.156 & 4.00 & 3.819 & 2.650 & 0.994 &0.0& 0.247 & 0.132 & 4.616 & 1.488 & 3.777 & 2.605 & 0.994 &0.0\\ 
 2$^*$ & 4.0 & 8.230 & 3.455 & 3.97 & 3.748 & 2.923 & 0.208 &0.0& 1.24 & 0.411 & 6.118 & 3.293 & 3.750 & 2.903 & 0.097 &0.0\\
 1 & 5.0 & 7.929 & 2.109 & 5.00 & 3.803 & 2.973 & 0.994 &0.0& 0.247 & 0.132 & 4.242 & 2.889 & 3.753 & 2.928 & 0.994 &0.0\\ 
 1 & 7.0 & 7.624 & 5.282 & 7.00 & 3.784 & 3.483 & 0.994 &0.0& 0.247 & 0.132 & 3.699 & 7.860 & 3.725 & 3.436 & 0.994 &0.0\\ 
 2 & 7.0 & 7.667 & 18.34 & 6.92 & 3.696 & 3.724 & 0.340 &0.0& 1.18 & 0.394 & 4.811 & 9.540 & 3.773 & 3.738 & 0.323 &0.0\\
 3 & 7.0 & 7.680 & 10.30 & 6.90 & 3.770 & 3.761 & 0.085 &0.0& 1.18 & 0.394 & 5.030 & 17.08 & 3.707 & 3.742 & 0.063 &0.0\\
 1 & 9.0 & 7.421 & 10.17 & 8.99 & 3.772 & 3.855 & 0.994 &0.0& 0.247 & 0.132 & 3.374 & 17.11 & 3.697 & 3.790 & 0.993 &0.0\\ 
 1 & 12.0 & 7.191 & 26.33 & 11.95 & 3.755 & 4.353 & 0.987 &0.0& 0.247 & 0.132 & 3.444 & 57.30 & 3.659 & 4.305 & 0.985 &0.0\\
 1 & 15.0 & 7.088 & 74.50 & 14.64 & 3.732 & 4.812 & 0.222 &0.0& 0.247 & 0.132 & 4.243 & 202.2 & 3.608 & 4.752 & 0.206 &0.0\\ 
& & \multicolumn{14}{c}{$\omega_{\rm ini} = 0.5$} \smallskip \\
 1 & 1.7 & 9.219 & 0.146 & 1.70 & 3.867 & 1.458 & 0.994 &97.9& 0.351 & 0.172 & 7.294 & 0.205 & 3.824 & 1.466 & 0.994 &79.9\\ 
 1 & 2.0 & 9.017 & 0.251 & 2.00 & 3.854 & 1.750 & 0.994 &75.5& 0.384 & 0.185 & 6.666 & 0.337 & 3.811 & 1.733 & 0.994 &62.3\\ 
 1 & 2.5 & 8.746 & 0.455 & 2.50 & 3.843 & 2.099 & 0.994 &53.6& 0.411 & 0.196 & 5.743 & 0.608 & 3.798 & 2.066 & 0.994 &44.4\\ 
 1 & 3.0 & 8.537 & 0.763 & 3.00 & 3.826 & 2.366 & 0.994 &41.3& 0.463 & 0.216 & 5.230 & 0.991 & 3.785 & 2.331 & 0.994 &34.1\\
 1 & 4.0 & 8.229 & 1.586 & 4.00 & 3.809 & 2.778 & 0.994 &29.0& 0.589 & 0.261 & 4.674 & 2.205 & 3.758 & 2.732 & 0.994 &22.3 \\
 2$^*$ & 4.0 & 8.289 & 4.626 & 3.96 & 3.739 & 3.033 & 0.215 &19.7& 3.31 & 0.796 & 6.003 & 4.640 & 3.740 & 3.038 & 0.131 &17.9 \\
 3$^*$ & 4.0 & 8.293 & 4.859 & 3.96 & 3.742 & 3.070 & 0.242 &16.1& 3.31 & 0.796 & 5.817 & 4.721 & 3.742 & 3.055 & 0.168 &16.1\\
 1 & 5.0 & 8.005 & 2.853 & 5.00 & 3.795 & 3.102 & 0.994 &21.9& 0.773 & 0.319 & 4.238 & 4.048 & 3.743 & 3.059 & 0.994 &16.5\\ 
 2$^*$ & 5.0 & 8.053 & 8.767 & 4.95 & 3.720 & 3.349 & 0.238 &14.0& 3.64 & 0.833 & 6.286 & 9.962 & 3.716 & 3.394 & 0.128 &9.8 \\ 
 3$^*$ & 5.0 & 8.062 & 11.07 & 4.94 & 3.714 & 3.439 & 0.217 &9.6& 3.64 & 0.833 & 5.821 & 10.52 & 3.716 & 3.419 & 0.105 &8.6 \\  
 1 & 7.0 & 7.696 & 7.140 & 7.00 & 3.775 & 3.606 & 0.994 &13.0& 1.17 & 0.423 & 3.683 & 10.97 & 3.715 & 3.560 & 0.994 &9.2\\ 
 2 & 7.0 & 7.723 & 24.64 & 6.93 & 3.679 & 3.797 & 0.514 &18.3& 4.49 & 0.908 & 4.520 & 11.70 & 3.765 & 3.816 & 0.507 &38.3\\
 3 & 7.0 & 7.744 & 14.38 & 6.89 & 3.766 & 3.912 & 0.026 &22.2& 4.49 & 0.908 & 4.605 & 28.28 & 3.686 & 3.889 & 0.018 &11.6\\
 1 & 9.0 & 7.495 & 13.95 & 8.99 & 3.766 & 3.990 & 0.992 &10.3& 1.65 & 0.521 & 3.495 & 25.84 & 3.683 & 3.926 & 0.991 &6.1 \\ 
 2 & 9.0 & 7.520 & 59.44 & 8.76 & 3.655 & 4.194 & 0.450 &14.1& 5.28 & 0.973 & 3.930 & 23.45 & 3.756 & 4.216 & 0.447 &47.5\\
 3 & 9.0 & 7.536 & 28.12 & 8.73 & 3.747 & 4.246 & 0.041 &20.9& 5.28 & 0.973 & 4.353 & 62.76 & 3.655 & 4.214 & 0.034 &7.0\\ 
 1 & 12.0 & 7.306 & 65.84 & 11.73 & 3.733 & 4.685 & 0.135 &3.4& 2.35 & 0.650 & 4.601 & 193.4 & 3.605 & 4.648 & 0.111 &2.1 \\
 1 & 15.0 & 7.165 & 107.2 & 14.40 & 3.727 & 4.949 & 0.144 &2.4& 2.80 & 0.749 & 4.500 & 336.9 & 3.598 & 4.922 & 0.121 &1.\\ 
& & \multicolumn{14}{c}{$\omega_{\rm ini} = 0.9$} \smallskip \\
 1 & 1.7 & 9.239 & 0.146 & 1.70 & 3.858 & 1.466 & 0.994 &179.7& 1.17 & 0.366 & 7.257 & 0.208 & 3.815 & 1.472 & 0.994 &143.9\\
 1 & 2.0 & 9.037 & 0.247 & 2.00 & 3.846 & 1.751 & 0.994 &143.5& 1.37 & 0.412 & 6.556 & 0.337 & 3.803 & 1.729 & 0.994 &114.7\\ 
 1 & 2.5 & 8.765 & 0.453 & 2.50 & 3.835 & 2.084 & 0.994 &103.7& 1.47 & 0.448 & 5.733 & 0.593 & 3.793 & 2.048 & 0.994 &84.5\\ 
 1 & 3.0 & 8.555 & 0.718 & 3.00 & 3.825 & 2.340 & 0.994 &81.0& 1.75 & 0.504 & 5.251 & 0.913 & 3.785 & 2.301 & 0.994 &67.2\\
 1 & 4.0 & 8.243 & 1.275 & 4.00 & 3.824 & 2.728 & 0.994 &61.6& 2.68 & 0.633 & 4.782 & 1.840 & 3.762 & 2.662 & 0.994 &46.0 \\ 
 2$^*$ & 4.0 & 8.300 & 3.561 & 3.98 & 3.748 & 2.936 & 0.374 &19.2& 8.85 & 1.02 & 6.838 & 4.321 & 3.745 & 3.020 & 0.156 &13.1\\
 1 & 5.0 & 8.017 & 2.162 & 5.00 & 3.810 & 3.022 & 0.994 &47.0& 4.11 & 0.765 & 4.344 & 3.157 & 3.749 & 2.961 & 0.994 &34.9 \\
 2 & 5.0 & 8.040 & 5.039 & 4.99 & 3.735 & 3.141 & 0.785 &37.9& 11.0 & 1.07 & 5.983 & 3.270 & 3.799 & 3.196 & 0.736 &50.2\\ 
 3 & 5.0 & 8.084 & 5.695 & 4.96 & 3.789 & 3.433 & 0.080 &25.7& 11.0 & 1.07 & 5.503 & 10.41 & 3.715 & 3.416 & 0.043 &15.1 \\ 
 1$^{*}$ & 7.0 & 7.708 & 6.540 & 7.00 & 3.784 & 3.598 & 0.986 &24.8& 8.07 & 0.967 & 4.868 & 6.964 & 3.782 & 3.622 & 0.980 &21.5\\
 2 & 7.0 & 7.767 & 16.66 & 6.94 & 3.766 & 3.986 & 0.026 &5.2& 8.88 & 0.994 & 4.641 & 36.62 & 3.674 & 3.959 & 0.015 & 6.4 \\
 1 & 9.0 & 7.565 & 37.05 & 8.94 & 3.739 & 4.345 & 0.014 &7.4& 14.6 & 1.13 & 4.190 & 89.23 & 3.637 & 4.306 & 0.008 & 6.3 \\
 1 & 12.0 & 7.348 & 77.78 & 11.58 & 3.733 & 4.757 & 0.312 &3.7& 23.5 & 1.33 & 4.770 & 238.8 & 3.599 & 4.717 & 0.277 & 2. \\
\hline
\end{tabular}
\leftline{$^*$ ~ Model returns to same side of IS before crossing next expected boundary.}
\label{tab:models_z06}
\end{table*}

\begin{table*}
\centering
\caption{Parameters of fundamental mode pulsation for models entering and exiting IS for $Z=0.002$ }
\begin{tabular}{@{}c@{\hspace{2mm}}r@{\hspace{6mm}}r@{\hspace{2mm}}r@{\hspace{2mm}}r@{\hspace{2mm}}r@{\hspace{2mm}}r@{\hspace{2mm}}r@{\hspace{2mm}}r@{\hspace{6mm}}r@{\hspace{2mm}}r@{\hspace{2mm}}r@{\hspace{2mm}}r@{\hspace{2mm}}r@{\hspace{2mm}}r@{\hspace{2mm}}r@{\hspace{2mm}}r@{}}
\hline
\vspace{2mm}  & & \multicolumn{7}{c}{{\bf Entering stage}} & \multicolumn{7}{c}{{\bf Exiting stage}}  \\
Xing & $M_{\rm{ini}}$ & $\log t$ & P~~ & $M$~~ & $\log T_{\rm eff}$ & $\log L$ & $Y_{\rm c}$ & $V_{\rm e}$~~~~ & N/C & N/O & $\log \Delta t$ & P~~  & $\log T_{\rm eff}$ & $\log L$ & $Y_{\rm c}$ & $V_{\rm e}$~~~~\\
 & [$M_\odot$] & [yr] & [d] & [$M_\odot$] & $[K]$ & [$L_\odot$] & & [km/s] & num & num & [yr] & [d] & [K] & [$L_\odot$] & & [km/s] \vspace{2mm} \\
\hline 
& & \multicolumn{14}{c}{$\omega_{\rm ini} = 0.0$}  \smallskip \\
 1 & 1.7 & 9.048 & 0.156 & 1.70 & 3.875 & 1.518 & 0.998 &0.0& 0.247 & 0.132 & 6.662 & 0.213 & 3.827 & 1.498 & 0.998 &0.0\\ 
 1 & 2.0 & 8.844 & 0.243 & 2.00 & 3.860 & 1.756 & 0.998 &0.0& 0.247 & 0.132 & 6.047 & 0.319 & 3.817 & 1.728 & 0.998 &0.0\\
 1 & 2.5 & 8.589 & 0.431 & 2.50 & 3.846 & 2.084 & 0.998 &0.0& 0.247 & 0.132 & 5.461 & 0.572 & 3.800 & 2.047 & 0.998 & 0.0\\ 
 1 & 3.0 & 8.395 & 0.680 & 3.00 & 3.836 & 2.351 & 0.998 &0.0& 0.247 & 0.132 & 5.067 & 0.914 & 3.788 & 2.310 & 0.998 &0.0\\
 2 & 3.0 & 8.460 & 1.697 & 2.99 & 3.770 & 2.560 & 0.467 &0.0& 1.22 & 0.371 & 6.886 & 1.233 & 3.824 & 2.620 & 0.306 &0.0\\ 
 3$^*$ & 3.0 & 8.487 & 1.378 & 2.98 & 3.822 & 2.670 & 0.110 &0.0& 1.22 & 0.371 & 6.208 & 1.543 & 3.820 & 2.720 & 0.231 &0.0\\
 4 & 3.0 & 8.491 & 1.422 & 2.98 & 3.822 & 2.684 & 0.128 &0.0& 1.22 & 0.371 & 6.260 & 2.189 & 3.766 & 2.672 & 0.055 & 0.0\\
 1 & 4.0 & 8.106 & 1.296 & 4.00 & 3.834 & 2.780 & 0.998 &0.0& 0.247 & 0.132 & 4.642 & 1.984 & 3.768 & 2.729 & 0.998 &0.0\\
 2 & 4.0 & 8.147 & 3.391 & 3.98 & 3.755 & 2.950 & 0.595 &0.0& 0.990 & 0.326 & 6.140 & 2.317 & 3.809 & 2.985 & 0.519 &0.0\\
 3 & 4.0 & 8.187 & 2.996 & 3.97 & 3.804 & 3.100 & 0.017 &0.0& 0.990 & 0.326 & 5.137 & 4.792 & 3.744 & 3.077 & 0.008 & 0.0\\
 1 & 5.0 & 7.898 & 2.601 & 5.00 & 3.803 & 3.098 & 0.998 &0.0& 0.247 & 0.132 & 4.109 & 3.672 & 3.753 & 3.064 & 0.998 &0.0\\  
 2 & 5.0 & 7.926 & 6.179 & 4.98 & 3.740 & 3.268 & 0.707 &0.0& 0.861 & 0.303 & 5.707 & 3.835 & 3.802 & 3.296 & 0.690 &0.0\\
 3 & 5.0 & 7.968 & 5.527 & 4.96 & 3.784 & 3.407 & 0.004 &0.0& 0.861 & 0.303 & 4.376 & 8.505 & 3.729 & 3.376 & 0.001 &0.0\\ 
 1 & 7.0 & 7.609 & 6.572 & 7.00 & 3.780 & 3.596 & 0.997 &0.0& 0.247 & 0.132 & 3.598 & 9.984 & 3.725 & 3.558 & 0.997 &0.0\\ 
 2 & 7.0 & 7.626 & 19.26 & 6.97 & 3.702 & 3.772 & 0.770 &0.0& 0.920 & 0.314 & 4.321 & 10.32 & 3.772 & 3.788 & 0.765 &0.0\\ 
 3 & 7.0 & 7.665 & 11.79 & 6.94 & 3.768 & 3.837 & 0.000 &0.0& 0.920 & 0.314 & 3.727 & 20.09 & 3.701 & 3.787 & 0.000 &0.0\\
 1 & 9.0 & 7.413 & 13.19 & 8.99 & 3.765 & 3.973 & 0.995 &0.0& 0.247 & 0.132 & 3.363 & 22.30 & 3.700 & 3.932 & 0.994 &0.0\\
 2 & 9.0 & 7.426 & 42.55 & 8.90 & 3.680 & 4.141 & 0.768 &0.0& 1.05 & 0.335 & 3.859 & 20.31 & 3.761 & 4.162 & 0.765 &0.0\\
 3 & 9.0 & 7.461 & 22.59 & 8.89 & 3.752 & 4.169 & 0.000 &0.0& 1.05 & 0.335 & 3.601 & 42.84 & 3.684 & 4.164 &  0.000 &0.0\\
 1 & 12.0 & 7.233 & 44.00 & 11.92 & 3.749 & 4.587 & 0.005 &0.0& 0.247 & 0.132 & 3.739 & 102.2 & 3.650 & 4.527 & 0.002 &0.0\\ 
 1 & 15.0 & 7.094 & 73.33 & 14.81 & 3.740 & 4.858 & 0.010 &0.0& 0.247 & 0.132 & 3.890 & 179.4 & 3.627 & 4.792 & 0.005 &0.0 \\
& & \multicolumn{14}{c}{$\omega_{\rm ini} = 0.5$}  \smallskip \\
 1 & 1.7 & 9.139 & 0.193 & 1.70 & 3.868 & 1.621 &  0.998 &76.9& 0.608 & 0.252 & 6.761 & 0.276 & 3.818 & 1.609 & 0.998 &61.8 \\ 
 1 & 2.0 & 8.935 & 0.305 & 2.00 & 3.857 & 1.877 & 0.998 &60.9& 0.702 & 0.284 & 6.194 & 0.428 & 3.808 & 1.855 & 0.998 &48.8\\
 1 & 2.5 & 8.671 & 0.562 & 2.50 & 3.839 & 2.204 & 0.998 &42.4& 0.808 & 0.320 & 5.482 & 0.770 & 3.792 & 2.174 & 0.998 &33.9\\ 
 1 & 3.0 & 8.473 & 0.823 & 3.00 & 3.841 & 2.478 & 0.998 &35.6& 0.992 & 0.378 & 5.197 & 1.256 & 3.778 & 2.434 & 0.998 &26.2\\   
 2$^*$ & 3.0 & 8.533 & 2.655 & 2.97 & 3.761 & 2.753 & 0.298 &13.1& 4.17 & 0.853 & 6.849 & 3.174 & 3.756 & 2.816 & 0.113 &3.9 \\
 3$^*$ & 3.0 & 8.543 & 3.317 & 2.97 & 3.756 & 2.839 & 0.152 &4.3& 4.17 & 0.853 & 5.801 & 3.280 & 3.754 & 2.827 & 0.110 &5.8 \\ 
 4$^*$ & 3.0 & 8.545 & 3.844 & 2.96 & 3.749 & 2.887 & 0.221 &23.9& 4.17 & 0.853 & 6.025 & 3.534 & 3.750 & 2.848 & 0.128 &24.5\\ 
 1 & 4.0 & 8.181 & 1.791 & 4.00 & 3.816 & 2.883 & 0.998 &23.9& 1.30 & 0.451 & 4.551 & 2.636 & 3.760 & 2.844 & 0.998 &17.8 \\ 
 2 & 4.0 & 8.206 & 4.334 & 3.99 & 3.746 & 3.032 & 0.728 &13.9& 3.90 & 0.832 & 6.243 & 2.940 & 3.801 & 3.079 & 0.641 &12.8\\ 
 3 & 4.0 & 8.242 & 4.720 & 3.96 & 3.794 & 3.288 & 0.070 &3.6& 3.90 & 0.832 & 5.560 & 8.407 & 3.729 & 3.287 & 0.039 &5.5\\
 1 & 5.0 & 7.971 & 3.171 & 5.00 & 3.804 & 3.208 & 0.998 &18.4& 1.72 & 0.542 & 4.184 & 4.884 & 3.745 & 3.169 & 0.997 &13.0\\
 2 & 5.0 & 7.983 & 7.320 & 4.99 & 3.733 & 3.322 & 0.882 &13.7& 4.15 & 0.863 & 5.651 & 4.610 & 3.793 & 3.353 & 0.852 &15.3\\  
 3 & 5.0 & 8.027 & 9.153 & 4.95 & 3.776 & 3.625 & 0.016 &3.6& 4.15 & 0.863 & 4.726 & 15.48 & 3.717 & 3.614 & 0.009 &3.8\\ 
 1 & 7.0 & 7.677 & 8.321 & 7.00 & 3.785 & 3.736 & 0.994 &10.1& 2.60 & 0.676 & 3.977 & 15.31 & 3.712 & 3.710 & 0.992 &6.5\\
 2 & 7.0 & 7.682 & 20.99 & 6.99 & 3.703 & 3.825 & 0.935 &16.9& 5.49 & 0.961 & 4.166 & 11.24 & 3.773 & 3.842 & 0.933 &30.3 \\
 3 & 7.0 & 7.722 & 16.81 & 6.94 & 3.764 & 3.996 &  0.000 &20.4& 5.49 & 0.961 & 3.748 & 33.51 & 3.690 & 3.985 & 0.000 &9.9\\ 
 1 & 9.0 & 7.524 & 31.83 & 8.97 & 3.751 & 4.338 & 0.000 &5.1& 3.52 & 0.780 & 3.774 & 63.90 & 3.668 & 4.324 & 0.000 &3.6\\ 
 1 & 12.0 & 7.302 & 62.44 & 11.89 & 3.743 & 4.725 & 0.158 &4.8& 4.06 & 0.849 & 3.794 & 147.4 & 3.637 & 4.656 & 0.000 & 2.6 \\
 1 & 15.0 & 7.166 & 95.53 & 14.75 & 3.731 & 4.943 & 0.000 &3.8& 4.25 & 0.895 & 4.101 & 271.4 & 3.616 & 4.943 & 0.000 & 1. \smallskip \\ 
& & \multicolumn{14}{c}{$\omega_{\rm ini} = 0.9$} \smallskip \\
 1 & 1.7 & 9.197 & 0.187 & 1.70 & 3.867 & 1.632 & 0.998 &136.6& 5.60 & 0.659 & 6.926 & 0.277 & 3.814 & 1.614 & 0.998 &104.9 \\
 1 & 2.0 & 8.958 & 0.300 & 2.00 & 3.850 & 1.862 & 0.998 &107.2& 4.10 & 0.653 & 6.320 & 0.415 & 3.804 & 1.836 & 0.998 &88.1\\
 1 & 2.5 & 8.688 & 0.509 & 2.50 & 3.840 & 2.168 & 0.998 &82.0& 4.82 & 0.756 & 5.660 & 0.700 & 3.792 & 2.132 & 0.998 &64.6\\ 
 1 & 3.0 & 8.495 & 0.821 & 3.00 & 3.830 & 2.433 & 0.998 &51.2& 6.65 & 0.880 & 5.150 & 1.127 & 3.781 & 2.393 & 0.998 &41.3\\ 
 2$^*$ & 3.0 & 8.545 & 2.050 & 2.98 & 3.769 & 2.654 & 0.532 &17.6& 15.4 & 1.11 & 7.208 & 3.057 & 3.756 & 2.800 & 0.200 &17.5\\ 
 1 & 4.0 & 8.195 & 1.522 & 4.00 & 3.815 & 2.797 & 0.998 &46.2& 10.5 & 1.07 & 4.553 & 2.051 & 3.769 & 2.756 & 0.997 &36.7 \\
 2 & 4.0 & 8.200 & 2.590 & 4.00 & 3.760 & 2.847 & 0.967 &53.8& 15.8 & 1.18 & 5.688 & 1.826 & 3.810 & 2.882 & 0.960 &66.9 \\
 3 & 4.0 & 8.263 & 4.576 & 3.97 & 3.790 & 3.266 & 0.0150 &37.9& 15.8 & 1.18 & 5.130 & 8.021 & 3.722 & 3.247 & 0.006 &24.5\\
 1 & 5.0 & 8.049 & 8.647 & 5.00 & 3.777 & 3.612 & 0.002 &22.7& 16.7 & 1.21 & 4.260 & 15.26 & 3.709 & 3.575 & 0.000 &13.6 \\
 1$^*$ & 7.0 & 7.759 & 21.04 & 6.99 & 3.761 & 4.093 & 0.000 &13.5& 33.5 & 1.42 & 4.260 & 21.84 & 3.761 & 4.114 & 0.000 &14.5\\ 
 2 & 7.0 & 7.760 & 20.62 & 6.99 & 3.763 & 4.093 & 0.000 &14.5& 34.1 & 1.42 & 4.260 & 38.83 & 3.679 & 4.029 & 0.000 &7.5\\ 
 1 & 9.0 & 7.577 & 37.59 & 8.97 & 3.755 & 4.434 & 0.000 &8.2& 53.6 & 1.58 & 4.156 & 97.40 & 3.649 & 4.465 & 0.000 &4.1\\ 
 1 & 12.0 & 7.351 & 79.21 & 11.84 & 3.741 & 4.821 & 0.003 &7.7& 55.7 & 1.59 & 3.799 & 196.0 & 3.627 & 4.751 & 0.000 &3.9 \\  
 1 & 15.0 & 7.213 & 125.3 & 14.61 & 3.735 & 5.066 & 0.075 &6.0& 48.1 & 1.70 & 4.103 & 329.3 & 3.628 & 5.039 & 0.066 &3.\\ 
\hline
\end{tabular}
\leftline{$^*$ ~Model returns to same side of IS before crossing next expected boundary.}
\label{tab:models_z02}
\end{table*}

\begin{table*}
\centering
\caption{Parameters of first overtone pulsation for models entering and exiting IS for $Z=0.014$ }
\begin{tabular}{@{}c@{\hspace{2mm}}r@{\hspace{6mm}}r@{\hspace{2mm}}r@{\hspace{2mm}}r@{\hspace{2mm}}r@{\hspace{2mm}}r@{\hspace{2mm}}r@{\hspace{2mm}}r@{\hspace{6mm}}r@{\hspace{2mm}}r@{\hspace{2mm}}r@{\hspace{2mm}}r@{\hspace{2mm}}r@{\hspace{2mm}}r@{\hspace{2mm}}r@{\hspace{2mm}}r@{}}
\hline
\vspace{2mm}  & & \multicolumn{7}{c}{{\bf Entering stage}} & \multicolumn{7}{c}{{\bf Exiting stage}}  \\
Xing & $M_{\rm{ini}}$ & $\log t$ & P~~ & $M$~~ & $\log T_{\rm eff}$ & $\log L$ & $Y_{\rm c}$ & $V_{\rm e}$~~~~ & N/C & N/O & $\log \Delta t$ & P~~  & $\log T_{\rm eff}$ & $\log L$ & $Y_{\rm c}$ & $V_{\rm e}$~~~~\\
 & [$M_\odot$] & [yr] & [d] & [$M_\odot$] & $[K]$ & [$L_\odot$] & & [km/s] & num & num & [yr] & [d] & [K] & [$L_\odot$] & & [km/s] \vspace{2mm} \\
\hline
                           & & \multicolumn{14}{c}{$\omega_{\rm ini} = 0.0$}  \smallskip \\
 1 & 2.0 & 9.010 & 0.104 & 2.00 & 3.880 & 1.491 & 0.986 & 0. & 0.247 & 0.132 & 7.211 & 0.142 & 3.835 & 1.488 & 0.986 & 0.   \\
 1 & 2.5 & 8.742 & 0.198 & 2.50 & 3.867 & 1.870 & 0.986 & 0. & 0.247 & 0.132 & 6.185 & 0.257 & 3.820 & 1.830 & 0.986 & 0.  \\
 1 & 3.0 & 8.518 & 0.322 & 3.00 & 3.852 & 2.140 & 0.986 & 0. & 0.247 & 0.132 & 5.573 & 0.407 & 3.807 & 2.089 & 0.986 & 0.  \\
 1 & 4.0 & 8.193 & 0.652 & 4.00 & 3.836 & 2.563 & 0.986 & 0. & 0.247 & 0.132 & 4.885 & 0.833 & 3.788 & 2.505 & 0.986 & 0.  \\
 1 & 5.0 & 7.955 & 1.135 & 5.00 & 3.821 & 2.882 & 0.986 & 0. & 0.247 & 0.132 & 4.458 & 1.530 & 3.767 & 2.822 & 0.986 & 0.  \\
 2 & 5.0 & 8.010 & 2.780 & 4.97 & 3.746 & 3.058 & 0.335 & 0. & 1.27  & 0.442 & 5.956 & 1.771 & 3.818 & 3.115 & 0.305 & 0.  \\
 3 & 5.0 & 8.024 & 1.951 & 4.97 & 3.810 & 3.133 & 0.196 & 0. & 1.27  & 0.442 & 6.349 & 3.179 & 3.743 & 3.113 & 0.998 & 0.  \\
 1 & 7.0 & 7.628 & 2.739 & 7.00 & 3.800 & 3.386 & 0.986 & 0. & 0.247 & 0.132 & 3.821 & 3.938 & 3.741 & 3.334 & 0.986 & 0.  \\
 2 & 7.0 & 7.634 & 7.757 & 6.93 & 3.721 & 3.596 & 0.307 & 0. & 1.30  & 0.453 & 5.038 & 4.741 & 3.788 & 3.619 & 0.283 & 0.  \\
 3 & 7.0 & 7.697 & 5.492 & 6.91 & 3.781 & 3.666 & 0.021 & 0. & 1.30  & 0.453 & 4.639 & 8.577 & 3.718 & 3.634 & 0.013 & 0.  \\
 1 & 9.0 & 7.426 & 5.539 & 8.99 & 3.783 & 3.766 & 0.986 & 0. & 0.247 & 0.132 & 3.443 & 8.858 & 3.709 & 3.704 & 0.986 & 0.  \\
 2 & 9.0 & 7.465 & 19.41 & 8.81 & 3.685 & 4.005 & 0.232 & 0. & 1.38  & 0.482 & 3.998 & 10.99 & 3.762 & 4.024 & 0.229 & 0.  \\
 3 & 9.0 & 7.479 & 8.563 & 8.80 & 3.788 & 4.006 & 0.002 & 0. & 1.38  & 0.482 & 3.923 & 14.00 & 3.716 & 3.957 & 0.000 & 0.  \\
 1 & 12.0 & 7.192 & 12.23 & 11.94 & 3.762 & 4.183 & 0.984 & 0. & 0.247 & 0.132 & 3.076 & 20.47 & 3.680 & 4.110 & 0.984 & 0.  \\
 1 & 15.0 & 7.066 & 50.80 & 14.63 & 3.692 & 4.723 & 0.634 & 0. & 0.247 & 0.132 & 3.976 & 83.19 & 3.618 & 4.698 & 0.625 & 0.  \\
                           & & \multicolumn{14}{c}{$\omega_{\rm ini} = 0.5$}  \smallskip \\
 1 & 2.0 & 9.106 & 0.116 & 2.00 & 3.877 & 1.556 & 0.986 & 180.0 & 0.287 & 0.148 & 7.308 & 0.182 & 3.823 & 1.593 & 0.986 & 80.8  \\
 1 & 2.5 & 8.833 & 0.261 & 2.50 & 3.856 & 1.994 & 0.986 & 69.5 & 0.300 & 0.153 & 6.226 & 0.351 & 3.807 & 1.965 & 0.986 & 56.1  \\
 1 & 3.0 & 8.613 & 0.405 & 3.00 & 3.854 & 2.291 & 0.986 & 55.8 & 0.319 & 0.161 & 5.714 & 0.575 & 3.794 & 2.242 & 0.986 & 42.9  \\
 1 & 4.0 & 8.283 & 0.887 & 4.00 & 3.828 & 2.712 & 0.986 & 37.2 & 0.364 & 0.179 & 4.976 & 1.265 & 3.767 & 2.652 & 0.986 & 28.2  \\
 1 & 5.0 & 8.046 & 1.562 & 5.00 & 3.815 & 3.039 & 0.986 & 28.0 & 0.435 & 0.206 & 4.531 & 2.369 & 3.746 & 2.976 & 0.986 & 20.2  \\
2* & 5.0 & 8.086 & 3.914 & 4.97 & 3.730 & 3.171 & 0.362 & 23.5 & 2.92  & 0.765 & 6.609 & 4.150 & 3.742 & 3.249 & 0.139 & 18.0  \\
3* & 5.0 & 8.100 & 4.461 & 4.96 & 3.738 & 3.271 & 0.228 & 16.9 & 2.92  & 0.765 & 5.901 & 4.461 & 3.718 & 3.253 & 0.121 & 11.4  \\
 1 & 7.0 & 7.715 & 3.895 & 7.00 & 3.792 & 3.539 & 0.986 & 16.8 & 0.626 & 0.271 & 3.879 & 6.303 & 3.718 & 3.481 & 0.986 & 11.7  \\
 2 & 7.0 & 7.748 & 10.85 & 6.92 & 3.706 & 3.712 & 0.392 & 29.0 & 3.46  & 0.825 & 4.750 & 5.970 & 3.784 & 3.732 & 0.382 & 44.3  \\
 3 & 7.0 & 7.765 & 7.057 & 6.89 & 3.784 & 3.811 & 0.041 & 30.9 & 3.46  & 0.825 & 4.948 & 13.76 & 3.694 & 3.782 & 0.026 & 17.7  \\
 1 & 9.0 & 7.503 & 8.407 & 8.99 & 3.767 & 3.915 & 0.986 & 12.6 & 0.892 & 0.349 & 3.430 & 13.75 & 3.690 & 3.853 & 0.986 & 8.3  \\
 1 & 12.0 & 7.273 & 22.64 & 11.90 & 3.739 & 4.409 & 0.982 & 14.9 & 1.33 & 0.480 & 3.310 & 43.02 & 3.641 & 4.357 & 0.981 & 8.1  \\
                           & & \multicolumn{14}{c}{$\omega_{\rm ini} = 0.9$}  \smallskip \\
 1 & 2.0 & 9.136 & 0.148 & 2.00 & 3.848 & 1.608 & 0.986 & 154.5 & 0.665 & 0.268 & 7.133 & 0.195 & 3.816 & 1.616 & 0.986 & 120.8  \\
 1 & 2.5 & 8.875 & 0.287 & 2.50 & 3.848 & 2.054 & 0.986 & 132.9 & 0.700 & 0.283 & 6.326 & 0.430 & 3.790 & 2.020 & 0.986 & 94.6  \\
 1 & 3.0 & 8.635 & 0.419 & 3.00 & 3.846 & 2.295 & 0.986 & 104.8 & 0.836 & 0.322 & 5.728 & 0.601 & 3.785 & 2.239 & 0.986 & 78.8  \\
 1 & 4.0 & 8.302 & 0.848 & 4.00 & 3.826 & 2.690 & 0.986 & 72.3 & 1.15  & 0.399 & 4.967 & 1.187 & 3.766 & 2.622 & 0.986 & 55.1  \\
 1 & 5.0 & 8.062 & 1.379 & 5.00 & 3.818 & 2.989 & 0.986 & 55.6 & 1.83  & 0.488 & 4.471 & 1.913 & 3.760 & 2.924 & 0.986 & 43.3  \\
2* & 5.0 & 8.117 & 3.835 & 4.97 & 3.731 & 3.166 & 0.253 & 35.2 & 6.21  & 0.939 & 6.470 & 4.216 & 3.731 & 3.213 & 0.103 & 18.3  \\
3* & 5.0 & 8.127 & 4.650 & 4.96 & 3.729 & 3.254 & 0.268 & 17.8 & 6.21  & 0.939 & 6.007 & 4.434 & 3.730 & 3.233 & 0.134 & 23.3  \\
 1 & 7.0 & 7.727 & 3.102 & 7.00 & 3.797 & 3.446 & 0.986 & 36.6 & 3.01  & 0.659 & 3.925 & 5.063 & 3.716 & 3.363 & 0.986 & 24.7  \\
 2 & 7.0 & 7.761 & 8.959 & 6.95 & 3.710 & 3.638 & 0.488 & 46.1 & 9.57  & 1.03  & 4.963 & 5.076 & 3.784 & 3.664 & 0.475 & 71.6  \\
 3 & 7.0 & 7.787 & 6.373 & 6.91 & 3.787 & 3.777 & 0.063 & 40.8 & 9.57  & 1.03  & 5.128 & 12.87 & 3.692 & 3.741 & 0.038 & 22.0  \\
 1 & 9.0 & 7.515 & 7.043 & 8.99 & 3.781 & 3.884 & 0.982 & 22.9 & 4.81  & 0.796 & 3.788 & 12.90 & 3.694 & 3.837 & 0.981 & 14.1  \\
 1 & 12.0 & 7.312 & 36.36 & 11.76 & 3.713 & 4.560 & 0.508 & 14.7 & 7.39 & 0.948 & 4.370 & 65.83 & 3.627 & 4.536 & 0.496 & 5.0  \\
\hline
\end{tabular}
\leftline{$^*$ ~Model returns to same side of IS before crossing next expected boundary.}
\label{tab:models_z14_FO}
\end{table*}

\begin{table*}
\centering
\caption{Parameters of first overtone pulsation models entering and exiting IS for $Z=0.006$ }
\begin{tabular}{@{}c@{\hspace{2mm}}r@{\hspace{6mm}}r@{\hspace{2mm}}r@{\hspace{2mm}}r@{\hspace{2mm}}r@{\hspace{2mm}}r@{\hspace{2mm}}r@{\hspace{2mm}}r@{\hspace{6mm}}r@{\hspace{2mm}}r@{\hspace{2mm}}r@{\hspace{2mm}}r@{\hspace{2mm}}r@{\hspace{2mm}}r@{\hspace{2mm}}r@{\hspace{2mm}}r@{}}
\hline
\vspace{2mm}  & & \multicolumn{7}{c}{{\bf Entering stage}} & \multicolumn{7}{c}{{\bf Exiting stage}}  \\
Xing & $M_{\rm{ini}}$ & $\log t$ & P~~ & $M$~~ & $\log T_{\rm eff}$ & $\log L$ & $Y_{\rm c}$ & $V_{\rm e}$~~~~ & N/C & N/O & $\log \Delta t$ & P~~  & $\log T_{\rm eff}$ & $\log L$ & $Y_{\rm c}$ & $V_{\rm e}$~~~~\\
 & [$M_\odot$] & [yr] & [d] & [$M_\odot$] & $[K]$ & [$L_\odot$] & & [km/s] & num & num & [yr] & [d] & [K] & [$L_\odot$] & & [km/s] \vspace{2mm} \\
\hline
                           & & \multicolumn{14}{c}{$\omega_{\rm ini} = 0.0$}  \smallskip \\
 1 & 1.7 & 9.123 & 0.0867& 1.70 & 3.886 & 1.368 & 0.994 & 0. & 0.247 & 0.132 & 7.273 & 0.121 & 3.838 & 1.364 & 0.994 & 0. \\			   
 1 & 2.0 & 8.922 & 0.139 & 2.00 & 3.874 & 1.639 & 0.994 & 0. & 0.247 & 0.132 & 6.505 & 0.184 & 3.828 & 1.611 & 0.994 & 0. \\
 1 & 2.5 & 8.658 & 0.238 & 2.50 & 3.868 & 1.983 & 0.994 & 0. & 0.247 & 0.132 & 5.871 & 0.324 & 3.812 & 1.933 & 0.994 & 0. \\
 1 & 3.0 & 8.454 & 0.385 & 3.00 & 3.852 & 2.248 & 0.994 & 0. & 0.247 & 0.132 & 5.354 & 0.509 & 3.801 & 2.196 & 0.994 & 0. \\
 1 & 4.0 & 8.150 & 0.772 & 4.00 & 3.837 & 2.667 & 0.994 & 0. & 0.247 & 0.132 & 4.763 & 1.064 & 3.780 & 2.610 & 0.994 & 0. \\
2* & 4.0 & 8.230 & 2.159 & 3.97 & 3.763 & 2.916 & 0.205 & 0. & 1.24  & 0.411 & 6.065 & 2.260 & 3.754 & 2.905 & 0.106 & 0. \\
 1 & 5.0 & 7.929 & 1.344 & 5.00 & 3.824 & 2.992 & 0.994 & 0. & 0.247 & 0.132 & 4.386 & 2.008 & 3.758 & 2.933 & 0.994 & 0. \\
 1 & 7.0 & 7.624 & 3.383 & 7.00 & 3.799 & 3.494 & 0.994 & 0. & 0.247 & 0.132 & 3.722 & 5.029 & 3.739 & 3.450 & 0.994 & 0. \\
 2 & 7.0 & 7.667 & 10.76 & 6.92 & 3.712 & 3.728 & 0.337 & 0. & 1.18  & 0.394 & 4.958 & 6.127 & 3.785 & 3.740 & 0.314 & 0. \\
 3 & 7.0 & 7.679 & 6.395 & 6.90 & 3.785 & 3.764 & 0.096 & 0. & 1.18  & 0.394 & 5.222 & 11.13 & 3.711 & 3.743 & 0.064 & 0. \\
 1 & 9.0 & 7.421 & 6.795 & 8.99 & 3.780 & 3.860 & 0.994 & 0. & 0.247 & 0.132 & 3.326 & 10.46 & 3.714 & 3.809 & 0.993 & 0. \\
 1 & 12.0& 7.191 & 18.74 & 11.95& 3.749 & 4.352 & 0.987 & 0. & 0.247 & 0.132 & 3.223 & 28.47 & 3.690 & 4.329 & 0.986 & 0. \\
                           & & \multicolumn{14}{c}{$\omega_{\rm ini} = 0.5$}  \smallskip \\
 1 & 1.7 & 9.215 & 0.100 & 1.70 & 3.880 & 1.445 & 0.994 & 106.1& 0.349 & 0.171 & 7.489 & 0.153 & 3.827 & 1.466 & 0.994 & 81.2 \\ 
 1 & 2.0 & 9.016 & 0.174 & 2.00 & 3.868 & 1.753 & 0.994 & 80.9 & 0.384 & 0.185 & 6.782 & 0.244 & 3.818 & 1.736 & 0.994 & 64.4 \\
 1 & 2.5 & 8.746 & 0.310 & 2.50 & 3.859 & 2.108 & 0.994 & 58.3 & 0.411 & 0.196 & 5.886 & 0.442 & 3.801 & 2.069 & 0.994 & 45.1 \\
 1 & 3.0 & 8.537 & 0.496 & 3.00 & 3.848 & 2.382 & 0.994 & 45.3 & 0.463 & 0.216 & 5.418 & 0.707 & 3.789 & 2.335 & 0.994 & 34.8 \\
 1 & 4.0 & 8.229 & 0.952 & 4.00 & 3.840 & 2.802 & 0.994 & 33.1 & 0.589 & 0.260 & 4.850 & 1.459 & 3.770 & 2.744 & 0.994 & 24.0 \\
2* & 4.0 & 8.293 & 3.307 & 3.96 & 3.746 & 3.068 & 0.242 & 17.8 & 3.31  & 0.796 & 5.746 & 3.218 & 3.746 & 3.055 & 0.178 & 16.8 \\
 1 & 5.0 & 8.005 & 1.728 & 5.00 & 3.821 & 3.122 & 0.994 & 24.6 & 0.773 & 0.319 & 4.409 & 2.756 & 3.749 & 3.065 & 0.994 & 17.1 \\
2* & 5.0 & 8.059 & 5.824 & 4.94 & 3.736 & 3.397 & 0.201 & 12.6 & 3.64  & 0.833 & 5.707 & 6.468 & 3.722 & 3.395 & 0.133 & 10.4 \\
3* & 5.0 & 8.062 & 6.911 & 4.94 & 3.725 & 3.439 & 0.216 & 10.6 & 3.64  & 0.833 & 5.803 & 6.894 & 3.720 & 3.419 & 0.108 &  9.0 \\
 1 & 7.0 & 7.696 & 4.625 & 7.00 & 3.788 & 3.614 & 0.994 & 14.0 & 1.17  & 0.423 & 3.742 & 7.277 & 3.720 & 3.566 & 0.994 &  9.5 \\
 2 & 7.0 & 7.723 & 11.79 & 6.93 & 3.719 & 3.809 & 0.511 & 28.8 & 4.49  & 0.908 & 4.354 & 7.877 & 3.770 & 3.817 & 0.506 & 39.4 \\
 3 & 7.0 & 7.744 & 9.281 & 6.89 & 3.775 & 3.914 & 0.028 & 23.3 & 4.49  & 0.908 & 4.577 & 15.34 & 3.708 & 3.897 & 0.020 & 14.8 \\
 1 & 9.0 & 7.495 & 9.585 & 8.99 & 3.768 & 3.991 & 0.992 & 10.4 & 1.65  & 0.521 & 3.302 & 14.47 & 3.709 & 3.957 & 0.991 &  7.2 \\
 2 & 9.0 & 7.520 & 32.25 & 8.76 & 3.662 & 4.198 & 0.450 & 16.2 & 5.28  & 0.973 & 3.911 & 15.47 & 3.760 & 4.216 & 0.447 & 48.8 \\
 3 & 9.0 & 7.536 & 18.95 & 8.73 & 3.746 & 4.245 & 0.041 & 20.6 & 5.28  & 0.973 & 4.206 & 29.40 & 3.684 & 4.229 & 0.036 & 11.2 \\
                           & & \multicolumn{14}{c}{$\omega_{\rm ini} = 0.9$}  \smallskip \\
 1 & 1.7 & 9.235 & 0.0997& 1.70 & 3.872 & 1.448 & 0.994 & 198.7 & 1.16 & 0.365 & 7.536 & 0.158 & 3.817 & 1.473 & 0.994 & 145.0 \\
 1 & 2.0 & 9.036 & 0.165 & 2.00 & 3.868 & 1.755 & 0.994 & 155.0 & 1.37 & 0.411 & 6.754 & 0.248 & 3.808 & 1.732 & 0.994 & 117.3 \\
 1 & 2.5 & 8.764 & 0.296 & 2.50 & 3.858 & 2.100 & 0.994 & 115.9 & 1.47 & 0.448 & 5.964 & 0.436 & 3.795 & 2.050 & 0.994 & 85.6 \\
 1 & 3.0 & 8.555 & 0.470 & 3.00 & 3.846 & 2.360 & 0.994 & 89.4 & 1.74  & 0.503 & 5.467 & 0.664 & 3.787 & 2.303 & 0.994 & 67.9 \\
 1 & 4.0 & 8.243 & 0.886 & 4.00 & 3.832 & 2.738 & 0.994 & 64.0 & 2.67  & 0.632 & 4.764 & 1.217 & 3.777 & 2.679 & 0.994 & 49.9 \\  
2* & 4.0 & 8.300 & 2.458 & 3.98 & 3.752 & 2.943 & 0.363 & 10.5 & 8.85  & 1.02  & 6.812 & 2.939 & 3.749 & 3.021 & 0.162 & 13.5 \\
 1 & 5.0 & 8.017 & 1.463 & 5.00 & 3.821 & 3.033 & 0.994 & 49.3 & 4.10  & 0.764 & 4.399 & 2.181 & 3.754 & 2.967 & 0.994 & 36.1 \\
 2 & 5.0 & 8.040 & 3.339 & 4.99 & 3.744 & 3.149 & 0.779 & 40.3 & 11.0  & 1.07  & 6.081 & 2.080 & 3.818 & 3.213 & 0.718 & 53.4 \\
 3 & 5.0 & 8.083 & 3.746 & 4.96 & 3.798 & 3.431 & 0.099 & 27.3 & 11.0  & 1.07  & 5.659 & 6.794 & 3.721 & 3.418 & 0.045 & 15.9 \\
1* & 7.0 & 7.708 & 4.514 & 7.00 & 3.787 & 3.598 & 0.986 & 25.3 & 8.07  & 0.967 & 4.928 & 4.699 & 3.789 & 3.627 & 0.979 & 22.1 \\
 2 & 7.0 & 7.767 & 11.18 & 6.94 & 3.769 & 3.987 & 0.026 &  5.0 & 8.86  & 0.994 & 4.636 & 20.77 & 3.681 & 3.963 & 0.016 & 6.5 \\
 1 & 9.0 & 7.565 & 22.26 & 8.94 & 3.746 & 4.347 & 0.014 &  7.6 & 14.6  & 1.13  & 4.127 & 41.49 & 3.658 & 4.324 & 0.010 & 6.7 \\
\hline
\end{tabular}
\leftline{$^*$ ~Model returns to same side of IS before crossing next expected boundary.}
\label{tab:models_z06_FO}
\end{table*}

\begin{table*}
\centering
\caption{Parameters of first overtone pulsation models entering and exiting IS for $Z=0.002$ }
\begin{tabular}{@{}c@{\hspace{2mm}}r@{\hspace{6mm}}r@{\hspace{2mm}}r@{\hspace{2mm}}r@{\hspace{2mm}}r@{\hspace{2mm}}r@{\hspace{2mm}}r@{\hspace{2mm}}r@{\hspace{6mm}}r@{\hspace{2mm}}r@{\hspace{2mm}}r@{\hspace{2mm}}r@{\hspace{2mm}}r@{\hspace{2mm}}r@{\hspace{2mm}}r@{\hspace{2mm}}r@{}}
\hline
\vspace{2mm}  & & \multicolumn{7}{c}{{\bf Entering stage}} & \multicolumn{7}{c}{{\bf Exiting stage}}  \\
Xing & $M_{\rm{ini}}$ & $\log t$ & P~~ & $M$~~ & $\log T_{\rm eff}$ & $\log L$ & $Y_{\rm c}$ & $V_{\rm e}$~~~~ & N/C & N/O & $\log \Delta t$ & P~~  & $\log T_{\rm eff}$ & $\log L$ & $Y_{\rm c}$ & $V_{\rm e}$~~~~\\
 & [$M_\odot$] & [yr] & [d] & [$M_\odot$] & $[K]$ & [$L_\odot$] & & [km/s] & num & num & [yr] & [d] & [K] & [$L_\odot$] & & [km/s] \vspace{2mm} \\
\hline
                           & & \multicolumn{14}{c}{$\omega_{\rm ini} = 0.0$}  \smallskip \\
1 & 1.7 & 9.048 & 0.114 & 1.70 & 3.885 & 1.521 & 0.998 & 0. & 0.247 & 0.132 & 6.742 & 0.161 & 3.831 & 1.500 & 0.998 & 0. \\ 
1 & 2.0 & 8.844 & 0.174 & 2.00 & 3.873 & 1.763 & 0.998 & 0. & 0.247 & 0.132 & 6.155 & 0.239 & 3.821 & 1.731 & 0.998 & 0. \\
1 & 2.5 & 8.589 & 0.302 & 2.50 & 3.861 & 2.095 & 0.998 & 0. & 0.247 & 0.132 & 5.571 & 0.420 & 3.805 & 2.051 & 0.998 & 0. \\
1 & 3.0 & 8.395 & 0.467 & 3.00 & 3.853 & 2.363 & 0.998 & 0. & 0.247 & 0.132 & 5.202 & 0.674 & 3.791 & 2.312 & 0.998 & 0. \\
2 & 3.0 & 8.460 & 1.209 & 2.98 & 3.774 & 2.564 & 0.451 & 0. & 1.22  & 0.371 & 7.298 & 0.884 & 3.845 & 2.689 & 0.225 & 0. \\
3 & 3.0 & 8.490 & 0.906 & 2.98 & 3.841 & 2.686 & 0.191 & 0. & 1.22  & 0.371 & 6.447 & 1.537 & 3.770 & 2.674 & 0.058 & 0. \\
1 & 4.0 & 8.106 & 0.894 & 4.00 & 3.846 & 2.789 & 0.998 & 0. & 0.247 & 0.132 & 4.707 & 1.414 & 3.772 & 2.733 & 0.998 & 0. \\
2 & 4.0 & 8.147 & 2.357 & 3.98 & 3.760 & 2.954 & 0.586 & 0. & 0.990 & 0.326 & 6.246 & 1.517 & 3.826 & 2.996 & 0.497 & 0. \\
3 & 4.0 & 8.187 & 1.941 & 3.97 & 3.820 & 3.105 & 0.021 & 0. & 0.990 & 0.326 & 5.253 & 3.192 & 3.753 & 3.081 & 0.009 & 0. \\
1 & 5.0 & 7.898 & 1.682 & 5.00 & 3.823 & 3.110 & 0.998 & 0. & 0.247 & 0.132 & 4.229 & 2.555 & 3.759 & 3.069 & 0.998 & 0. \\
2 & 5.0 & 7.926 & 4.046 & 4.98 & 3.751 & 3.273 & 0.733 & 0. & 0.861 & 0.303 & 5.738 & 2.575 & 3.813 & 3.301 & 0.680 & 0. \\
3 & 5.0 & 7.968 & 3.428 & 4.96 & 3.805 & 3.415 & 0.005 & 0. & 0.861 & 0.303 & 4.495 & 5.391 & 3.742 & 3.386 & 0.002 & 0. \\ 
1 & 7.0 & 7.609 & 4.279 & 7.00 & 3.795 & 3.604 & 0.997 & 0. & 0.247 & 0.132 & 3.610 & 6.367 & 3.738 & 3.570 & 0.997 & 0. \\
2 & 7.0 & 7.627 & 11.35 & 6.97 & 3.714 & 3.777 & 0.769 & 0. & 0.920 & 0.314 & 4.266 & 7.035 & 3.777 & 3.789 & 0.765 & 0. \\
3 & 7.0 & 7.665 & 8.313 & 6.94 & 3.769 & 3.837 & 0.000 & 0. & 0.920 & 0.314 & 3.675 & 12.03 & 3.709 & 3.794 & 0.000 & 0. \\
1 & 9.0 & 7.413 & 9.126 & 8.99 & 3.769 & 3.974 & 0.995 & 0. & 0.247 & 0.132 & 3.327 & 13.85 & 3.708 & 3.940 & 0.994 & 0. \\
2 & 9.0 & 7.426 & 19.49 & 8.90 & 3.718 & 4.156 & 0.766 & 0. & 1.05  & 0.335 & 3.532 & 13.85 & 3.763 & 4.162 & 0.765 & 0. \\
3 & 9.0 & 7.461 & 16.97 & 8.89 & 3.740 & 4.169 & 0.000 & 0. & 1.05  & 0.335 & 3.426 & 23.64 & 3.694 & 4.166 & 0.000 & 0. \\
                           & & \multicolumn{14}{c}{$\omega_{\rm ini} = 0.5$}  \smallskip \\
1 & 1.7 & 9.138 & 0.140 & 1.70 & 3.877 & 1.622 & 0.998 & 79.3 & 0.607 & 0.251 & 6.834 & 0.205 & 3.822 & 1.611 & 0.998 & 63.1 \\
1 & 2.0 & 8.934 & 0.217 & 2.00 & 3.868 & 1.881 & 0.998 & 64.0 & 0.702 & 0.284 & 6.289 & 0.316 & 3.811 & 1.857 & 0.998 & 49.6 \\
1 & 2.5 & 8.671 & 0.355 & 2.50 & 3.867 & 2.219 & 0.998 & 47.8 & 0.807 & 0.320 & 5.700 & 0.557 & 3.796 & 2.177 & 0.998 & 34.9 \\
1 & 3.0 & 8.473 & 0.594 & 3.00 & 3.847 & 2.481 & 0.998 & 36.5 & 0.992 & 0.378 & 5.224 & 0.907 & 3.780 & 2.436 & 0.998 & 26.7 \\
2*& 3.0 & 8.534 & 1.876 & 2.97 & 3.766 & 2.762 & 0.285 & 12.1 & 4.17  & 0.853 & 6.750 & 2.113 & 3.763 & 2.812 & 0.132 & 5.45 \\
3*& 3.0 & 8.543 & 2.248 & 2.97 & 3.762 & 2.837 & 0.151 & 5.0  & 4.17  & 0.853 & 5.581 & 2.190 & 3.762 & 2.827 & 0.125 & 5.04 \\
4*& 3.0 & 8.545 & 2.384 & 2.96 & 3.763 & 2.876 & 0.220 & 26.9 & 4.17  & 0.853 & 5.907 & 2.295 & 3.761 & 2.850 & 0.148 & 26.9 \\
1 & 4.0 & 8.181 & 1.200 & 4.00 & 3.831 & 2.892 & 0.998 & 25.5 & 1.30  & 0.451 & 4.605 & 1.791 & 3.770 & 2.852 & 0.998 & 18.9 \\
2 & 4.0 & 8.208 & 2.841 & 3.98 & 3.757 & 3.043 & 0.708 & 13.1 & 3.90  & 0.832 & 6.392 & 1.859 & 3.824 & 3.102 & 0.591 & 12.5 \\
3 & 4.0 & 8.242 & 2.915 & 3.96 & 3.812 & 3.289 & 0.085 & 4.5  & 3.90  & 0.832 & 5.602 & 5.104 & 3.744 & 3.288 & 0.046 &  6.2 \\
1 & 5.0 & 7.970 & 2.098 & 5.00 & 3.818 & 3.215 & 0.998 & 19.7 & 1.72  & 0.542 & 4.238 & 3.296 & 3.752 & 3.175 & 0.997 & 13.8 \\
2 & 5.0 & 7.983 & 4.889 & 4.99 & 3.739 & 3.326 & 0.879 & 13.8 & 4.15  & 0.863 & 5.759 & 2.922 & 3.811 & 3.362 & 0.839 & 15.5 \\
3 & 5.0 & 8.027 & 6.372 & 4.95 & 3.778 & 3.625 & 0.016 & 3.6  & 4.15  & 0.863 & 4.608 & 8.725 & 3.737 & 3.619 & 0.011 &  3.8 \\
1 & 7.0 & 7.677 & 6.301 & 7.00 & 3.778 & 3.735 & 0.994 & 9.7  & 2.60  & 0.676 & 3.754 & 9.003 & 3.730 & 3.721 & 0.993 &  7.1 \\
2 & 7.0 & 7.682 & 11.77 & 6.99 & 3.723 & 3.833 & 0.934 & 21.5 & 5.49  & 0.961 & 4.014 & 7.899 & 3.775 & 3.843 & 0.933 & 30.6 \\
3 & 7.0 & 7.723 & 10.75 & 6.94 & 3.777 & 4.013 & 0.000 & 23.2 & 5.49  & 0.961 & 3.972 & 16.05 & 3.720 & 3.984 & 0.000 & 15.2 \\
1 & 9.0 & 7.524 & 20.49 & 8.97 & 3.752 & 4.338 & 0.000 & 5.1  & 3.52  & 0.780 & 3.575 & 33.93 & 3.685 & 4.321 & 0.000 &  4.3 \\
1 & 12.0& 7.302 & 36.48 & 11.89& 3.747 & 4.725 & 0.002 & 4.9  & 4.06  & 0.849 & 3.660 & 69.58 & 3.653 & 4.684 & 0.000 &  2.6 \\
1 & 15.0& 7.166 & 49.30 & 14.75& 3.745 & 4.948 & 0.000 & 4.1  & 4.23  & 0.893 & 4.105 & 141.4 & 3.616 & 4.943 & 0.000 &  1.5 \\
                           & & \multicolumn{14}{c}{$\omega_{\rm ini} = 0.9$}  \smallskip \\
1 & 1.7 & 9.197 & 0.137 & 1.70 & 3.877 & 1.632 & 0.998 &141.6 & 5.59  & 0.659 & 7.004 & 0.209 & 3.817 & 1.616 & 0.998 & 106.1 \\
1 & 2.0 & 8.958 & 0.205 & 2.00 & 3.868 & 1.869 & 0.998 &116.6 & 4.04  & 0.651 & 6.477 & 0.305 & 3.809 & 1.839 & 0.998 & 90.3 \\
1 & 2.5 & 8.688 & 0.348 & 2.50 & 3.857 & 2.179 & 0.998 & 88.6 & 4.81  & 0.756 & 5.785 & 0.510 & 3.796 & 2.135 & 0.998 & 66.6 \\
1 & 3.0 & 8.495 & 0.548 & 3.00 & 3.849 & 2.447 & 0.998 & 55.1 & 6.65  & 0.880 & 5.281 & 0.803 & 3.787 & 2.399 & 0.998 & 42.5 \\
2*& 3.0 & 8.547 & 1.485 & 2.98 & 3.772 & 2.666 & 0.499 & 17.3 & 15.4  & 1.11  & 7.136 & 2.021 & 3.764 & 2.793 & 0.235 & 20.2 \\
1 & 4.0 & 8.195 & 0.996 & 4.00 & 3.835 & 2.812 & 0.998 & 50.4 & 10.5  & 1.07  & 4.714 & 1.477 & 3.772 & 2.759 & 0.997 & 37.3 \\
2 & 4.0 & 8.200 & 1.825 & 4.00 & 3.765 & 2.851 & 0.967 & 55.4 & 15.8  & 1.18  & 5.909 & 1.183 & 3.831 & 2.898 & 0.955 & 71.9 \\
3 & 4.0 & 8.263 & 2.801 & 3.97 & 3.811 & 3.270 & 0.020 & 41.2 & 15.8  & 1.18  & 5.217 & 4.696 & 3.745 & 3.257 & 0.008 & 30.5 \\
1 & 5.0 & 8.049 & 5.964 & 5.00 & 3.782 & 3.613 & 0.002 & 23.4 & 16.7  & 1.21  & 4.216 & 9.285 & 3.720 & 3.586 & 0.000 & 15.2 \\
1*& 7.0 & 7.759 & 14.38 & 6.99 & 3.760 & 4.094 & 0.000 & 13.4 & 33.5  & 1.42  & 4.251 & 14.78 & 3.760 & 4.114 & 0.000 & 14.4 \\
2 & 7.0 & 7.760 & 13.98 & 6.99 & 3.762 & 4.093 & 0.000 & 14.5 & 34.1  & 1.42  & 4.172 & 22.48 & 3.692 & 4.051 & 0.000 &  8.3 \\
1 & 9.0 & 7.577 & 22.23 & 8.97 & 3.764 & 4.438 & 0.000 &  8.8 & 53.6  & 1.58  & 4.082 & 56.12 & 3.658 & 4.473 & 0.000 &  4.2 \\
1 &12.0 & 7.351 & 44.15 &11.84 & 3.744 & 4.822 & 0.003 &  7.9 & 55.7  & 1.59  & 3.720 & 91.55 & 3.638 & 4.766 & 0.000 &  4.4 \\
\hline
\end{tabular}
\leftline{$^*$ ~Model returns to same side of IS before crossing next expected boundary.}
\label{tab:models_z02_FO}
\end{table*}

\end{appendix}

\end{document}